\documentclass[longauth]{aa} % for the long lists of affiliations 
\usepackage{graphicx}
\usepackage{txfonts}
\usepackage{xcolor}
\usepackage{multicol}
\usepackage{graphicx,subcaption}
\usepackage{tikz}
\def \1s{$1\,\sigma$}
\def \t0{T$_0$}

\begin{document}

   \title{Characterizing planetary systems with SPIRou: Detection of a sub-Neptune in a 6-day period orbit
   around the M~dwarf Gl~410\thanks{Based on observations obtained with the spectropolarimeter SPIRou at the Canada-France-Hawaii Telescope (CFHT).
	  }
          }

%   \subtitle{
  %        }

     \author{A. Carmona
          \inst{1}
          \and
           X. Delfosse \inst{1}
 	 \and
           M. Ould-Elhkim \inst{2} 
         \and
           P. Cort\'es-Zuleta \inst{3}
           \and
          N. C. Hara \inst{4}
         \and
           E. Artigau \inst{5,6}
         \and
	   C. Moutou \inst{2}
          \and
           A. C. Petit \inst{7}
           \and 
           L. Mignon \inst{1}
          \and
          J.F. Donati \inst{2}
          \and
          N.J. Cook \inst{5} 
           \and
          J. Gagn\'e \inst{8,5}
          \and
          T. Forveille \inst{1}
          \and
          R.F. Diaz \inst{9,10}
          \and  
          E. Martioli \inst{11,12} 
          \and
          L. Arnold \inst{13} 	
          \and
          C. Cadieux \inst{5} 
          \and
          I. Boisse \inst{4}
          \and 
          J. Morin \inst{14}
          \and 
          P. Petit \inst{2} 
          \and
          P. Fouqu\'e \inst{2}
          \and
          X. Bonfils \inst{1}
           \and 
          G. H\'ebrard \inst{12}
          \and
          L. Acu\~{n}a \inst{4,15}
          \and
          J.-D.~do Nascimento, Jr. \inst{16,17}
          \and
          R. Cloutier \inst{18}
          \and
          N. Astudillo-Defru\inst{19}
          \and
          F. Bouchy \inst{20}
          \and
          V. Bourrier \inst{20}
          \and
          S. Dalal \inst{12}
          \and
          M. Deleuil \inst{4}
          \and
          R. Doyon \inst{5}
          \and 
          X. Dumusque \inst{20}
          \and 
          S. Grouffal \inst{4}
          \and
          N. Heidari \inst{12}
          \and 
          S. Hoyer \inst{4}
          \and    
          P. Larue \inst{1} 
          \and
          F. Kiefer \inst{12}
          \and
          A. Santerne \inst{4}
          \and
         D. S\'egransan \inst{20}
         \and
         J. Serrano Bell\inst{10}
         \and 
         M. Stalport \inst{21}
         \and
         S. Sulis \inst{4}
         \and
         S. Udry \inst{20}
         \and
         H. G. Vivien \inst{4}
        }
 
   \institute{Univ. Grenoble Alpes, CNRS, IPAG, 38000 Grenoble, France\\
    \email{andres.carmona@univ-grenoble-alpes.fr}     
    \and
      Univ. de Toulouse, CNRS, IRAP, 14 avenue Belin, 31400 Toulouse, France
      \and
     SUPA School of Physics and Astronomy, University of St Andrews, North Haugh, St Andrews KY16 9SS, UK
      \and
     Aix Marseille Universit\'e, CNRS, CNES, Institut Origines, LAM, Marseille, France  
     \and
      Trottier Institute for Research on Exoplanets, Universit\'e de Montr\'eal, D\'epartement de Physique, C.P. 6128 Succ. Centre-ville, Montr\'eal, QC H3C 3J7, Canada
     \and 
     Observatoire du Mont-M\'egantic, Universit\'e de Montr\'eal, D\'epartement de Physique, 
     C.P. 6128 Succ. Centre-ville, Montr\'eal, QC H3C 3J7, Canada
     \and 
     Universit\'e C\^ote d'Azur, Observatoire de la C\^ote d'Azur, CNRS, Laboratoire Lagrange, France
     \and
     Plan\'etarium de Montr\'eal, Espace pour la Vie, 4801 av. Pierre-de Coubertin, Montr\'eal, Qu\'ebec, Canada 
     \and
     Instituto Tecnol\'ogico de Buenos Aires (ITBA), Buenos Aires C1437, Argentina
     \and
      International Center for Advanced Studies and ICIFI (CONICET),
    ECyT-UNSAM, Campus Miguelete, 25 de Mayo y Francia, (1650), Buenos Aires, Argentina
     \and 
     Laborat\'{o}rio Nacional de Astrof\'{i}sica, Rua Estados Unidos 154, 37504-364, Itajub\'{a} - MG, Brazil
        \and
      Institut d'astrophysique de Paris, CNRS, UMR 7095, Sorbonne Universit\'{e}, 98 bis bd Arago, 75014 Paris, France
      \and
     Canada France Hawaii Telescope Corporation (CFHT), UAR2208 CNRS-INSU, 65-1238 Mamalahoa Hwy, Kamuela 96743 HI, USA
     \and
      Universit\'e de Montpellier, CNRS, LUPM, F-34095 Montpellier, France
          \and   
     Max-Planck-Institut f\"ur Astronomie, K\"onigstuhl 17, D-69117 Heidelberg, Germany
    \and
    Center for Astrophysics, Harvard $\&$ Smithsonian, 60 Garden Street, Cambridge, MA 02138, USA
    \and
    Dep. de F\'isica, Univ. Federal do Rio Grande do Norte - UFRN, Natal, RN, 59078-970, Brazil
         \and
     Department of Physics \& Astronomy, McMaster University, 1280 Main St West, Hamilton, ON, L8S 4L8, Canada
     \and
     Departamento de Matem\'atica y F\'isica Aplicadas, Universidad Cat\'olica de la Sant\'isima Concepci\'on, Alonso de Rivera 2850, Concepci\'on, Chile
     \and
      Observatoire Astronomique de l'Universit\'e de Gen\`eve, Chemin Pegasi 51b, 1290 Versoix, Switzerland
    \and
    Space sciences, Technologies and Astrophysics Research (STAR) Institute, Universit\'e de Li\`ege, All\'ee du Six-Ao\^ut 19C, 4000 Li\`ege, Belgium
 }

   \date{}

\titlerunning{A 6-day period planet around the M dwarf Gl~410}
% \abstract{}{}{}{}{} 
% 5 {} token are mandatory
 
 \abstract
   {The search for exoplanets around nearby M dwarfs represents a crucial milestone in  
     the census of planetary systems in the vicinity of our Solar System.
    } 
   {
   Since 2018 our team has been conducting a blind search program for planets around nearby M dwarfs 
   with the near-IR spectro-polarimeter and velocimeter SPIRou at the Canada-France-Hawaii Telescope and with 
   the optical velocimeter SOPHIE at the Haute-Provence Observatory in France.
   The aim of this paper is to present our results on Gl~410, a 0.55 M$_\odot$ 480$\pm$150 Myr old active M dwarf distant 12 pc.
   }
   {
   We searched for planetary companions using radial velocities (RVs).
   We used the line-by-line (LBL) technique to measure the RVs with SPIRou and the template matching method with SOPHIE.
   Three different methods were employed, 
   two based on principal component analysis (PCA),
    to clean the SPIRou RVs for systematics.
   We applied Gaussian processes (GP) modeling to correct the SOPHIE RVs for stellar activity.
   The $\ell_1$ and apodized sine periodogram analysis was used to search for planetary signals in the SPIRou 
   data taking into account activity indicators.
   We analyzed TESS data and searched for planetary transits.
   }
   {
   We report the detection of a $M$\,sin$(i)$=8.4$\pm$1.3 M$_\oplus$ sub-Neptune planet at a period of 6.020$\pm$0.004 days 
   in circular orbit with SPIRou.
   The same signal, although with lower significance, was also retrieved in the SOPHIE RV data after correction for activity
   using a GP trained on SPIRou's longitudinal magnetic field ($B_\ell$) measurements. 
   The TESS data indicate that the planet is not transiting.
   Within the SPIRou wPCA RVs,
   we find tentative evidence for two additional planetary signals at 2.99 and 18.7 days.
   }
   {
   Infrared RVs are a powerful method to detect extrasolar planets around active M dwarfs. 
   Care should be taken, however, to correct or filter 
   systematics generated by residuals of the telluric correction or small structures in the detector plane. 
   The LBL technique combined with PCA offers a promising way to reach this objective.
   Further monitoring of Gl~410 is necessary.}


   \maketitle
%-------------------------------------------------------------------
\section{Introduction}

One of the most interesting open questions in current Astronomy is the census of extrasolar planets in the vicinity of the Solar System.
The identification of the closest extrasolar planetary systems is a crucial milestone for the study of exoplanet atmospheres and biomarkers 
using future instrumentation combining high-dispersion spectroscopy with high contrast imaging \citep[e.g.,][]{Kasper2021,Chauvin2024}.
As the most stars in the Galaxy, and thus in the Solar neighbourhood, are low-mass  stars ($M\le0.5$ M$_\odot$),
in recent years, there has been growing interest in extending the planet search effort from solar-like stars to M dwarfs.

Radial velocity \citep[RV;][]{Bonfils2013,Pinamonti2022,Mignon2025} 
and transit surveys  
\citep[][]{DressingCharbonneau2013,DressingCharbonneau2015,Mulders2015,Hsu2020,MentCharbonneau2023}
have revealed that a large fraction of M dwarfs,
if not all, host a planetary system, and these surveys have provided strong evidence supporting the
hypothesis that the rocky planet ratio per star is indeed larger than one.
Extrapolating the results from field stars to our stellar vicinity
is extremely encouraging,  
as it suggests that it is highly likely that a large fraction of nearby M dwarfs host multi-planetary systems.
Nearby stars are distributed all over the sky at random orientations.
A small fraction of sources are expected to present planets in transit.
Missions such as TESS, and in the future, PLATO will detect a significant fraction of those
transiting planets.
Nevertheless, to establish the planetary systems census of our closest stellar neighbors, 
techniques such as astrometry or RVs need to be used in addition to transits.

Surveys for planetary systems around nearby bright M dwarfs with high-precision velocimeters 
have been carried so far mainly in the optical domain.
Recently, 
the search using RVs has been extended to the near-infrared (IR) domain
with the start of operations of new velocimeters such
as HPF\footnote{Habitable Zone Planet Finder.} \citep[][]{HPF2014}, 
CARMENES\footnote{Calar Alto high-Resolution search for M dwarfs with Exoearths with Near-infrared and optical Echelle Spectrographs.} \citep[][]{Quirrenbach2018}, 
SPIRou\footnote{SPectropolarim\`etre Infra-ROUge.} \citep[][]{Donati2020} 
and NIRPS\footnote{Near Infra Red Planet Searcher.} \citep[][]{Bouchy2017}.
High-precision IR RVs enable the exploration of lower mass stars, 
which are sometimes inaccessible for precise RV work in the optical due to their faintness.
Furthermore, 
IR RVs have the potential of being less affected by stellar activity
than optical RVs, especially for highly or moderately active stars \citep[e.g.,][]{Carmona2023}.

With more than two thousand low-mass M dwarfs in the 25 pc vicinity,
the effort to monitor all of these objects is colossal.
Thanks to large programs with the near-IR spectra-polarimeter and velocimeter SPIRou at CFHT 
(the SPIRou Legacy Survey, the SPICE large program) 
and a series of PI SPIRou follow-up programs,
our team has been monitoring a select sub-sample of around 60 nearby M dwarfs.
At the time of writing, 
we have obtained more than 100 visits per source for around 45 of them. 

In this paper, 
we report the detection of a sub-Neptune mass planet in a 6-day period around the nearby ($\sim$12 pc) 
0.55 M$_\odot$ { 480 Myr old} M dwarf Gl~410 with SPIRou,
and the tentative detection of  two additional planetary signals at 2.99 and 18.7 days.
We complement our near-IR data set with quasi simultaneous 
optical observations obtained with the velocimeter SOPHIE\footnote{
Spectrographe pour l'Observation des Ph\'enom\`enes des Int\'erieurs stellaires et des Exoplan\`etes
(spectrograph for the observation of the phenomena of the stellar interiors and of the exoplanets).}.
The paper is organized as follows.
We describe the physical properties of Gl~410 in Sect. 2.
Then, in Sect. 3, we present the details of the optical and infrared RV measurements.
In Sect. 4, we analyze our RV measurements and test the robustness of the planet detection.
In Sect. 5, we search for the 6-day period signal detected with SPIRou in the SOPHIE data using 
Gaussian processes (GP) activity filtering techniques.
We derive the orbital and physical properties of the detected 6.02 day planet using 
Markov chain Monte Carlo (MCMC) techniques in Sect. 6. 
In Sect. 7, we investigate the presence of other planetary signals within the SPIRou wPCA RVs.
In Sect. 8, we present TESS observations of Gl~410 and investigate whether Gl~410b is transiting.
Finally, we discuss our findings in Sect. 9, and in Sect. 10, we provide our conclusions.

\begin{table}
\begin{center}
\caption{Stellar properties of Gl~410.}
\small
\begin{tabular}{lcccc}
	\hline
	\hline
	\\[-5pt]
        Parameter 		& Value 				& Ref. \\[0.6mm]
        \hline
        \\[-5pt]
        RA (J2000)			& 11:02:38.34			& 1, 2 \\[0.4mm]
        DEC (J2000)			&  +21 58 01.70		& 1, 2 \\[0.4mm]
        Proper motion [mas]		& 142.771 -51.661		& 1, 2 \\[0.4mm]
        Parallaxes				& 83.7642$\pm$0.0204	& 1, 2 \\[0.4mm]
       Distance [pc] 			& 11.9327$^{+0.0026}_{-0.0030}$ 	&  3, 2 \\[0.4mm]
       Effective temperature [K] 	&  3842$\pm$31  				& 4 \\[0.4mm]
        log $g$ 				& 4.87$\pm$0.05				& 4 \\[0.4mm]
        [M/H] 				& 0.05$\pm$ 0.10				& 4 \\[0.4mm]
        [$\alpha$/H]			& 0.05$\pm$ 0.04				& 4 \\[0.4mm]
        Mass [M$_\odot$] 		&  0.55$\pm$0.02  				& 4 \\[0.4mm]
        Radius [R$_\odot$] 		&  0.543$\pm$0.009 				& 4 \\[0.4mm]
	Luminosity [ log $(L_{*}/L_\odot)$ ] 	&  -1.239$\pm$0.002  			& 4 \\[0.4mm]
        Rotational period$^\dagger$ [days] 	
        						& 13.87$\pm$0.08				& 5 \\[0.4mm]
						& 13.91$\pm$0.09 				& 6 \\[0.4mm]
         					& 13.93$\pm$0.09				& this work \\[0.4mm]
        Age [Myr]				&  890$\pm$120				& 5 \\[0.4mm]
        						&  480$\pm$150				& this work \\[0.4mm]
       Brightness [mag]		& $V =  9.572 $ 				& 1, 7 \\[0.4mm]
         	      				& $R =  8.638 $ 				& 1, 7\\[0.4mm]
         	       				& $J =  6.522 $ 				& 1, 8\\[0.4mm]
                					& $H = 5.899 $					& 1, 8\\[0.4mm]
           	     				& $K = 5.688 $ 					& 1, 8\\[0.4mm]
\\[-5pt]
\hline
\end{tabular}
\tablefoot{ 
$^\dagger$ Rotational period derived from SPIRou's longitudinal magnetic field ($B_\ell$).
References:
(1) SIMBAD;
(2) Gaia DR3 \citep[][]{Vallenari2023};
(3) \citet[][]{Bailer-Jones2021} based in Gaia DR3;
(4) \citet[][]{Cristofari2022};
(5) \citet[][]{Fouque2023};
(6) \citet[][]{Donati2023};
(7) \citet[][]{Koen2010};
(8) 2MASS catalogue \citet[][]{Cutri2003}
}
\label{stellar-properties}
\end{center}
\end{table}

\section{Gl410}
\label{gl410_properties}
We summarize the basic physical properties of Gl~410 (DS Leo, HD 95650) in Table~\ref{stellar-properties}.
Gl~410 is a nearby star located at a distance of 11.9327$^{+0.0026}_{-0.0030}$ pc \citep[Gaia DR3, ][]{Bailer-Jones2021,Vallenari2023}.
It has a spectral type M1.0V and a mass of 0.55$\pm$0.02 M$_\odot$ \citep[][]{Cristofari2022}.
Gl~410 is known to exhibit cyclic photometric variability in the optical with periods of about 14 days,
which is interpreted as being caused by rotational modulation of stellar spots \citep[][]{FekelHenry2000}.
The rotational period of Gl~410 has been tightly constrained using spectropolarimetry observations 
in the optical with  ESPaDOnS, NARVAL, and HARPS \citep[][]{Donati2008,Hebrard2016},
and in the near-IR with SPIRou \citep[][]{Fouque2023,Donati2023}.
The latest estimation of $P_{\rm rot}$ 
is of 13.91$\pm$0.09 days \citep[][]{Donati2023}. 
Gl~410 displays surface differential rotation with periods at the equator and pole 
of 13.37$\pm$0.86 and 14.96$\pm$1.25 days, 
respectively \citep[][]{Hebrard2016}.
Based on gyrochronologly \citet[][]{Fouque2023} derived an age of 0.89$\pm$0.1 Gyr  for Gl~410.
However, there is the possibility that Gl~410 is in fact younger.
In Sect.~\ref{age_determination} in the Appendix,
we provide further details.
We find that the empirical rotation-temperature sequences in clusters of known ages indicate that Gl~410 is 480$\pm$150 Myr,
thus placing it among the youngest stars in the solar neighborhood \citep[][]{SilvaAguirre2018,Chen2020}
and suggesting that Gl~410 might be related to the
Ursa Major (UMA) cluster which is approximately 400 Myr old. 

Based on SPIRou observations,
\citet[][]{Cristofari2023} measured an average magnetic small scale field strength <$B$> of 0.71$\pm$ 0.03 kG for Gl~410.
Together with the M dwarfs GJ~1289 and GJ~1286, 
Gl~410 is one of the most magnetic stars among the M dwarfs studied in the <$B$> surveys of \citet[][]{Reiners2022} and \citet[][]{Cristofari2023}.
The magnetic activity of Gl~410 is further indicated by the presence of $H_\alpha$ emission 
in the optical spectra. Gl~410 is part of the top one-third of the sample in terms of $H_\alpha$ emission 
strength among the 331 M stars surveyed by \citet[][]{Schoefer2019}.

Zeeman Doppler imaging (ZDI) investigations by \citet[][]{Donati2008}, 
\citet[][]{Hebrard2016}, and \citet[][]{Bellotti2024} 
have shown that the large-scale magnetic field of Gl~410 rapidly evolves both in strength and in topology.
Between 2007 and 2008, the field had a predominantly toroidal geometry with an average strength of 100 G \citep[][]{Donati2008}.
In 2014, the field remained mostly toroidal with axisymmetric geometry, but the average field strength decreased to 60 G \citep[][]{Hebrard2016}.
In 2020b to 2021a the magnetic energy became almost equally distributed between the poloidal and the toroidal components
and the average field strength decreased to 44 G \citep[][]{Bellotti2024}.
In the latest measurements from 2021b to 2022a, 
the  poloidal component became much stronger than the toroidal component with 73\% of the magnetic energy,
and the average field strength further decreased to 18 G \citep[][]{Bellotti2024}.

\section{Observations}
\subsection{SOPHIE}
\begin{figure}[t]
\centering
\includegraphics[width=0.49\textwidth]{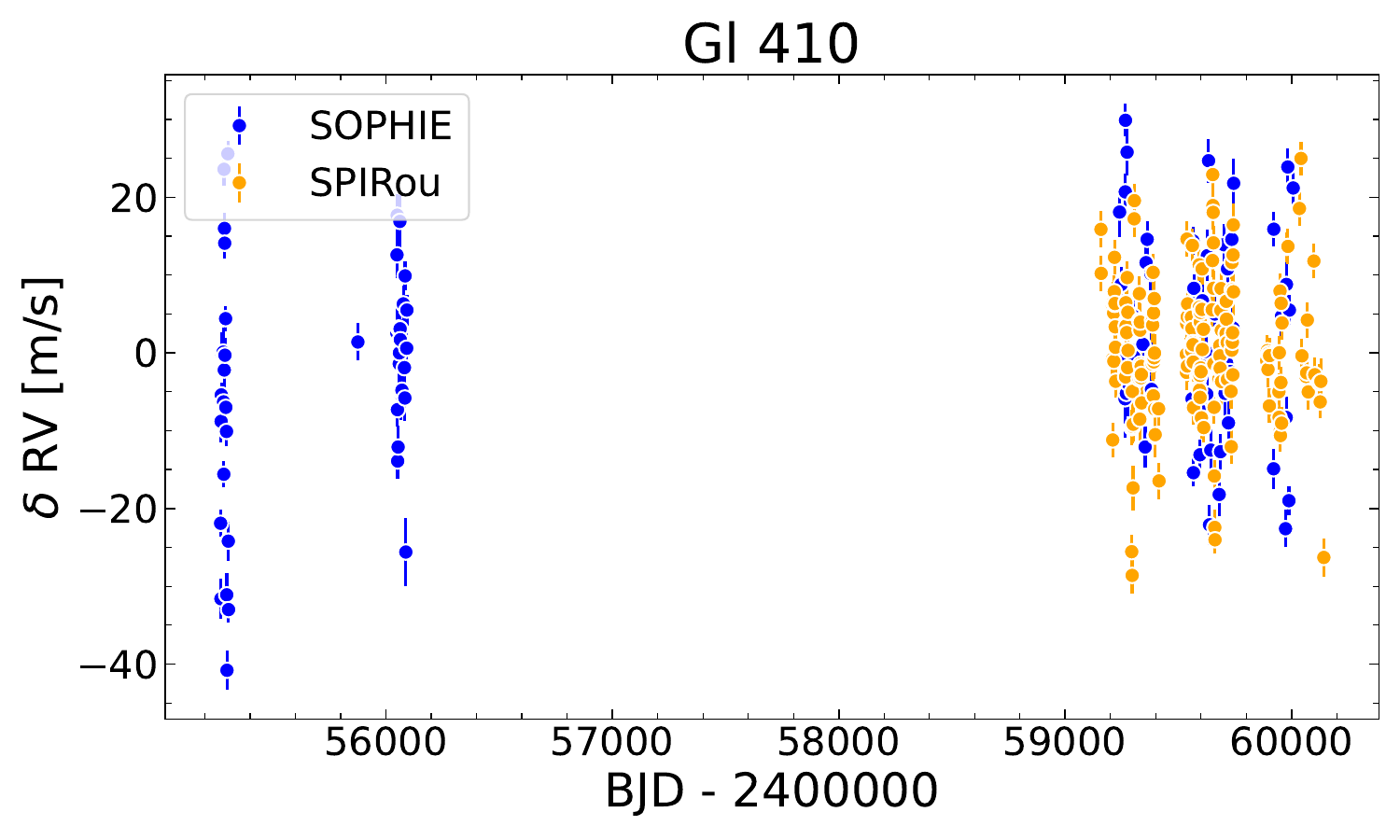}
\caption{Radial velocity time series measured in the optical with SOPHIE (blue dots) and in the near-IR with SPIRou (orange dots).
}
\label{Gl410_RV_data}
\end{figure}

Gl~410 was observed during two periods 9 years apart with the high-resolution optical velocimeter 
SOPHIE
\citep[][]{Perruchot2008,Bouchy2013} 
at the 1.93m Telescope of the Haute-Provence Observatory in the South of France.
The first period with 45 visits was observed from March 14, 2010, to June 14, 2012.
A second observation period was planned to coincide with the SPIRou observations,
with  58 visits performed from January 26, 2021, to  March 04, 2023.
Spectra were reduced using SOPHIE's data reduction system \citep[][]{Bouchy2009a,Heidari2022thesis}.
The RV measurements were obtained using a template-matching technique 
implemented in the code \texttt{NAIRA} \citep[][]{Astudillo2015}, 
which has been adapted for SOPHIE \citep[][]{Hobson2018}.
The SOPHIE RV measurements are given in Table \ref{GL410_SOPHIE_RVtable} 
(the full table is available at the CDS). 
In Table~\ref{table:SOPHIE_RV_periodogram}, 
we provide a summary of the properties of the RV time series.
Figure~\ref{Gl410_RV_data} displays the median subtracted SOPHIE RV data.

\subsection{SPIRou}
Gl~410 was observed with the near-IR velocimeter and polarimeter SPIRou \citep[][]{Donati2020}, 
mounted at the Canada-France-Hawaii Telescope atop Mauna-Kea during 156 visits from April 2018 to July 2023.
Data were obtained in the context of the SPIRou's commissioning in 2018, 
the SPIRou Legacy Survey from 2019 to 2022A,
and a series of PI follow-program programs (PI. Carmona, PI. Artigau) in 2022B, 2023A and 2023B.
Nightly SPIRou observations typically consist of a Stokes V sequence of four exposures,
each taken with different  positions of the Fresnel rhombs inside the polarimeter.
The combination of the channel A and channel B of the four  sub-exposures enables 
the recovery of  the Stokes V (and Stokes I) spectra
for magnetic field analysis.
For each science observation, 
a simultaneous Fabry-P\'erot exposure was obtained in the calibration channel C.
In the context of the RV analysis, 
we used the combined extraction of the science channels A and B and 
each sub-exposure of the Stokes V sequence was treated independently.

For the RV analysis, 
we used the data reduced using the \texttt{APERO} \citep[][]{Cook2022}
pipeline version v07275 installed at the SPIRou Data Centre hosted at the Laboratoire d'Astrophysique de Marseille. 
We note that \texttt{APERO} extracts the combined science channel A and channel B spectra, performs the 
wavelength calibration using nightly observations of the Uranium-Neon lamp and Fabry-P\'erot exposures,
and conducts the telluric correction.
The telluric correction uses a combination of an atmospheric absorption model generated by \texttt{TAPAS} \citep[][]{Bertaux2014},
an empirical template of the star deduced from observations at different barycentric earth radial velocity (BERV)
and a library of nightly observations of telluric standard stars taken since the commissioning of the instrument in 2018.
Further details on the telluric correction are given in 
\citet[][]{Artigau2014},
\citet[][]{Cook2022}
and Artigau et al. (in prep.).
Individual RV measurements were obtained using the LBL technique \citep[][]{Artigau2022},
and they were corrected by the BERV and the drift of the Fabry-P\'erot at each exposure.
A further correction for the instrumental RV zero point was done using a GP model
based on observations of a sample of RV standard stars observed by SPIRou \citep[][]{Cadieux2022}.
The SPIRou LBL RV measurements are given in Table \ref{GL410_SPIROU_RVtable} 
(full table available at the CDS). 

For analysis of the stellar activity, we used the simultaneous longitudinal magnetic ($B_\ell$) field measurements 
obtained with the  \texttt{Libre-ESpRIT} pipeline\footnote{The nominal ESPaDOnS data reduction pipeline \citep[][]{Donati1997} adapted for the SPIRou data \citep[][]{Donati2020}.}
previously published in \citet[][]{Donati2023}.

\begin{table}[t]
\tiny
\caption{SOPHIE RV measurements of Gl~410.}
\centering
\begin{tabular}{c c c c c c}
\hline 
\hline
DATE             & RJD   & RV     & $\sigma_{\rm RV}$ & $\delta$ RV & $\sigma_{\delta{\rm RV}}$   \\[1pt]
[yyyy-mm-dd] & [days] & [km/s] & [km/s]        & [m/s]            & [m/s]         \\[1pt]
\hline
2010-03-14 & 55269.43680 &        -13.8604 &    0.0017 &      -21.9 & 1.7 \\ 
2010-03-15 & 55270.60680 &        -13.8701 &    0.0026 &      -31.6 & 2.6 \\ 
2010-03-16 & 55271.58010 &        -13.8473 &    0.0027 &       -8.8 & 2.7 \\ 
...   &  		...	   & 	...	 &  		...   &	 ...	      &      ... \\
\hline
\end{tabular}
\tablefoot{Full table available at the CDS.}
\label{GL410_SOPHIE_RVtable}
\end{table}

\begin{table*}
\tiny
\caption{SPIRou LBL and Wapiti RV measurements of Gl~410.}
\centering
\begin{tabular}{c c c | c c c c |c c c c}
\hline
\hline 
		     &					& & \multicolumn{4}{c|}{LBL} &  \multicolumn{4}{c}{Wapiti} \\
DATE             & RJD    & $V_{\rm tot}$ 	& RV     & $\sigma_{\rm RV}$ & $\delta$ RV & $\sigma_{\delta{\rm RV}}$   & RV     & $\sigma_{\rm RV}$ & $\delta$ RV & $\sigma_{\delta{\rm RV}}$\\[1pt]
[yyyy-mm-dd] & [days] 			& [km/s]		& [km/s] & [km/s]        	      & [m/s]            & [m/s]     & [km/s] & [km/s]        & [m/s]            & [m/s]             \\[1pt]
\hline
2020-11-04 & 59158.11239 & -40.58  & -13.8818   & 0.0023   &   15.88 & 2.3      & -13.9924   & 0.0030   &   11.00 & 3.0  \\
2020-11-05 & 59159.10674 & -40.80  & -13.8875   & 0.0023   &   10.22 & 2.3      & -14.0006   & 0.0030   &    2.83 & 3.0  \\
2020-12-24 & 59208.05450 & -40.68  & -13.8916   & 0.0028   &    6.16 & 2.8      & -13.9993   & 0.0036   &    4.10 & 3.6  \\
...		  &  		...	   & 	...	 &  		...   &	 ...	      &      ...     &   ...     	  &   ...   	    &    ...	     &      ...       &  ... \\
\hline
\end{tabular}
\tablefoot{Full table available at the CDS.}
\label{GL410_SPIROU_RVtable}
\end{table*}

\subsection{TESS}
The Transit Exoplanet Survey \citep[TESS;][]{Ricker2014} observed Gl~410
(TIC 97488127) with a two-minute cadence during Sector 22, from February
18 to March 18, 2020. We obtained the Presearch Data Conditioning
(PDC) flux time series \citep{Smith2012,Stumpe2012,Stumpe2014} processed
by the TESS Science Processing Operations Center
\citep[][]{jenkinsSPOC2016} from the Mikulski Archive for Space
Telescopes (MAST)\footnote{\url{mast.stsci.edu}}.

\begin{table}
\caption{SOPHIE RV time series and periodogram properties.}
\centering
\tiny
\begin{tabular}{l c c c c c}     % 7 columns 
\hline\hline      
                       		 	& SOPHIE1   		& SOPHIE2 	     	& SOPHIE 1+2     \\ 
                      			& $(2010-2012)$ 	& $(2021-2023)$ 	& $(2010 - 2023)$  \\
\hline   
Time series \\
\hline
N					& 45				& 58				& 103		\\[0.4mm]
<RV>  [km s$^{-1}$]  	& -13.8399	 	& -13.8347		& -13.8385 	\\[0.4mm]
<$\sigma$> [m s$^{-1}$]	& 2.5				& 2.4				& 2.4 \\[0.4mm]	
rms    [m s$^{-1}$]		& 14.6			& 12.7			& 13.9 \\[0.4mm]
\hline
LS periodogram \\[0.4mm]
\hline
$P$  [days]              	 	& 6.99 			& 6.91			& 6.92\\[0.4mm]
norm. power		  	& 0.37			& 0.42			& 0.33\\[0.4mm]
FAP$_{\rm bootstrap}$  		& 0.2		& 3.8$\times$10$^{-3}$		& 4.0$\times$10$^{-4}$  \\[0.4mm]
\hline
\end{tabular}
\label{table:SOPHIE_RV_periodogram} 
\end{table}

\section{Radial velocity analysis} 

For the RV analysis, 
we used Generalized Lomb-Scargle (GLS) periodograms \citep[][]{ZK09}.
To determine the false alarm probability levels (FAP), 
we used two methods.
For peaks with a significance lower than 10$^{-5}$ and the 10\%, 1\%, and 0.1\% FAP levels in the periodograms,
we used a bootstrap algorithm with 10$^5$ iterations.
The method consists in generating a time series with dates equal to the real series of measurements, 
but assigning to each of them a RV measurement chosen randomly from the original data series.
We labeled these FAP levels FAP$_{\rm bootstrap}$.
For peaks with a significance higher than 10$^{-5}$ (i.e., inferior to 1 / number of boostrap iterations),
we used the GLS FAP analytical implementation of the package \texttt{pyAstronomy} \citep[][]{pya}.
We labeled these FAP levels FAP$_{\rm analytical}$.

\subsection{SOPHIE radial velocities}
\begin{figure}[t]
\centering
\includegraphics[width=0.49\textwidth]{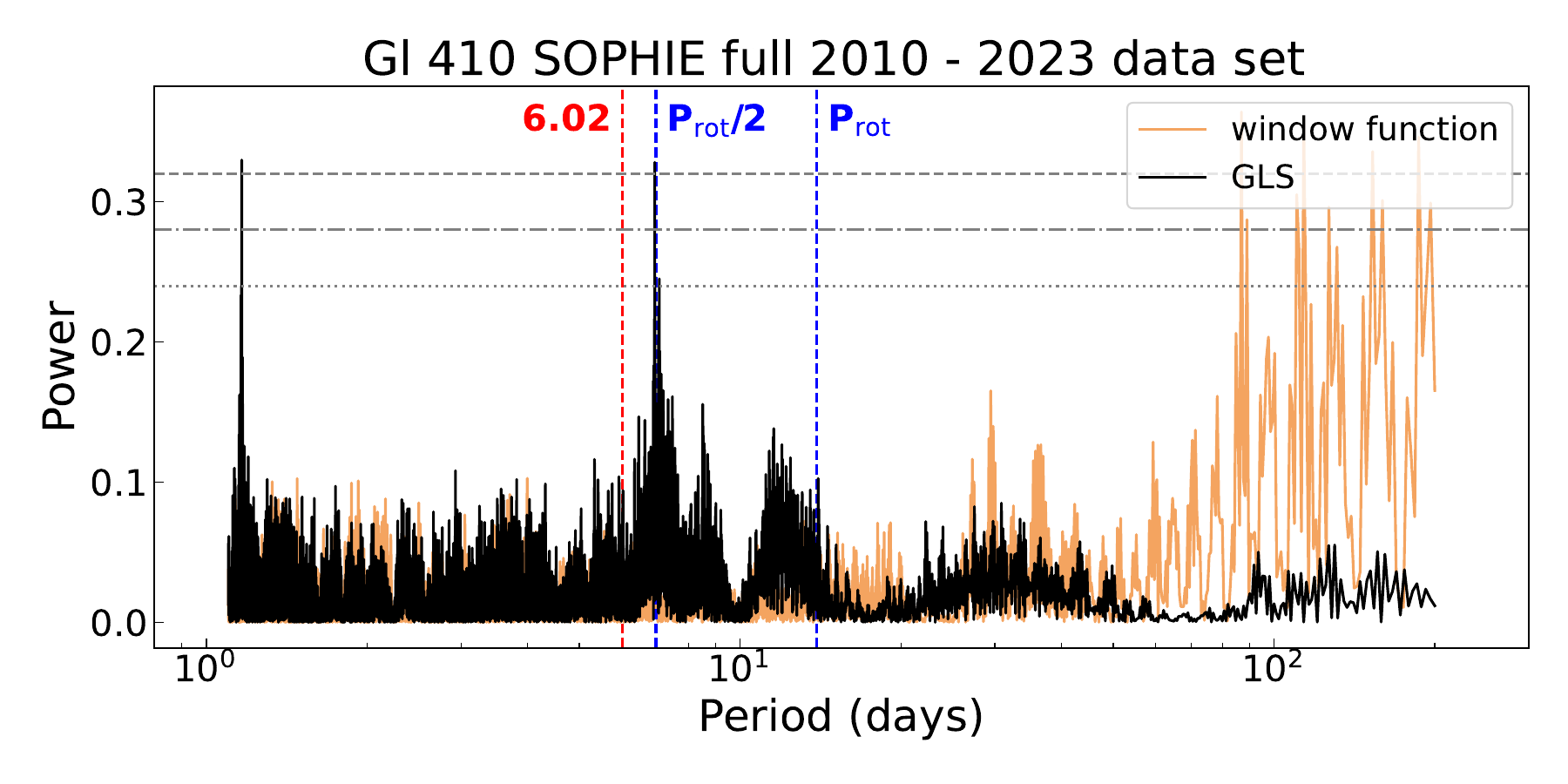}
\includegraphics[width=0.49\textwidth]{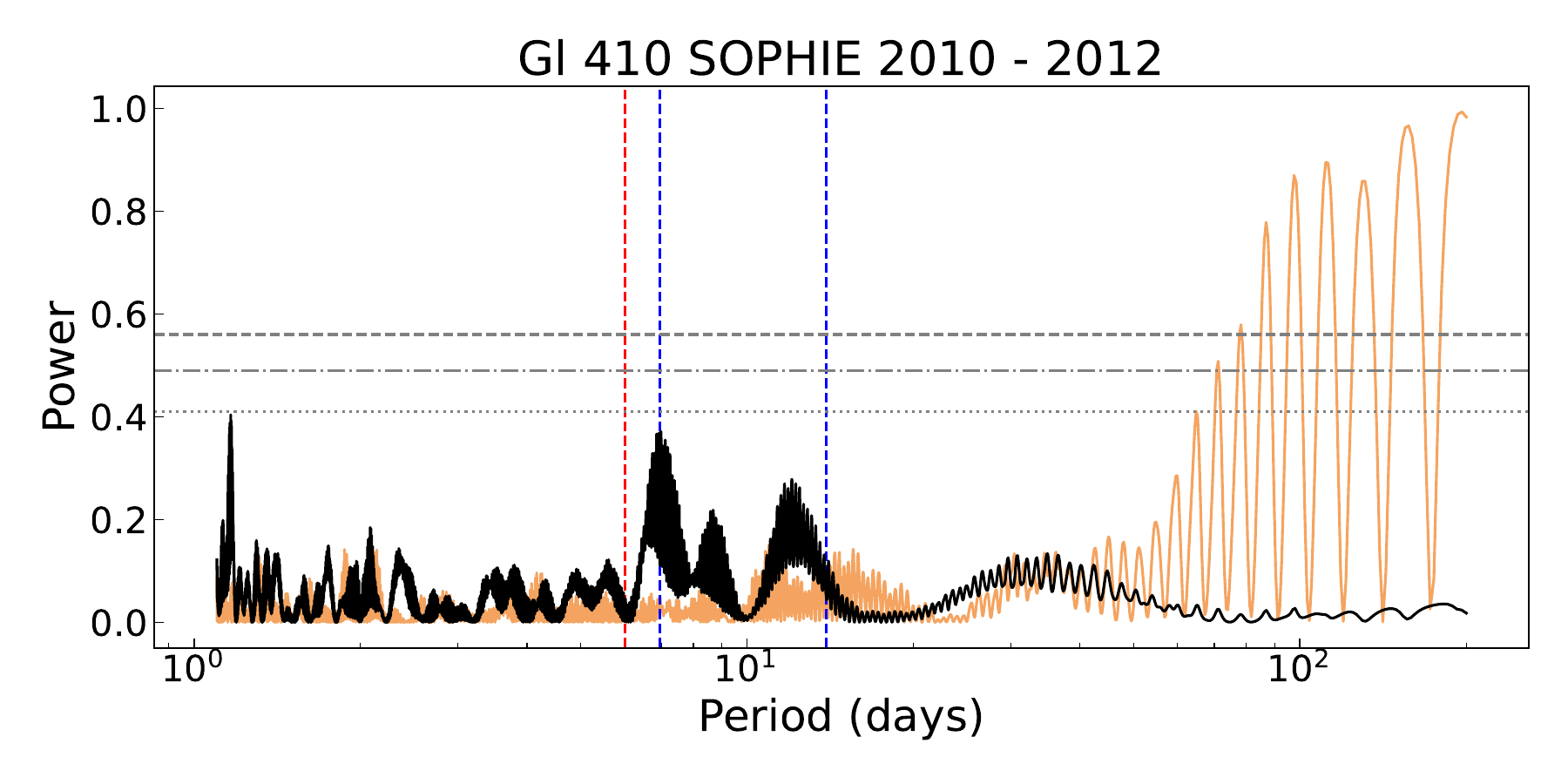}
\includegraphics[width=0.49\textwidth]{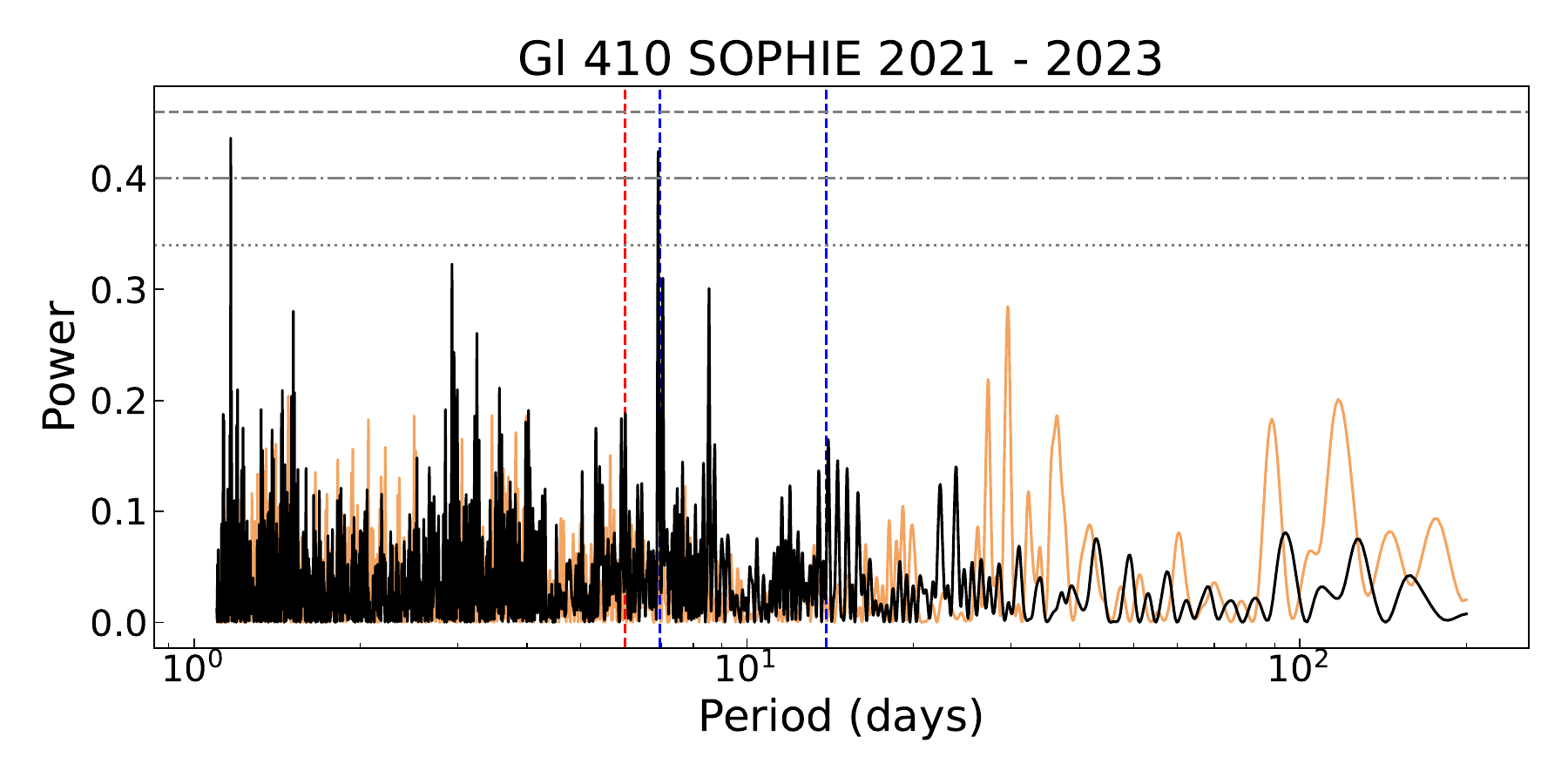}
 \caption{Generalized Lomb-Scargle Periodograms of the SOPHIE RV time series for the full 
 2010$-$2023 data set (top),
 the 2010$-$2012 period (middle),
and the 2021$-$2023 period (bottom).
The location of the rotational period of $P_{\rm rot}=13.9$ days and its $P_{\rm rot}$/2 harmonic
are indicated with blue vertical lines.
The position of $P=6.02$ days is displayed with a red vertical line.
The horizontal lines indicate the bootstrap calculated 
at 0.1\%, 1\%, and 10\% FAP levels.
The window function is shown with a light orange color.
 }
\label{Gl410_SOPHIE_periodograms}
\end{figure}

We achieved with SOPHIE an average RV precision per measurement of 2.4 m s$^{-1}$  
and an rms of  13.9 m s$^{-1}$  in the full time series.
Figure~\ref{Gl410_SOPHIE_periodograms} displays the GLS periodograms obtained
for the full data set, the 2010$-$2012, and 2021$-$2023 periods, respectively.
Table~\ref{table:SOPHIE_RV_periodogram} summarizes the properties 
of the time series and the strongest peak in the periodogram.

The full data set displays a peak at $P=6.92$ days with an FAP$_{\rm bootstrap}$ of 4.0$\times10^{-4}$.
This peak corresponds to the $P$/2 harmonic of the stellar rotation period of $P=13.91$ days
\citep[][]{Donati2023}.
Apart from the 1 day alias, there are no other peaks in the full data set periodogram.
No signal is observed at the stellar rotation period.
The 2010$-$2012 period shows no significant periodicities.  
A broad peak is observed close to $P=6.99$ days but its significance is low (FAP$_{\rm bootstrap}=0.2$).
In the 2021$-$2023 period, 
a peak at $P=6.91$ days appears at an FAP$_{\rm bootstrap}$ level of 3.8$\times$10$^{-3}$.
Except for the 1 day alias, 
no peaks are present in the periodogram with a significance below an FAP$_{\rm bootstrap}$=10\%.

In summary,
in the raw SOPHIE RV data, there is a signal at $P=6.9$ days, 
but it is most likely related to the stellar rotation.

\begin{table*}
\caption{SPIRou LBL wPCA RV measurements of Gl~410.} 
\scriptsize
\begin{tabular}{c c c c c c r r r r r r c c}
\hline \hline 
DATE              & RJD    & RV     		& $\sigma_{\rm RV}$ 	& $\delta$ RV 		& $\sigma_{\delta{\rm RV}}$   
& \multicolumn{1}{c}{\texttt{d2v}} & \multicolumn{1}{c}{$\sigma_{\texttt{d2v}}$}     
& \multicolumn{1}{c}{\texttt{d3v}} & \multicolumn{1}{c}{$\sigma_{\texttt{d3v}}$}    
&\multicolumn{1}{c}{\texttt{dLW}} & \multicolumn{1}{c}{$\sigma_{\texttt{dLW}}$} 
&\multicolumn{1}{c}{$\texttt{dTEMP}_{4000}$}  &\multicolumn{1}{c}{$\sigma_{\texttt{dTEMP}_{4000}}$}
\\[1pt]
[yyyy-mm-dd]  & [days] 
& [km/s]    & [km/s]        		
& [m/s]          & [m/s]       
& \multicolumn{1}{c}{[m$^{2}$/s$^{2}$]} & \multicolumn{1}{c}{[m$^{2}$/s$^{2}$]}
& \multicolumn{1}{c}{[m$^{3}$/s$^{3}$]} &\multicolumn{1}{c}{[m$^{3}$/s$^{3}$]} 
& \multicolumn{1}{c}{[m/s]}  &\multicolumn{1}{c}{[m/s]}  
& \multicolumn{1}{c}{[K]} &\multicolumn{1}{c}{[K]}  \\[1pt]
\hline
2020-11-04 & 59158.11239 & -13.7958   & 0.0023   &  12.10 & 2.3  &  -3.05e+04 &   5.87e+03 &   1.39e+07 &   1.17e+07 &  -1.69e+05 &   3.25e+04 & -17.4 &   0.5     \\
2020-11-05 & 59159.10674 & -13.7984   & 0.0022   &   9.50 & 2.2  &  -4.46e+04 &   5.85e+03 &   5.51e+06 &   1.16e+07 &  -2.47e+05 &   3.24e+04 & -22.9 &   0.5     \\
2020-12-24 & 59208.05450 & -13.7967   & 0.0027   &  11.12 & 2.7  &   5.45e+04 &   7.18e+03 &  -3.94e+06 &   1.43e+07 &   3.02e+05 &   3.98e+04 &   6.1 &   0.6     \\
...		  &  		...	   & 	...	 	&  	...   &	 ...	    & ...     &   ...     	  &    ...   	        &    ...	     &      ...       &  ...    	     &      ...       &  ...       & ... \\
\hline
\end{tabular}
\tablefoot{Full table available at the CDS.}
\label{GL410_SPIROU_WPCA}
\end{table*}

\subsection{SPIRou radial velocities}
\label{Sec:SPIRou_radial_velocities}
\begin{figure*}
\centering
\setlength{\tabcolsep}{0pt} 
\begin{tabular}{ccc}
\includegraphics[width=0.33\textwidth]{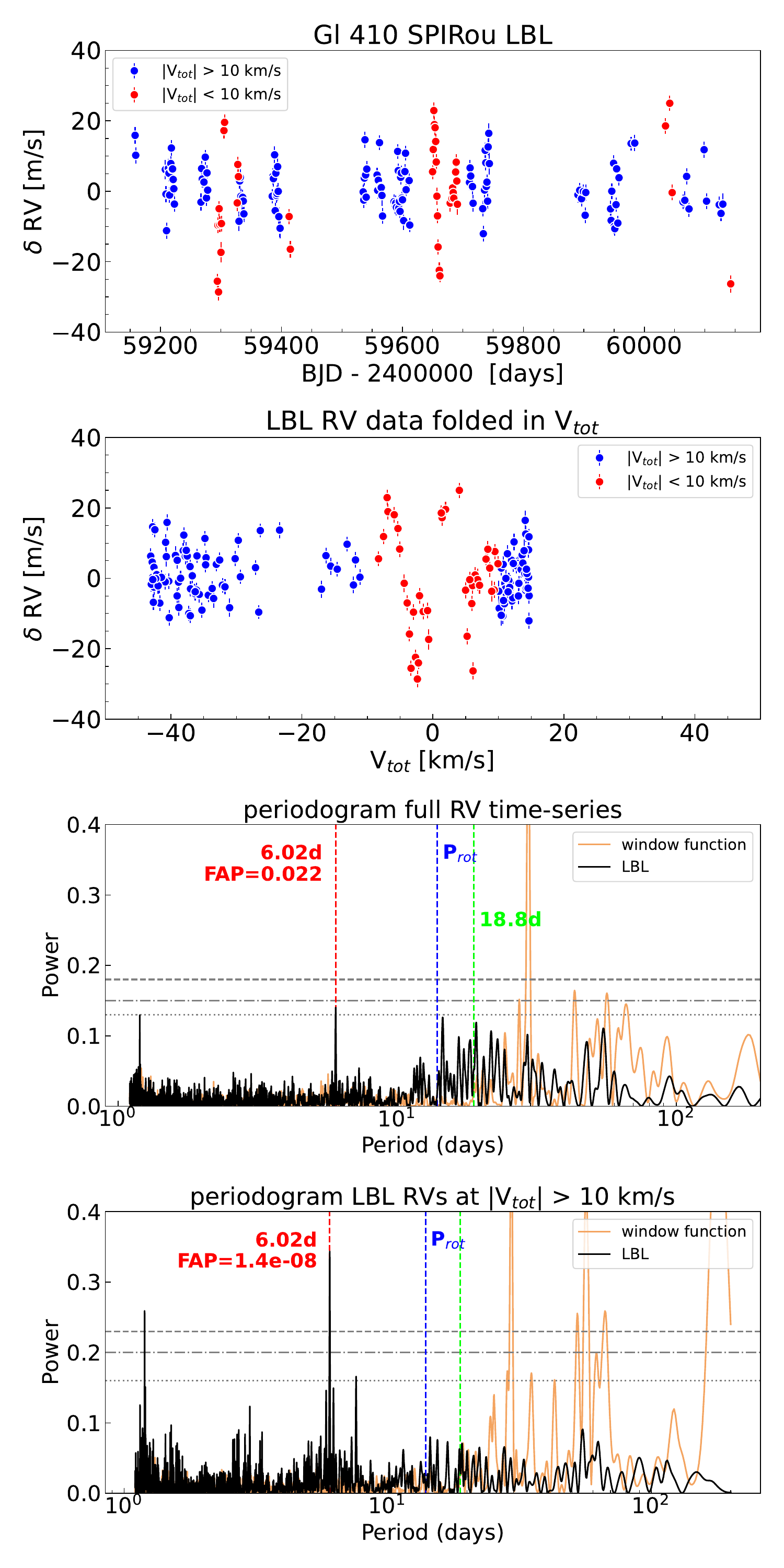} &
\includegraphics[width=0.33\textwidth]{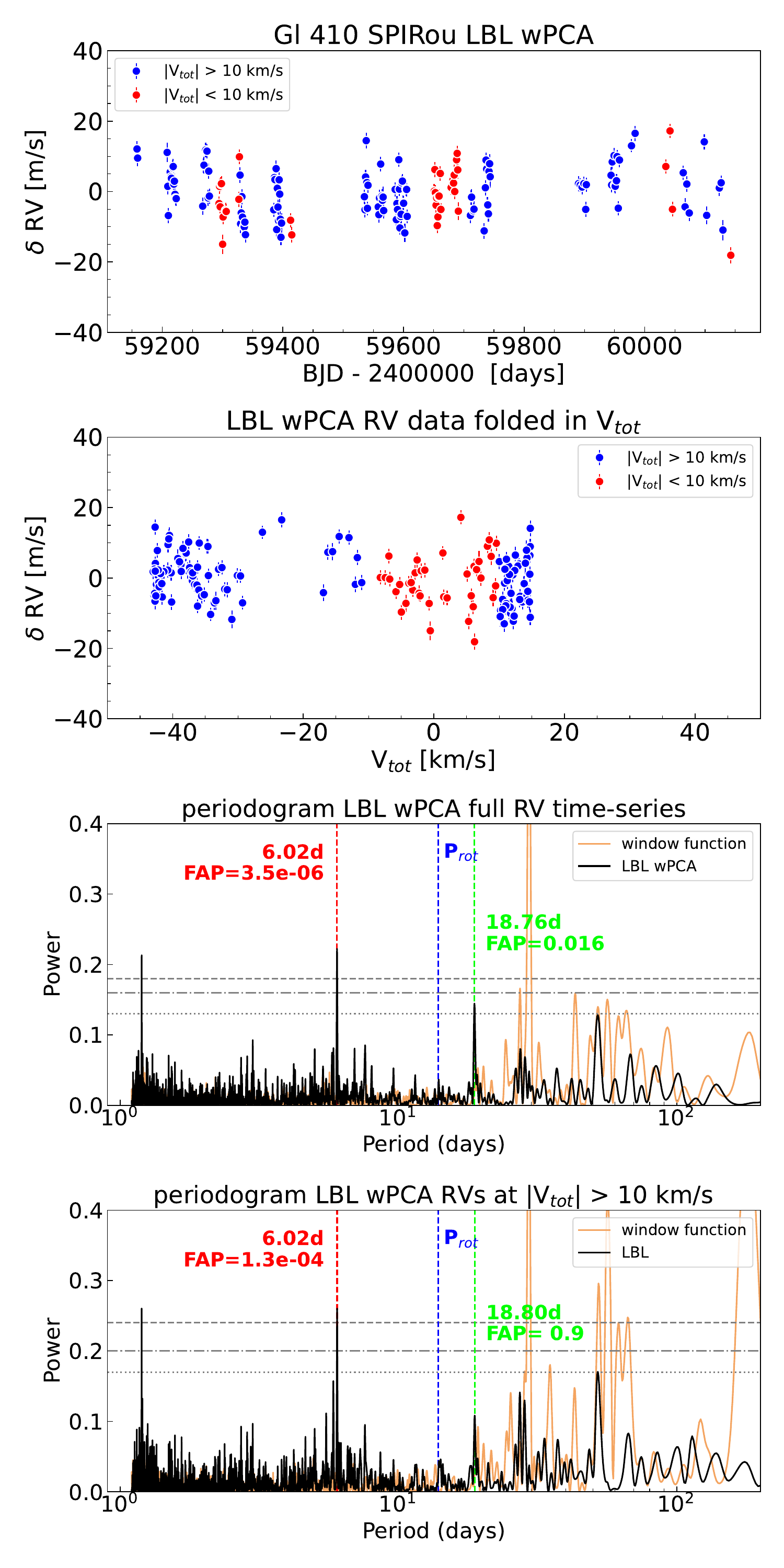} &
\includegraphics[width=0.33\textwidth]{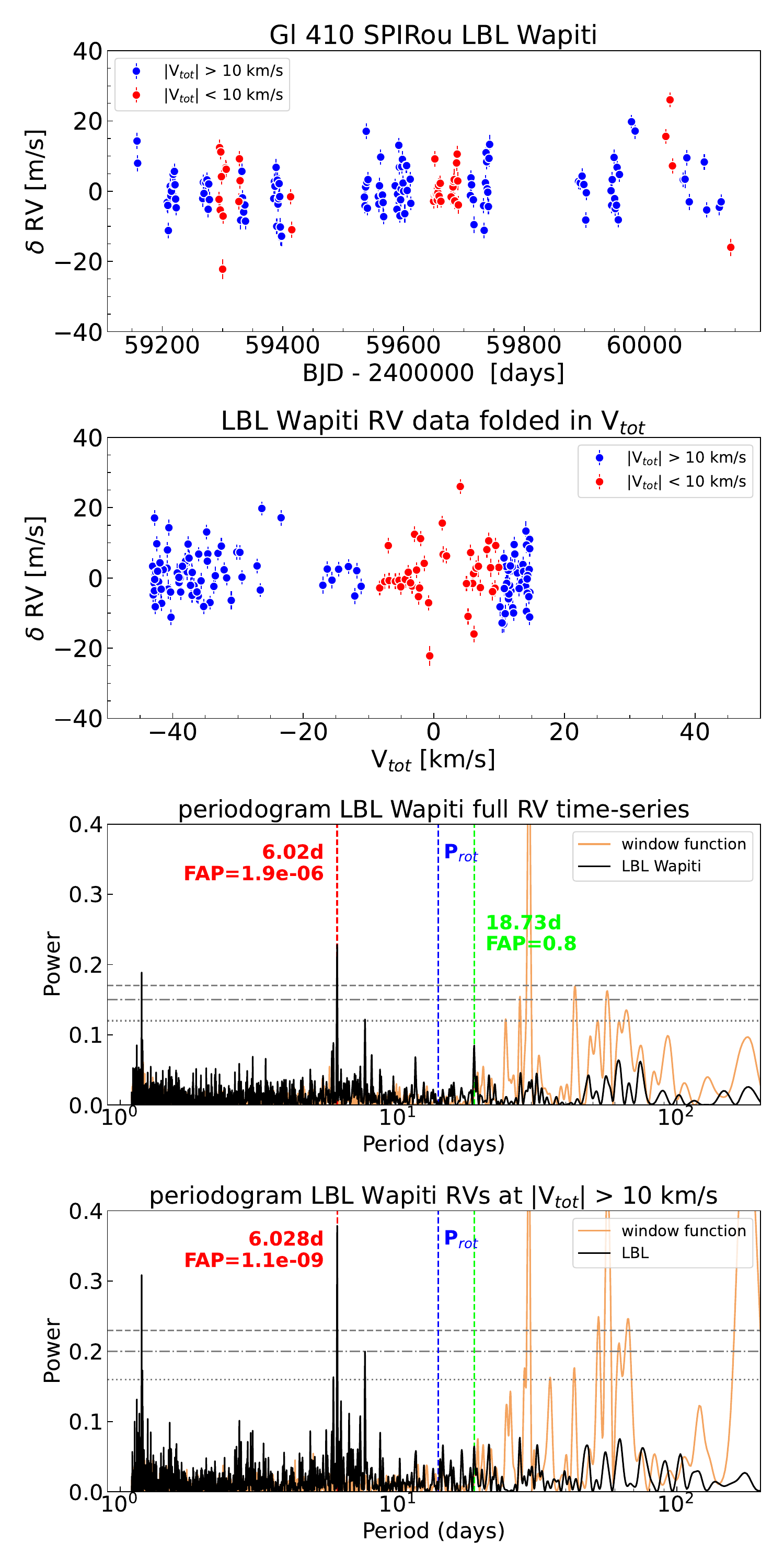} \\
\end{tabular}

\caption{SPIRou RV measurements as a function of time and GLS periodograms for the ``raw'' LBL measurements
and the PCA corrected LBL measurements obtained with wPCA (Artigau et al., in prep) and Wapiti \citep[][]{Wapiti2023}.
Blue dots indicate the measurements taken at $|V_{\rm tot}|>10$ km s$^{-1}$. 
Red dots are the measurements obtained when $|V_{\rm tot}|<10$ km s$^{-1}$  (i.e., moment of the highest influence of the atmosphere on the 
stellar spectrum).
In the periodograms,
the gray horizontal lines indicate the { bootstrap-calculated} 10\%, 1\%, and 0.1\% FAP levels.
A summary of the statistics of the each of the time series is provided in Table~\ref{table:SPIRou_RV_periodogram}.}
\label{fig:SPIRou_RV_periodogram}
\end{figure*}

With SPIRou, 
we achieved an average RV precision per measurement of 2.3 m s$^{-1}$  
and an rms of  9.1 m s$^{-1}$  in the full LBL time series.
We display in the left panels of  Fig.~\ref{fig:SPIRou_RV_periodogram},
the SPIRou LBL RV measurements as a function of time and their GLS periodogram.
The periodogram displays a peak at $P=6.02$ days with a { FAP$_{\rm bootstrap}$ level of 2.2\%}.
No peaks are observed at the rotational period of the star nor its harmonics.
The SPIRou LBL RV measurements are listed
in Table~\ref{GL410_SPIROU_RVtable} (the full table is available at the CDS).
A summary of the properties of the LBL RV time series is given
in Table~\ref{table:SPIRou_RV_periodogram}.

Due to the reflex motion of the Earth in its orbit (i.e., the BERV),
observations acquired at different moments of  the year
have different velocity  shifts with respect to the center of telluric absorption lines.
If a star has an intrinsic average systemic velocity $V_{\rm sys}$ with respect to the Sun,
the total velocity $V_{\rm tot}$ of the star at the time of the observations is given by
\begin{equation}
V_{\rm tot} = V_{\rm sys} - {\rm BERV}
\end{equation}

In the near-IR,
the telluric absorption lines are abundant,
thus the stellar photospheric absorption lines are affected by tellurics.
The first effect of telluric absorption is the deformation of the photospheric line profiles.
After telluric correction, 
residuals from atmospheric lines, 
at a much lower level than original, may remain,
and even for a perfect correction, 
the noise is higher in the affected velocity channels.
The moment the stellar spectra is most affected by tellurics is 
when the atmospheric absorption is at the core of the stellar photospheric lines,
that is when $V_{\rm tot}$ is between $-$10 and $+$10~km~s$^{-1}$. 
In other words, this is when the absorption lines of chemical species common to the Earth's atmosphere and the stellar photosphere (e.g., H$_2$O, CO) 
are on the top of each other, particularly in regions where the optical depth of the atmosphere is the greatest.

We have color-coded in Fig.~\ref{fig:SPIRou_RV_periodogram} 
the RV measurements based $V_{\rm tot}$. 
We display  the observations obtained when $|V_{\rm tot}|>10$ km~s$^{-1}$ in blue 
and the observations taken when $|V_{\rm tot}|<10$ km~s$^{-1}$ in red.
The telluric correction of \texttt{APERO} is good, 
the LBL algorithm is efficient at filtering outliers,
and the zero-point correction in time and BERV space further corrects residuals of telluric lines.
However, 
as shown in Fig.~\ref{fig:SPIRou_RV_periodogram}, 
at the meter per second RV precision level, 
the systematic noise of tellurics is evident.
Fig.~\ref{fig:SPIRou_RV_periodogram}
shows that a large fraction of the outliers in the RV time series are in red.
The effect of telluric lines becomes evident when the RV time series 
is plotted as a function of $V_{\rm tot}$ (see Fig.~\ref{fig:SPIRou_RV_periodogram}).
The systematics due to the telluric absorption can be seen,
as an up and down correlated-noise pattern.

\begin{table}[t]
\caption{SPIRou RV time series and GLS periodogram properties.}
\scriptsize
\begin{center}
\begin{tabular}{l c c c c c c c }     % 7 columns 
\\[-15pt]
\hline\hline      
                       		 	&  full LBL			& |V$_{\rm tot}$| $>$ 10 km s$^{-1}$ 	& LBL wPCA	 		 & LBL Wapiti     \\ 
%:
					&  data set			& LBL				&  	  &  	\\
\hline   
\\[-5pt]
Time series \\
\hline
visits					& 160			&121								& 157 		&  157 \\[0.4mm]	
<RV>  [km s$^{-1}$]  	& $-$13.898	 	& $-$13.897						& $-$13.807	& - \\[0.4mm]	
<$\sigma$> [m s$^{-1}$]	& 2.3				& 2.3								& 2.2			& 2.2 \\[0.4mm]	
rms    [m s$^{-1}$]		& 9.1				& 6.5								& 6.8			& 6.3\\[0.4mm]
\\[-4pt]
\hline
\\[-5pt]
{GLS} periodogram \\

\hline
\\[-4pt]
{\bf Gl~410b}\\[1pt]
$P_b$  [days]              	 & 6.02 						& 6.02							& 6.02						& 6.02 \\[0.4mm]
norm. power		  	& 0.14						& 0.34							& 0.22						& 0.23\\[0.4mm]
FAP$^\dagger$  		& $2.2\times10^{-2}$ 			& $1.4\times10^{-8}$					& $3.5\times10^{-6}$ 	                 & $1.9\times10^{-6}$ \\[0.4mm]
\\[-4pt]
\multicolumn{3}{l}{\bf candidate signal at 18.7 days }\\[1pt]
$P_c$  [days]              	& 							&								& 18.76				& 18.73 \\[0.4mm]
norm. power		  	& 							&								& 0.14						& 0.09 \\[0.4mm]
FAP$^\dagger$  		& 							&								& 0.016				&  0.8 \\[0.4mm]

\hline
\end{tabular}
\tablefoot{$^\dagger$ FAP levels $>$10$^{-5}$ were calculated using a bootstrap algorithm. 
FAP levels $<$10$^{-5}$ were calculated with the GLS analytical approximation.}
\end{center}
\label{table:SPIRou_RV_periodogram} 
\end{table}

If the RV time series is filtered from the data-points taken at $|V_{\rm tot}| <10$ km s$^{-1}$
(i.e., the red points in Fig.~\ref{fig:SPIRou_RV_periodogram}),
the rms of the time series (see Table~\ref{table:SPIRou_RV_periodogram}),
improves: from 9.1 m s$^{-1}$ for the full time series to 6.5 m s$^{-1}$  for the filtered time series.
If the periodogram is made only with observations taken at $|V_{\rm tot}| >10$ km~s$^{-1}$,
we not only recover the peak at 6.02 days detected in the full time series, 
but its significance increases from an FAP$_{\rm bootstrap}$=0.022 in the full time series
to  FAP$_{\rm analytical}$=1.4$\times10^{-8}$ in the filtered time series.
The case of Gl~410 dramatically shows the effect that telluric residuals 
can have adding RV noise, ultimately blurring a planet detection.

Filtering  data-points taken at $|V_{\rm tot}| <10$ km s$^{-1}$  is a robust and
fast way of finding promising planet candidate signals.
However, apart the obvious fact that this approach takes out a large of number data points (39 measurements),
it also has the limitation that it can boost the power at certain frequencies
because specific and periodic times are missing in the time sampling.
The RV systematics and correlated noise generated both by 
telluric residuals and detector-based glitches and imperfections
can be corrected. The LBL framework produces a 2D data set,
a time series for each of the thousands of LBL line measurements.
In this 2D data set, 
data-driven techniques such as principal component analysis  (PCA)
can be employed to further filter, mitigate, and correct systematic noise in the RV measurements.

In the context of the SPIRou collaboration, 
two PCA-based LBL RV correcting algorithms have been developed: 
Wapiti \citep[][]{Wapiti2023} and a weighted PCA method (wPCA; Artigau et al. in prep). 
We refer to those papers for details on each methodology.
The main difference between these two approaches is that Wapiti performs a PCA in the stellar velocity space, 
whereas in the second method, 
the PCA is applied in the detector pixel space \citep[inspired on][]{Cretignier2023}.
The Wapiti and wPCA RV measurements are given in Tables \ref{GL410_SPIROU_RVtable} 
and \ref{GL410_SPIROU_WPCA},
respectively (both tables are available at the CDS).
We re-analyzed the RV data of Gl~410 with both of these methods.
In the central and right panels of Fig.~\ref{fig:SPIRou_RV_periodogram},
we display the RV measurements and GLS periodograms obtained with wPCA and Wapiti.
In Table~\ref{table:SPIRou_RV_periodogram}, 
we present the summary of the properties of the RV time series.

Figure~\ref{fig:SPIRou_RV_periodogram} illustrates how
wPCA and Wapiti correct for the systematic noise at $|V_{\rm tot}| <10$ km s$^{-1}$ .
The rms of the full time series decreases from 9.1 m s$^{-1}$ in the raw LBL 
to  6.8 m s$^{-1}$ with wPCA
and 6.3 m s$^{-1}$ with Wapiti.
The rms of the points at $|V_{\rm tot}|<10$ km s$^{-1}$ (red dots in Fig.~\ref{fig:SPIRou_RV_periodogram}) 
decreases from 14.3 m s$^{-1}$ in the raw LBL,
to 7.1 m s$^{-1}$ with wPCA
and 8.3 m s$^{-1}$ with Wapiti.
We also observed that the peak at $P=6.02$ days becomes stronger in the full time series and that it reaches 
an FAP$_{\rm bootstrap}$  level of 3.5$\times10^{-6}$.
Furthermore  in the wPCA RV data,  
a new peak at $P=18.76$ days with an FAP$_{\rm bootstrap}$  level of 1.6\% becomes visible.
A peak at 18.73 days is also present  in the Wapiti time series,  
however, it has a negligible significance (FAP$_{\rm bootstrap}$=80\%).

In summary, the SPIRou RV data shows a clear periodic signal at a $P=6.02$ days.
The signal is not a harmonic or alias of the rotation period.
There is evidence for a second, tentative, candidate signal at $P=18.7$ days.

\subsection{A test of robustness of the 6-day SPIRou signal}

No periodicities were detected at 6.02 nor at 18.7 days 
in the activity indicators time series of SOPHIE and SPIRou (details are provided in  Sect.~\ref{Sect_activity_analysis} in the Appendix).
We performed an additional test of robustness to exclude the possibility 
that the signal at 6 days is due to activity.
If a candidate signal is due to a planet (and not activity),
it is expected that as more RV data points are accumulated
the significance of the detection increases, 
and this should not depend on the way the time-sampling 
of the measurements is performed.

Using the $|V_{\rm tot}|>10$ km s$^{-1}$  { SPIRou wPCA} time series, 
we performed a test that consists in checking the evolution 
of the FAP level of the $P=6$ days peak in 100 different randomly sampled time series.
The construction of an RV time series starts by randomly selecting ten measurements, 
calculating the GLS periodogram, 
and measuring the FAP of the peak at $P=6$ days.
One randomly selected measurement is then included in the time series,
the periodogram is recalculated, and the FAP of the peak at $P=6$ days is measured.
The procedure is repeated until the complete number of visits is reached for a time series.
In this way, we constructed 100 independent evolutions of the FAP as a function 
of the number of measurements.
This test implies the calculation of around twelve thousand FAPs. 
To keep the computation time short,  
we calculated the FAP level using the analytical approximation implemented in \texttt{pyAstronomy} \citep[][]{pya}.
The analytical FAPs are a factor of a few lower than the FAPs calculated using a bootstrap method.

\begin{figure}
\centering
\includegraphics[width=0.45\textwidth]{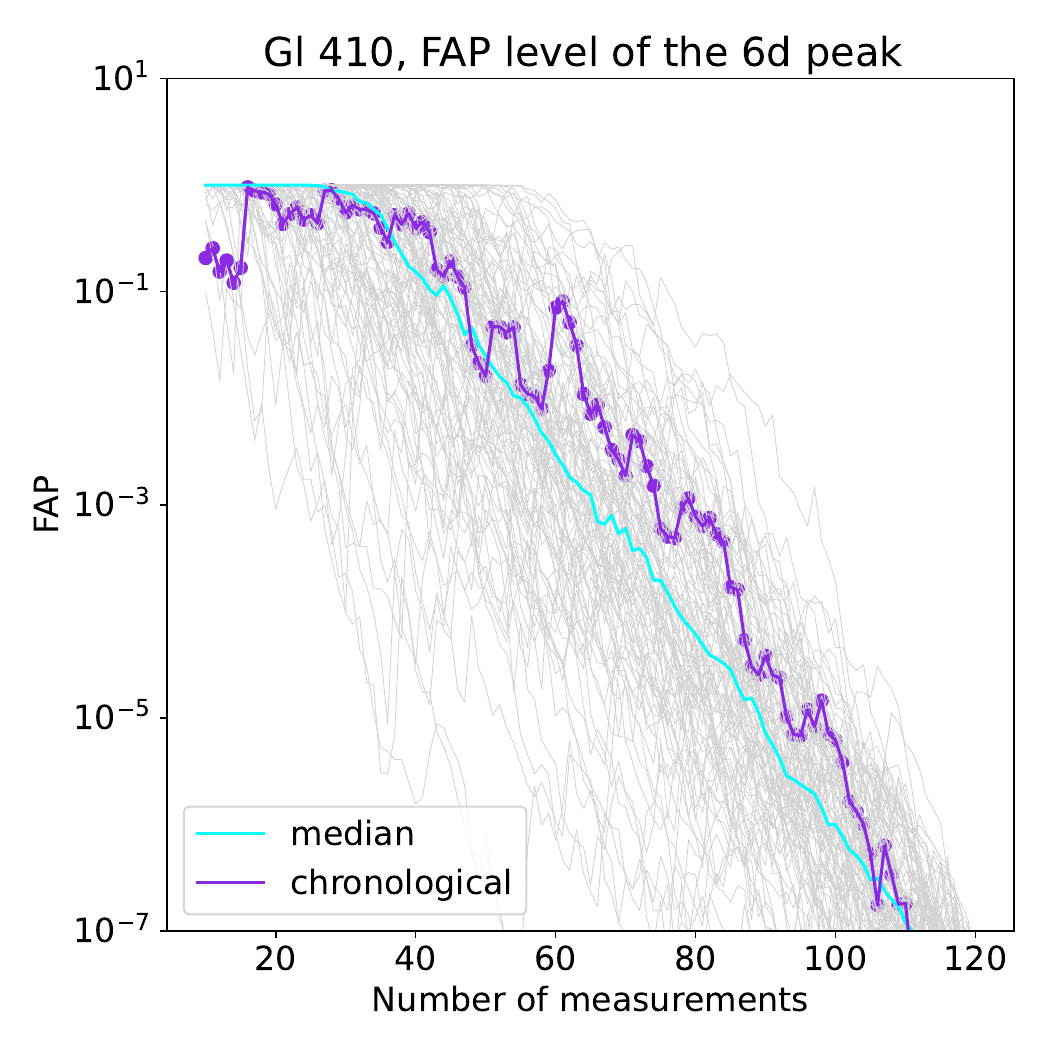}
\caption{Test on the change of the FAP level of the peak on the periodogram at 6 days as 
a function of the number of visits for 100 randomly selected time series. 
The FAP level of the $P=6$ days peak decreases  as a function of the number of measurements,
as expected for a planetary origin of the signal.
The random time series were sampled from the $|V_{\rm tot}|>10$ km s$^{-1}$  RV time series.
The FAP level in this test was calculated using the analytical approximation implemented in \texttt{pyAstronomy} \citep[][]{pya}.
}
\label{Gl410_FAP_test}
\end{figure}

The result of the test is displayed in Fig.~\ref{Gl410_FAP_test}.
One can see that for all the time series, 
the FAP level of the $P=6$ days peak decreases as 
the number of measurements increases.
In Fig.~\ref{Gl410_FAP_test}, 
we include the plot of the  FAP level in the chronological time series
and the median FAP for each bin of number of measurements.
The test indicates that the peak at $P=6$ days in the periodogram is physical
and that it does not depend on the sampling of the time series.
This test supports the argument that the $P=6$ days peak in the periodogram
is indeed of Keplerian origin.

\section{Stellar activity filtering of the SOPHIE radial velocities and search for the Gl~410b signal}

As opposed to a signal due to activity, 
a Keplerian reflex motion caused by the presence of a planet is achromatic.
There is a hint of the 6-day period signal in the SOPHIE RV data of Gl~410.
If the SOPHIE and SPIRou RVs are plotted  phase-folded,
both data sets are consistent.
However, the SOPHIE RVs have a much larger dispersion (14 m s$^{-1}$) than the SPIRou RVs (7 m s$^{-1}$).
One interesting question is whether Gl~410b's RV periodic signal
could be retrieved in the SOPHIE data set if the RV jitter induced by
stellar activity is corrected.

We tested two methods to search for the 6-day signal within the SOPHIE RV data. 
In the first method, the SOPHIE RVs time series is analyzed with a model that
is the sum of a quasi-periodic GP and a Keplerian model.
In the second method, 
we used the measurements of the longitudinal magnetic field ($B_\ell$) 
from SPIRou to inform a GP that is used to correct the SOPHIE RVs for activity jitter,
and then we searched for the planet signal in the RV residuals.

\subsection{Method 1: SOPHIE Gaussian process modeling using the SOPHIE radial velocity data alone}

As the rotational period of Gl~410 is well constrained by spectropolarimetry,
the SOPHIE RV data can be analyzed using a GP plus Keplerian model.
The goal here is to determine what model is more probable
between a model without a planet (i.e., only a GP) 
and a model with a GP plus a planet.
To perform the analysis, 
we used \texttt{RadVel} \citep[][]{radvel2018} and assumed a circular orbit.
A Gaussian prior was used  for the GP period, 
with center at 13.9 days and $\sigma$=0.1 days, 
which corresponds to the stellar rotation period derived from the
measurements of the longitudinal magnetic field $B_\ell$ \citep[][]{Donati2023}.
Uniform priors were set for
the planet period $P_b$ (0 to 20 days),
the planet semi-amplitude $K_b$ (0 to 20 m s$^{-1}$),
the time of inferior conjunction $T{\rm conj}_b$ (2459648.0 $\pm$ 10 days),
the GP amplitude $A$ (0 to 75 m s$^{-1}$),
the GP decay time $l$ (0 to 2000 days),
the GP smoothing length $\Gamma$ (0 to 2 days),
and the uncorrelated noise $\sigma_{\rm SOPHIE}$ (0.0 to 10 m s$^{-1}$).

Table~\ref{comparison_SOPHIE_radvel} provides a summary of the small sample Akaike information criterion (AICc) 
tests comparing the models.
The model with a GP model alone (i.e., without planet) was statistically ``ruled out'' by \texttt{RadVel}. 
In Table~\ref{tab:comp} in the Appendix, we present the posterior distributions obtained for
the model\footnote{The stellar rotation period is well constrained from spectropolarimetry 
and provides a stringent prior on the cycle length of the GP, which is recovered, as expected, in the posterior distribution.
We tested two models with a broader Gaussian distribution prior for the cycle length (13.9$\pm$3.5 and  13.9$\pm$7 days).
All models converged to a GP cycle length of 13.84$\pm$0.03 days.
}.
The Keplerian signal at $P=6.02$ days is retrieved in the model.
This test provides independent evidence for the existence of the 6-day planet inside the SOPHIE RV data.
In Sect.~\ref{SOPHIE_GP_SPIRou_radvel_1p}, 
we discuss the coherence of the Keplerian signals obtained with SPIRou and SOPHIE.

\begin{table}[t]
\begin{center}
\small
\caption[]{\texttt{RadVel} model comparison table for the GP plus Keplerian model on the SOPHIE RVs alone.}
\begin{tabular}{llccccc}
\\[-15pt]
\hline
\hline
{Model} & {$N_{\rm free}$} &{RMS} & {BIC} & {AICc} & $\Delta$AICc \\
 \hline
 \\[-5pt]
{AICc Favored Model}\\
$K_{b}$, GP$_{\rm }$, {$\sigma$}, {$\gamma$} 		    & 9  	& 4.9  & 450 & 435  	   & 0.0\\[5pt]
Ruled Out \\
GP$_{\rm }$, {$\sigma$}, {$\gamma$} 				   & 6 	 & 23 & 866 & 855	     & 420  \\
\hline \\
\end{tabular}
 \label{comparison_SOPHIE_radvel}
\end{center}
\end{table}

\subsection{Method 2: SOPHIE Gaussian process modeling using SPIRou's $B_\ell$ as an activity proxy}
\label{SOPHIE_GP_Bl}
In this second method, 
the search of the signal is done using a two step approach.
In step one,  a quasi-periodic GP model is fit to the SPIRou's longitudinal magnetic field ($B_\ell$) time series,
and in step two, the posterior distribution of the hyper parameters (period, decay time, and smoothing length) of the $B_\ell$ GP are used 
as prior for a quasi-periodic GP model of the SOPHIE RVs (other hyperparamers, 
such as ln $A$, and ln $s$, priors are taken as uniform).
The aim of this method is to filter the SOPHIE RVs for activity using $B_\ell$ as a proxy of the activity signal 
and search for the signal at 6 days in the residuals. 

\begin{figure}
\centering
\includegraphics[width=0.5\textwidth]{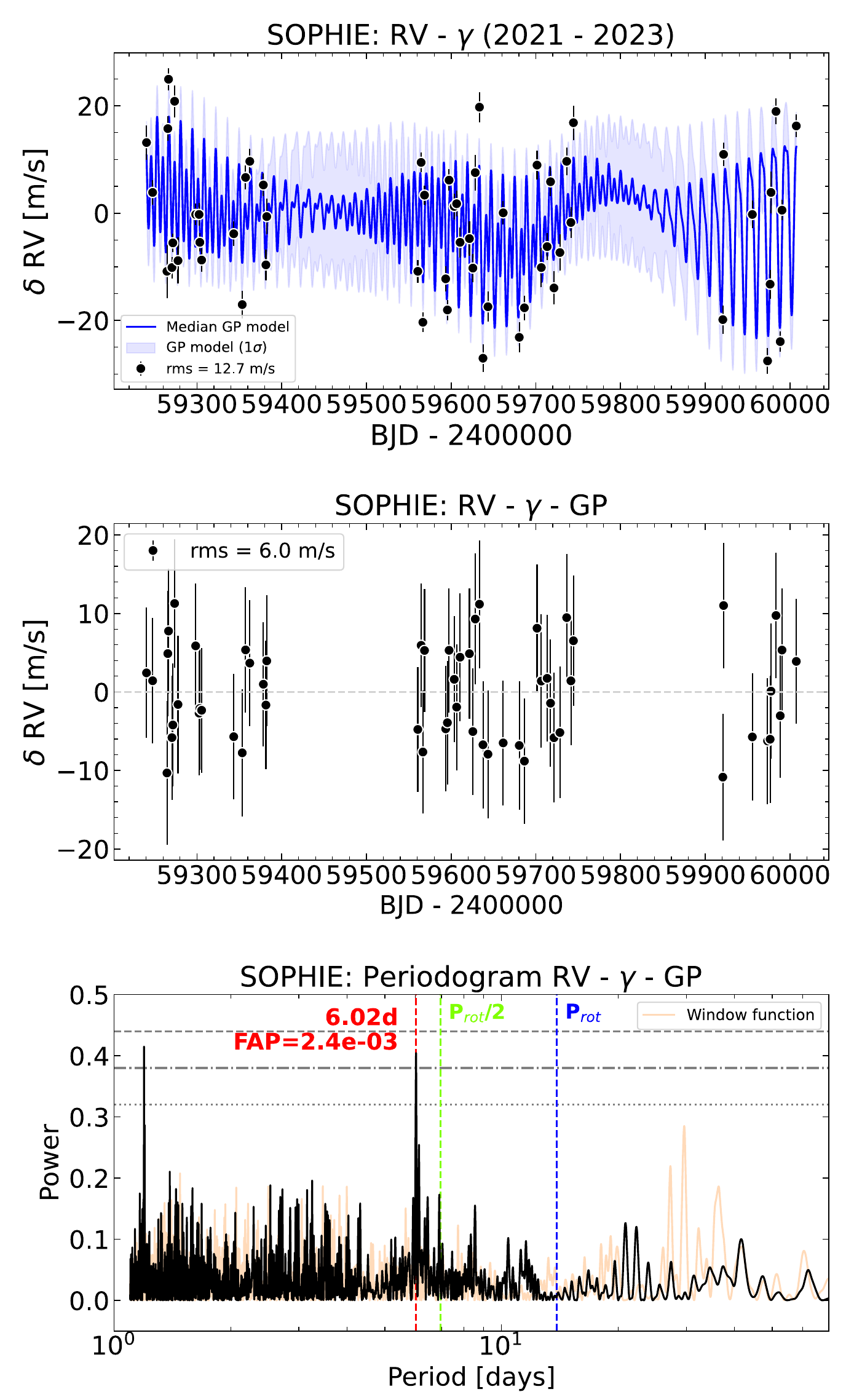}
\caption{
{\it Upper panel:} SOPHIE RVs minus the RV offset $\gamma$ (zoom-in of the period 2021$-$2023) together with  
the quasi-periodic GP model (blue lines) obtained using SPIRou's $B_\ell$ GP posterior distributions of the period, decay time, and smoothing length
as priors for the SOPHIE RVs GP. 
{\it Middle panel:} SOPHIEs RVs after correction of the RV offset and the GP.
The rms of the RV time series decreases from 12.7 m s$^{-1}$  to 6.0 m s$^{-1}$  in the GP-corrected data. 
{\it Lower panel:} Periodogram of the GP-corrected SOPHIE RVs. 
The peaks related to P$_{\rm rot}$/2 disappeared after GP correction.
A peak at the period of 6.02 days (FAP$_{\rm bootstrap}$=$2.4\times10^{-3}$),
the same period found in the SPIRou RVs,
is retrieved.
Horizontal lines are the 10\%, 1\%, and 0.1\% FAP levels (calculated using a bootstrap algorithm).
}
\label{SOPHIE_GP}
\end{figure}

We used the SPIRou $B_\ell$ measurements from \citet[][]{Donati2023}.
In Sect.~\ref{Bl_GP_modeling}  in the Appendix, 
we describe the quasi-periodic GP model we used.
In Fig.~\ref{fig:corner_bell}, also in the Appendix, we display the MCMC corner plots of the $B_\ell$ GP.
We retrieved in our GP model of $B_\ell$ a stellar rotational period of $13.93\pm0.09$ days,
which is consistent with the previous determinations of the rotational period by \citet[][]{Donati2023} and \citet[][]{Fouque2023}.
We modeled the SOPHIE RVs  with the same quasi-periodic GP model and implementation used for $B_\ell$.
We focused on  the SOPHIE RVs from 2021 to 2023,  
as this period has the quasi-simultaneous 
SPIRou $B_\ell$ measurements\footnote{\citet[][]{Bellotti2023PhD} investigated in detail the SPIRou $B_\ell$ 
measurements of Gl~410 using GPs. One of their conclusions was that the decay time differed in the 
2020$-$2021 and 2021$-$2022 data sets. Using ZDI, \citet[][]{Bellotti2023PhD} found
that the surface differential rotation was stronger in 2021 to 2022 
inducing a faster surface variability of activity tracers. 
Here, we use  a single GP for the complete $B_\ell$ time series. 
Therefore, our GP 
represents an ``average'' that does not capture these fluctuating activity levels.}. 
The prior and posterior distributions for the RV GP are given in Table~\ref{GP_table} in the Appendix.
In Figure~\ref{fig:corner_sophie_rv}, also in the Appendix,
we show the MCMC corner plots of the RV GP model.
Figure~\ref{SOPHIE_GP} summarizes the model results.
The upper panel shows the GP model (in blue) and the SOPHIE RVs (black points).
The middle panel displays the GP-corrected SOPHIE RVs.
The lower panel shows the GLS periodogram of the GP-corrected SOPHIE RVs.
After correcting for the activity jitter,
the RV rms decreases from 12.7 m~s$^{-1}$ to 6.0 m~s$^{-1}$,
and the GLS periodogram shows a clear and strong peak at $P=6.02$ days with a 
FAP$_{\rm bootstrap}$ of 2.4$\times10^{-3}$.
This period is the same as that obtained in the SPIRou RV measurements.

\begin{figure}
\centering
\includegraphics[width=0.5\textwidth]{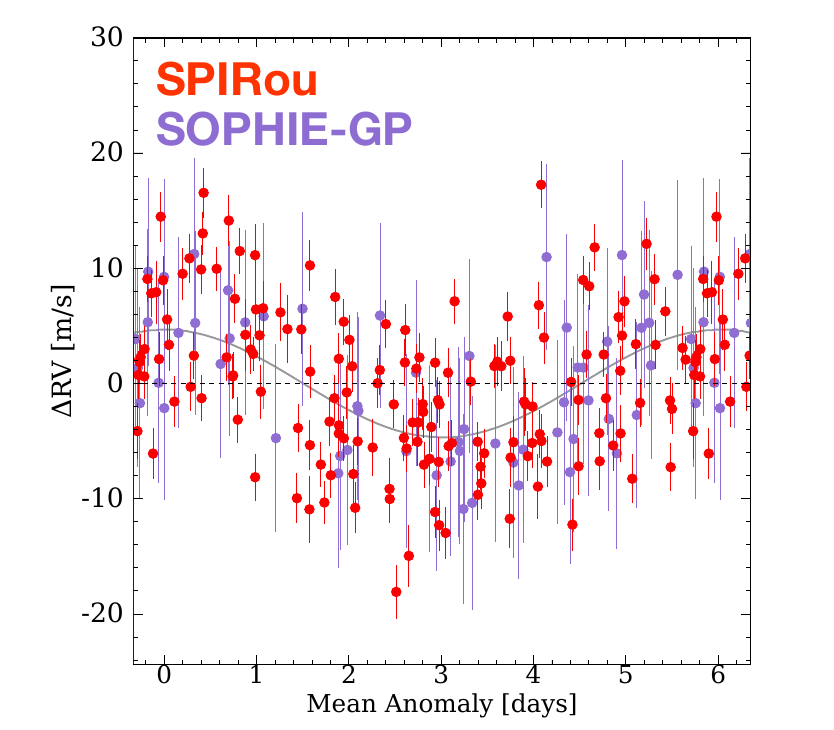}
\caption{
Phase-folded SPIRou (wPCA) and GP-corrected (using B$_\ell$ as an activity proxy) SOPHIE RV measurements.
The model in gray is a circular Keplerian fit to the SPIRou data with $K=5$ m s$^{-1}$ and $P=6$ days.
}
\label{SPIRou_SOPHIE_phase_folded}
\end{figure}

In Figure~\ref{SPIRou_SOPHIE_phase_folded},
we display the phase-folded SPIRou RV time series together with the SOPHIE GP-corrected RVs,
while overlaying a Keplerian model fit with $K=5$ m s$^{-1}$ and $P=6$ days. 
The data of both instruments are compatible.
This, 
provided us, 
one more time, 
with further arguments to support the idea that the 
periodic signal at 6 days detected with SPIRou indeed has a Keplerian origin.

\begin{figure}
\centering
\includegraphics[width=0.5\textwidth]{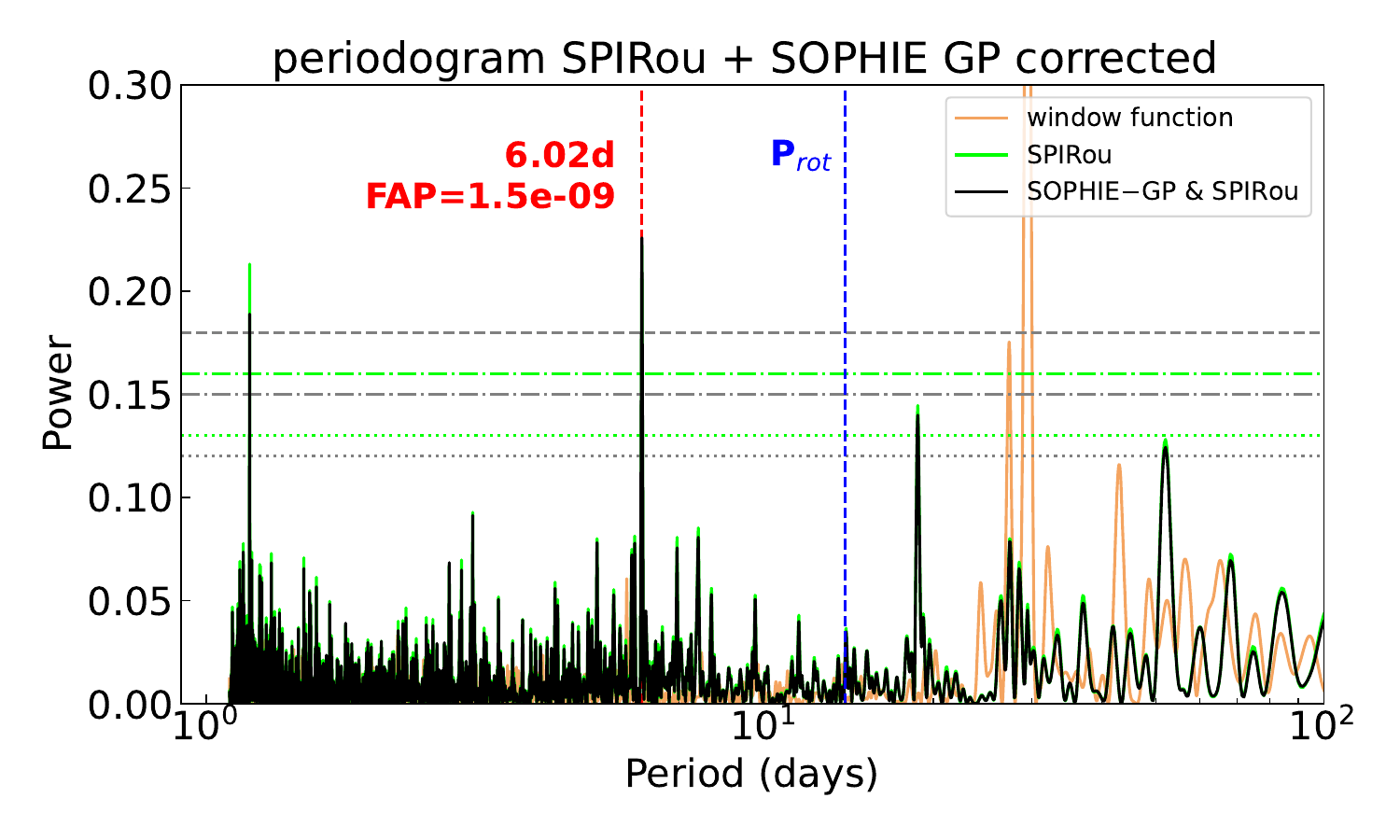}
\caption{Generalized Lomb-Scargle periodogram of the combined SPIRou (wPCA) 
and SOPHIE GP-corrected (using B$_\ell$ as an activity proxy) RV measurements.
The gray horizontal lines display the bootstrap-calculated FAP levels of 10\%, 1\%, and 0.1\%.
For comparison, 
we overplot in green the periodogram and FAP levels  of the SPIRou data alone.
The FAP of the 6-day period peak improves  in the  SOPHIE plus SPIRou time series 
with respect to the SPIRou data alone (see Fig.~\ref{fig:SPIRou_RV_periodogram}).
}
\label{SPIRou_SOPHIE_GP_periodogram}
\end{figure}

We calculated a periodogram for the combined 
SPIRou (wPCA) and SOPHIE minus GP  RV time series.
It is shown in 
Fig.~\ref{SPIRou_SOPHIE_GP_periodogram}.
The periodogram of the combined SPIRou and SOPHIE minus GP data set (black lines)
is very similar to that of the wPCA SPIRou data alone (green lines).
However, 
the addition of the SOPHIE minus GP RVs makes the FAP levels of the periodogram 
decrease to lower powers 
(we emphasize the difference between the horizontal gray and green lines in the plot). 
The significance of the peak at $P=6.02$ days increases 
from an FAP$_{\rm analytical}$=3.5$\times10^{-6}$ in the wPCA SPIRou data alone
to an FAP$_{\rm analytical}$=1.5$\times10^{-9}$ in the combined near-IR plus optical data set.

We find that the properties of the RV GP trained on $B_\ell$ are different 
from those of the GP based on the RV data alone.
The period of both GPs is the same, as we imposed in the priors the rotational period of the star. 
The amplitude of each GP is slightly different:
21$^{+19}_{-8}$ m s$^{-1}$ for the GP based on the SOPHIE RVs alone
and 12$\pm$3 m s$^{-1}$ for the GP trained on $B_\ell$.
The decay time of each GPs is significantly different,
from 602$^{+370}_{-230}$ days from the GP based on the SOPHIE RVs alone
to 60$\pm$8 days from the GP informed on $B_\ell$.
These differences in the GPs will
produce different amplitudes of the recovered Keplerian signal.
In Sect.~\ref{Gl410b_physical_properties}, 
we discuss in detail the properties of the Keplerian signal retrieved in the SPIRou data.
We show in Section~\ref{Gl410b_physical_properties},
that the $K_b$ recovered in the SOPHIE data set corrected with the GP using $B_\ell$
as a proxy of activity is almost the same as the  $K_b$ derived 
from the SPIRou wPCA and Wapiti data sets. 
Our analysis suggests that a GP informed on $B_\ell$ as 
proxy of activity provides a good correction for the stellar activity RV jitter in the optical.

\section{Gl~410b physical properties}
\label{Gl410b_physical_properties}

\subsection{Gl~410b physical properties derived from the SPIRou radial velocities}
We derived the Gl~410b planet candidate properties using \texttt{RadVel}.
We calculated \texttt{RadVel} models for each of the SPIRou RV data sets:
LBL  $|V_{\rm tot}|>10$ km s$^{-1}$, wPCA, and Wapiti.
Models with circular and eccentric orbits were considered.
In all models, 
we started performing a maximum-likelihood fit to define the prior centers of the fit manually
and then we performed the MCMC model exploration.
Uniform bounds priors were utilized for
the planet's period $P_b$ (between 0 and 20 days, with the initial guess at 6 days);
the planet's semi-amplitude $K_b$ 
(between 0 and 20 km s$^{-1}$, with the initial guess 6 m s$^{-1}$);
the SPIRou uncorrelated noise $\sigma_{\rm SPIRou}$ 
(between 0 and 10 m s$^{-1}$, with the  initial guess 1 m s$^{-1}$);
and $T{\rm conj}_{b}$ (2459648.0$\pm$10 days and an initial guess of 2459648.0).
The slope (\texttt{dvdt}) and  the curvature (\texttt{curv}) were set to zero.
We fed \texttt{RadVel} the median-subtracted RV points,
and thus, the initial guess for the velocity zero-point was set to zero.
The eccentricity was left to vary from zero to 0.99.
We used a stellar mass of 0.55$\pm$0.02 M$_{\odot}$.
The MCMC \texttt{RadVel} modeling used 50 walkers, 
10000 steps, eight ensembles
and a minimum autocorrelation factor of 40.

Table~\ref{radvel_SPIRou_model_comparison} show the results of the  \texttt{RadVel} AICc tests.
In the $|V_{\rm tot}|>10$ km~s$^{-1}$ data,
the circular orbit is "somewhat disfavored" with respect to the eccentric orbit.
In the wPCA data, 
the eccentric solution is favored, 
but the circular orbit is "nearly indistinguishable" from the eccentric orbit.
In the Wapiti data, 
the circular solution is favored,
but the eccentric solution is "nearly indistinguishable" from the circular solution.
The eccentricity distributions retrieved by \texttt{RadVel}
are $e_b=0.27\pm0.14$ for the $|V_{\rm tot}|>10$ km~s$^{-1}$ data,
$e_b=0.26\pm0.17$ for the wPCA data, 
and $e_b=0.21^{+0.17}_{-0.14}$ for the Wapiti data.
In all data sets, the eccentricity, $e_b$, is constrained to be lower than 0.4.
As the circular and eccentric orbits provide solutions that statistically are nearly indistinguishable
and the center of the eccentricity distributions is below 2$\sigma$ away from zero,
we favor a circular orbit model. This model has a lower number of free parameters,
and it provides a good description of the data.

\begin{figure*}
\centering
\includegraphics[width=\textwidth]{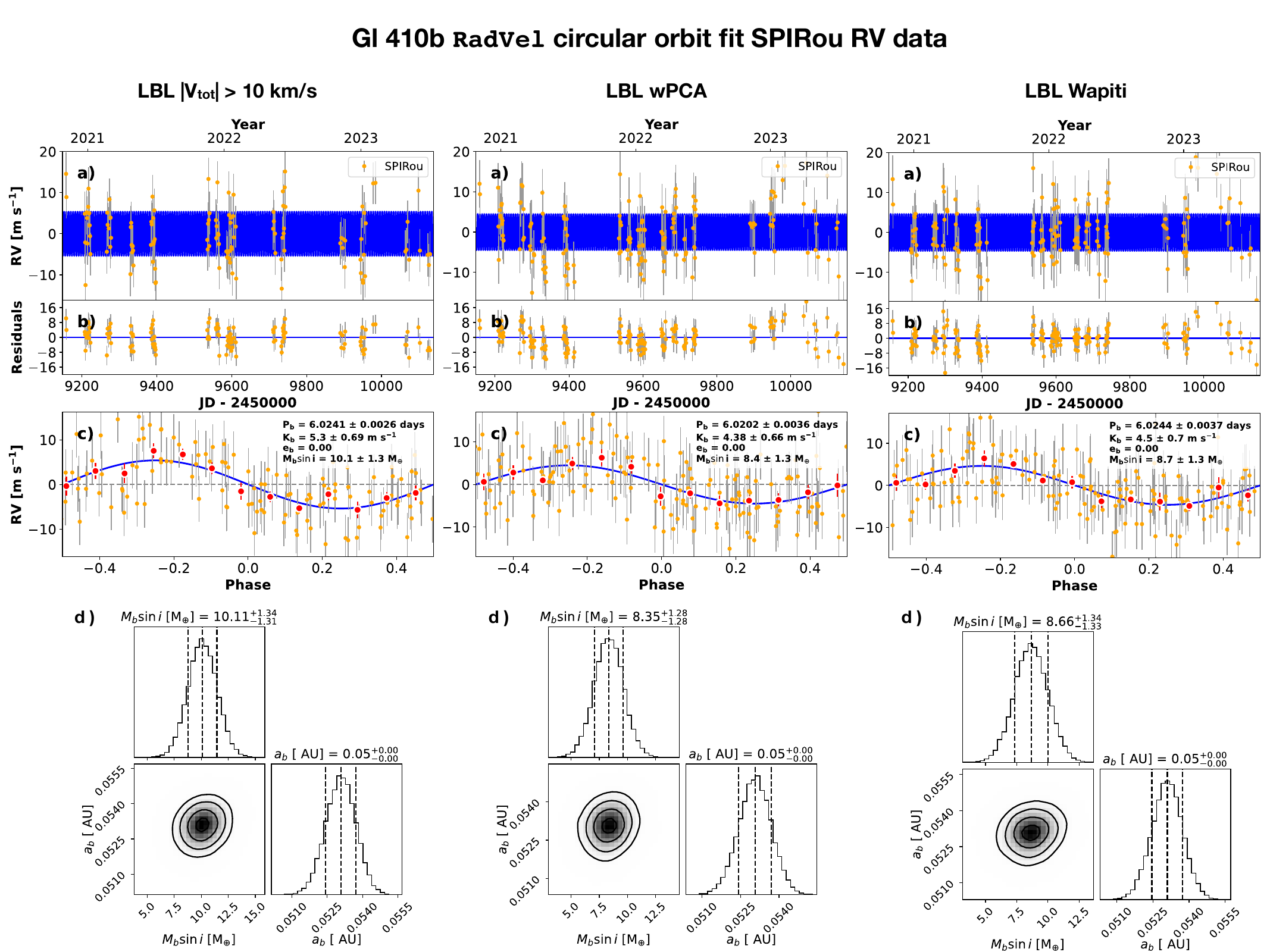}
\caption{Results of the \texttt{RadVel}  MCMC Keplerian  circular orbital model with one planet of the SPIRou
RV time series.
Each column represents the results for each set of data reductions:
LBL at $|V_{\rm tot}|>10$  km s$^{-1}$, wPCA, and Wapiti.
Panel $a$ shows the RV time series,
panel $b$ displays the residuals as a function of time,
panel $c$ plots the phase-folded RVs together with the best fit model (blue line),
and the panel $d$ displays the corner plots of $M_b\sin (i)$  and semimayor axis of the planet detected.
The red circles in panel $c$ are the RVs binned in 0.08 units of the orbital phase.
In all RVs, the \texttt{RadVel} RV jitter $\sigma_{\rm SPIRou}$ described in Table~\ref{planet-properties}
is added in quadrature to the RV uncertainties.
The orbital, planet, and fit parameters of the models are summarized in Table~\ref{planet-properties}.
Corner plots of the circular fit are provided in the Appendix (Figs.~\ref{SPIRou_corner_plot_radvel},
\ref{SPIRou_corner_plot_radvel_wPCA}, and \ref{SPIRou_corner_plot_radvel_wapiti}).
Our baseline solution is the model based on the SPIRou wPCA data set.
 }
 \label{radvel_SPIRou_LBL_wPCA_Wapiti_e0}
\end{figure*}

Table~\ref{planet-properties} summarizes the retrieved orbital and planet parameters
for the circular orbit model in each data set.
Figure~\ref{radvel_SPIRou_LBL_wPCA_Wapiti_e0} displays the phase-folded SPIRou RVs together
with the best-fit circular orbit model for each of the SPIRou RV reductions and
the MCMC corner plots for the planet's mass and semimajor axis.
The \texttt{RadVel} models further confirm the detection of Gl~410b.
As for all SPIRou data sets, the model without a planet is ``ruled out''.
Across all the SPIRou data sets and models the period of the planet converges to $P=6.02$ days;
thus, 
a semi-mayor axis of the planet, $a$, equal to 0.0531 au.
The wPCA and Wapiti data sets give similar properties for the planet.
The semi-amplitude retrieved in the LBL $|V_{\rm tot}|>10$ km s$^{-1}$ data is larger, 
5.3$\pm$0.7 m s$^{-1}$,  which translates to a $M_b\sin (i)$ of $10.1\pm 1.3$ M$_{\oplus}$.

We note that the PCA methods filtering the LBL data do in fact 
decrease the global dispersion of the all the RVs 
(not only those at  $|V_{\rm tot}|<10$ km s$^{-1}$).
This is likely because of the diminution of the correlated noise 
and telluric correction residuals.
However,
the estimations of the planet mass with and without PCA correction methods are 
consistent within their 1$\sigma$ error bar.
As the wPCA displays the second peak in the periodogram,
and the time series has the best correction of the  $|V_{\rm tot}|<10$ km s$^{-1}$ data,
we favor the wPCA model. 
In the following, the wPCA circular orbital model is our baseline solution.

Taking the results altogether,
the SPIRou RVs analysis indicate that the candidate planet Gl~410b 
has a $M \sin (i)$ of $8.4\pm1.3$ Earth masses,
orbits at a distance of 0.0531$\pm$0.0006 au from of its central star,
with a period of $P=6.020\pm0.004$ days 
in a circular orbit.
Gl~410b is a sub-Neptune planet.

\begin{table}
\begin{center}
\small
  \caption[]{ \texttt{RadVel} model comparison table for the SPIRou RVs data sets (one-planet model).}
\begin{tabular}{llccccc}
\\[-15pt]
  \hline
  \hline
{Model} & {$N_{\rm free}$} &{RMS} & {BIC} & {AICc} & $\Delta$AICc \\
 \hline
 \\[-5pt]
 \multicolumn{6}{c}{\bf LBL  $|V_{\rm tot}|>10$ km s$^{-1}$}\\
  \hline
  \\[-7pt]
{AICc Favored Model}\\
 $e_{b}$, $K_{b}$,  {$\sigma$}, {$\gamma$}     	& 5 	& 5.1 & 770  & 751	& 0.0  \\[5pt]
 Somewhat Disfavored \\
 $K_{b}$, {$\sigma$}, {$\gamma$}			& 7  	& 5.2 & 767 & 753  	& 2.1  \\[5pt]
  Ruled Out \\
  {$\sigma$}, {$\gamma$} 					& 2 	& 6.5 & 1868 & 1863	& 1112  \\
  \hline 
   \\[-5pt]
\multicolumn{6}{c}{\bf wPCA} \\
 \hline 
   \\[-7pt]
 {AICc Favored Model}\\
  $e_{b}$, $K_{b}$, {$\sigma$}, {$\gamma$} 		& 7 	& 5.8 & 1034 & 1013 	& 0.0  \\[5pt]
 Nearly Indistinguishable \\
   $K_{b}$,  {$\sigma$}, {$\gamma$} 				& 5  	& 5.9  & 1029 & 1014  	& 0.5 \\[5pt]
  Ruled Out \\
  {$\sigma$}, {$\gamma$} 						& 2 	& 6.8  & 2422 & 2416	& 1402  \\
  \hline
   \\[-5pt]
 \multicolumn{6}{c}{\bf Wapiti}\\ 
 \hline
   \\[-7pt]
 {AICc Favored Model}\\
 $K_{b}$,  {$\sigma$}, {$\gamma$} 				& 5  	& 6.0  & 1033 & 1018  	& 0.0\\[5pt]
 Nearly Indistinguishable\\
  $e_{b}$, $K_{b}$, {$\sigma$}, {$\gamma$} 		& 7 	& 5.9 & 1040 & 1019 	& 1.0 \\[5pt]
  Ruled Out \\
  {$\sigma$}, {$\gamma$} 						& 2 	& 6.9 & 2422 & 2416       & 1398 \\
  \hline
  \hline
\end{tabular}
%\\[-12pt]
 \tablefoot{ 
$N_{\rm data}=121$ for the LBL  $|V_{\rm tot}|>10$ km~s$^{-1}$ data set,
and 157 for the wPCA and Wapiti data sets. 
 }
\label{radvel_SPIRou_model_comparison}
\end{center}
\end{table}

\begin{table*}
\small
\begin{center}
\caption{Orbital parameters and physical properties of Gl~410b derived from the SPIRou RV data 
using a \texttt{RadVel} MCMC model of a planet in a circular orbit.}
\begin{tabular}{llrrrrrr}
\hline
\hline
\\[-5pt]
&									& \multicolumn{1}{c}{LBL  $|V_{\rm tot}|>10$ km s$^{-1}$} & \multicolumn{1}{c}{wPCA} 						& \multicolumn{1}{c}{Wapiti}\\[5pt]	
&									& 											& \multicolumn{1}{c}{(baseline solution)}		  		& 	\\
\hline
\\[-5pt]
{\bf Orbital parameters} &$P_{b}$	[days] 						&	$6.024\pm0.003$ 			& $6.020\pm0.004$ 	 	& $6.024\pm0.004$ 		\\[3pt]
&$K_{b}$	[m s$^{-1}$] 					& 	$5.3\pm0.7$				& $4.4\pm0.7$		 	& $4.5\pm0.7$			\\[3pt]	
&$T{\rm peri}_{b}$ [JD]  					&	$59646.67\pm0.12$			& $59646.59\pm0.17$	& $59646.61\pm0.16$ 	\\[5pt]					
{\bf Planet parameters} &$M_b\sin (i)$  [M$_{\oplus}$]  				& 	$10.1\pm1.3$				& $8.4\pm1.3$			 & $8.7\pm1.3$			\\[3pt]
&$a_b$ [au]				     			& 	$0.0531\pm0.0006$ 			& $0.0531\pm 0.0006$     	 & $0.0531\pm0.0006$ 	\\[3pt]
{\bf Other parameters}&$\sigma_{\rm SPIRou}$ [m s$^{-1}$] 			& 	$4.8\pm0.4$ 				& $5.6\pm0.4$ 			& $5.7\pm0.4$			\\
\hline
\end{tabular}
\tablefoot{ 
Epochs given in JD - 2400000.0
}
\label{planet-properties}
\end{center}
\end{table*}

\subsection{Gl~410b properties retrieved from the SOPHIE radial velocities corrected by activity}
\label{SOPHIE-GP-results}

As discussed in Sect.~\ref{SOPHIE_GP_Bl},
two methods were used to correct for the stellar activity RV jitter in the SOPHIE RV data
and search for the RV signal of Gl~410b.
In the first method, a \texttt{RadVel} Keplerian plus GP model was fit to the SOPHIE RV data 
using the SOPHIE RV data alone.
In the second method,
a GP model trained using SPIRou's longitudinal magnetic field measurements was fit to the SOPHIE RV time series, 
and then a \texttt{RadVel} Keplerian model is fit to the RV residuals.
In both methods, the Keplerian signal of Gl~410b is recovered at 6 days.
The aim of this section is to check the derived planet properties, 
in particular $K_b$, from the two activity-correction methods and compare them 
to properties retrieved from SPIRou RVs.

When GPs are used to correct for the stellar activity RV jitter,
there is uncertainty as to how good the activity correction is
and, therefore, how reliable the  derived planet's properties are.
Using the planet properties derived from SPIRou wPCA RVs as benchmark,
we aim to determine which activity correction method provides the 
best match.
As the SPIRou data suggest a circular orbit, 
we only used circular orbits for  this analysis.
We limited the analysis to the  2021 to 2023 SOPHIE observations,
as these data have quasi-simultaneous SPIRou $B_\ell$ measurements.

\begin{figure}
\centering
\includegraphics[width=0.47\textwidth]{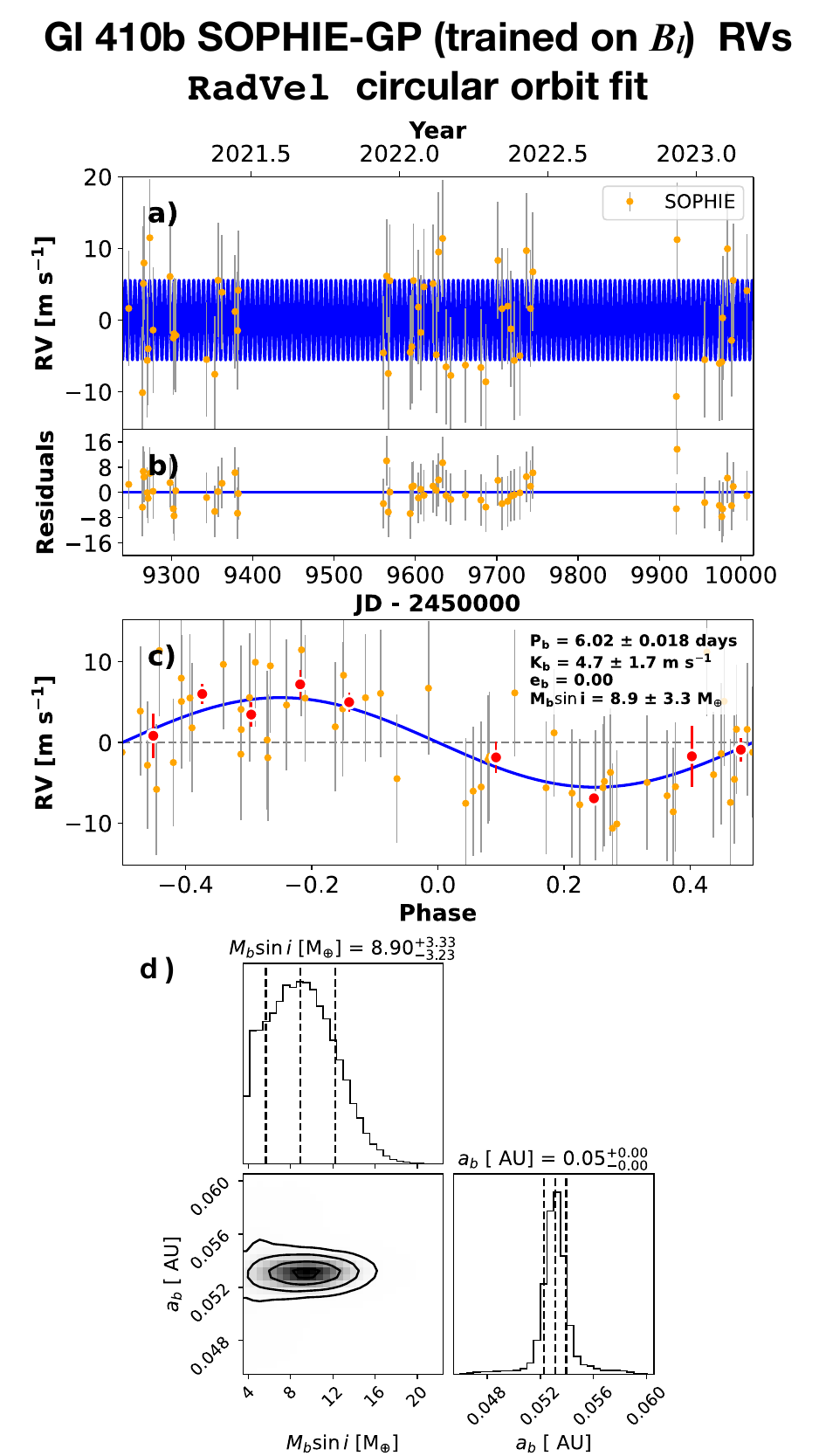}
\caption{Results of the \texttt{RadVel} model of one planet in circular orbit for 
the SOPHIE RVs corrected for activity using a GP that used SPIRou's $B_\ell$ measurements
as an activity proxy. We note that only the 2021$-$2023 data was used, 
as these data have quasi-simultaneous SPIRou $B_\ell$ data. 
The planet parameters obtained from the SOPHIE$-$GP RVs are very similar to those retrieved 
from the wPCA and Wapiti SPIRou data.
}
\label{SOPHIE_1planet_e0}
\end{figure}

\begin{table}
\begin{center}
\small
  \caption[]{\texttt{RadVel} model comparison table for the SOPHIE minus GP (trained on $B_\ell$) data set (one-planet model).}
  \label{comparison_SOPHIE_radvel_Bell}
\begin{tabular}{llccccc}
  \hline
  \hline
{Model} & {$N_{\rm free}$} &{RMS} & {BIC} & {AICc} & $\Delta$AICc \\
 \hline
 \\[-5pt]
{AICc favored model}\\
 $K_{b}$, {$\sigma$}, {$\gamma$} 			& 5  	& 4.6  & 382 & 372 	& 0.0\\[5pt]
  Ruled out \\
  {$\sigma$}, {$\gamma$} 					   & 2     & 6.1 & 883 & 880	& 507  \\
\hline
\end{tabular}
\\[-12pt]
\end{center}
\end{table}

As previously discussed, for the \texttt{RadVel} Keplerian plus GP model using the SOPHIE RV data alone, 
we used a uniform distribution prior 
for the planet's orbital period (0 to 20 days; i.e., we left \texttt{RadVel} free to find the period),
the planet's semi-amplitude $K_b$ (0.0 to 20.0 m s$^{-1}$),
the time of inferior conjunction \texttt{tc} (2459648.0$\pm$10 days),
and the uncorrelated noise $\sigma_{\rm SOPHIE}$ (0.0 to 15.0 m s$^{-1}$).

We used the same broad priors on the \texttt{RadVel}  model of the RV data corrected 
by activity with the GP trained on $B_{\ell}$. 
\texttt{RadVel} retrieved the  expected period of 6 days.
However, to have a better determination of the planet parameters,
we narrowed down the uniform priors to
$P_b$  (5  to 7 days)\footnote{This is justified because in the SOPHIE RV-GP (trained on B$_\ell$) data set, 
the periodogram shows a strong peak at 6.02 days
and the 6-day period is retrieved using the broad prior.},
\texttt{tc} (2459648.0$\pm$3 days),
and $K_b$ (2.0 to 20.0 m s$^{-1}$).

Table~\ref{comparison_SOPHIE_radvel} already presents the results of the \texttt{RadVel} AICc
tests for  the Keplerian plus GP model for the GP using the SOPHIE RV data alone.
Table~\ref{comparison_SOPHIE_radvel_Bell} describes the results of the AICc
tests for the SOPHIE RV minus GP informed on the $B_\ell$ time series.
For both activity-correction methods,
the \texttt{RadVel} AICc tests ``rule out'' the model without a planet.
Table~\ref{planet-properties-SOPHIE-GP} summarizes the orbital solutions
found by \texttt{RadVel} for both activity methods. 
Both methods of activity filtering suggest the presence of a sub-Neptune mass planet 
with a period of 6.02 days and semimajor axis 0.0531 au.

\begin{table}
\small
\begin{center}
\caption{Circular orbit parameters and physical properties of Gl~410b  derived from the activity-corrected SOPHIE RV measurements
using a GP based on the SOPHIE RVs alone and a GP using $B_\ell$ as an activity proxy. }
\begin{tabular}{lrrrrrr}
\hline
\hline
\\[-5pt]
&  \multicolumn{1}{c}{SOPHIE$-$GP} &  \multicolumn{1}{c}{SOPHIE$-$GP} \\
&  \multicolumn{1}{c}{GP on RV$_{\rm SOPHIE}$ alone} &  \multicolumn{1}{c}{GP informed on $B_{\ell~{\rm SPIRou}}$} \\[5pt]	
\hline
\\[-5pt]
\multicolumn{2}{l}{\bf Orbital parameters}\\[3pt]
$P_{b}$	[days] 					&	$6.020^{+0.005}_{-0.006}$ 			&   $6.02^{+0.02}_{-0.01}$		 	 	\\[3pt]
$K_{b}$	[m s$^{-1}$] 				& 	$7.6\pm1.4$						&   $4.7\pm1.7$			\\[3pt]
$T{\rm peri}_{b}$ [JD]  				& 	$59646.53\pm0.19$					&   $59646.62^{+0.38}_{-0.42}$			\\[3pt]
\\[1pt]
\multicolumn{2}{l}{\bf Planet parameters}\\[1mm]
$M_b\sin (i)$  [M$_{\oplus}$]  			& 	$14.5\pm2.6$ 						 &   $8.9\pm3.3$			\\[3pt]
$a_b$ [au]				     		& 	$0.0531\pm0.0007$    				 &   $0.0531\pm0.0008$ 			\\[3pt]
\hline
\end{tabular}
\tablefoot{ 
Epochs given in JD - 2400000.0
}
\label{planet-properties-SOPHIE-GP}
\end{center}
\end{table}

\begin{table}
\begin{center}
\small
  \caption[]{\texttt{RadVel} model comparison table of the one-planet model of the combined 
  SPIRou wPCA and SOPHIE$-$GP (trained on $B_\ell$) data sets.}
\begin{tabular}{llccccc}
\\[-15pt]
  \hline
  \hline
{Model} & {$N_{\rm free}$} &{RMS} & {BIC} & {AICc} & $\Delta$AICc \\
  \hline
  \\[-7pt]
\multicolumn{2}{l}{AICc Favored Model}\\
  $K_{b}$, {$\sigma$}, {$\gamma$}			& 7  	& 5.6 & 1403 & 1380.3 	& 0.0 \\[5pt]
\multicolumn{2}{l}{Nearly Indistinguishable} \\
 $e_{b}$, $K_{b}$,  {$\sigma$}, {$\gamma$}     	& 5 	& 5.6 & 1410  & 1380.6	& 0.34  \\[5pt]
  Ruled Out \\
  {$\sigma$}, {$\gamma$} 					& 2 	& 6.6 & 3308 & 3295		& 1915 \\
  \hline 
  \end{tabular}
\\[-12pt]
\label{AICtests_comparison_1planet-model_SPIRou_SOPHIE-GP}
\end{center}
\end{table}

The SOPHIE RV \texttt{RadVel} GP plus Keplerian model retrieves a $K_{b}$ of 7.6$\pm1.4$ m s$^{-1}$,
thus a $M_b\sin (i)$ of 14.5$\pm2.6$ M$_{\oplus}$.
The SOPHIE$-$GP informed on the $B_\ell$ time series model retrieves
a $K_{b}$ of $4.7\pm1.7$  m s$^{-1}$ and 
$M_b\sin (i)$ of $8.9\pm3.3$ M$_{\oplus}$.
While both methods provide a $K_{b}$, 
and thus a $M_b\sin (i)$, 
consistent between them at the 1$\sigma$ level,
the activity-correction method that provides the closest match to the $M_b\sin (i)$ derived from the SPIRou RVs
is the correction by the GP trained on SPIRou's $B_\ell$.
We display the model in Figure~\ref{SOPHIE_1planet_e0}.

Our analysis suggests that the measurements of $B_\ell$ performed in the near-IR with SPIRou 
could be a good activity proxy to correct for the stellar RV activity jitter in 
quasi-simultaneous measurements taken in the optical. 
The independent detection of the same signal in both the SPIRou and SOPHIE data sets
gives solid support to a planet being responsible of the periodic variation
seen in near-IR and optical RVs.

\subsection{A joint model for Gl~410b using the SPIRou wPCA and the activity-corrected SOPHIE data sets}
\label{SOPHIE_GP_SPIRou_radvel_1p}
As the planet signal is recovered independently in the near-IR and in the optical,
a joint model using both RV data sets can be performed.
Merging RV data from different wavelengths and instruments has risks and can be problematic, 
as instruments have different RV systematic noise, and stellar activity is chromatic.
Nevertheless, 
to check the properties derived from a combined optical and near-IR RV data set,
we used \texttt{RadVel} and calculated a joint solution for the orbital parameters of Gl~410b using 
the wPCA SPIRou RVs
and the activity-corrected (using $B_\ell$ as an activity proxy) SOPHIE RVs.
We used a uniform distribution priors
for the planet's orbital period $P_b$ (0 to 20 days),
the planet's semi-amplitude $K_b$ (0.0 to 20.0 m s$^{-1}$),
the time of inferior conjunction \texttt{tc} (2459648.0$\pm$10 days),
the uncorrelated noise $\sigma_{\rm SPIRou}$ (0.0 to 10.0 m s$^{-1}$),
and $\sigma_{\rm SOPHIE}$ (0.0 to 10.0 m s$^{-1}$).
We tested models with circular and eccentric orbits.

Table~\ref{AICtests_comparison_1planet-model_SPIRou_SOPHIE-GP} summarizes 
the results of the \texttt{RadVel}  AICc tests.
The model without a planet was ``ruled out'' by \texttt{RadVel}.
The model with a planet in circular orbit is favoured,
although a model with an eccentric orbit is ``nearly indistinguishable'' from the model with a circular orbit. 
We favor the circular solution because it describes correctly the data with a model with fewer free parameters.
We display the retrieved planet properties in Fig.~\ref{SOPHIE_wPCA_1planet} in the Appendix.
The combination of the wPCA SPIRou and the activity-corrected SOPHIE$-$GP data 
gives essentially the same planet properties as the
wPCA SPIRou data alone.
The only (small) differences are 
a decrease in the period uncertainty  from 0.004 to 0.003 days 
and an increase in $M_b\sin (i)$ 
from $8.4\pm1.3$ M$_{\oplus}$ to $8.7\pm1.2$ M$_{\oplus}$ 
($K_b=4.4\pm0.7$ m s$^{-1}$ to $K_b=4.6\pm0.6$ m s$^{-1}$) in the combined data set.
The larger dispersion of the SOPHIE$-$GP RVs with respect of wPCA SPIRou RVs
could be responsible of the small increase on the planet's semi-amplitude.
The uncertainty in the period becomes smaller in the combined 
data set because in both data sets the planet signal is time and phase coherent.

\begin{figure}
    \centering
    \includegraphics[width=0.45\textwidth]{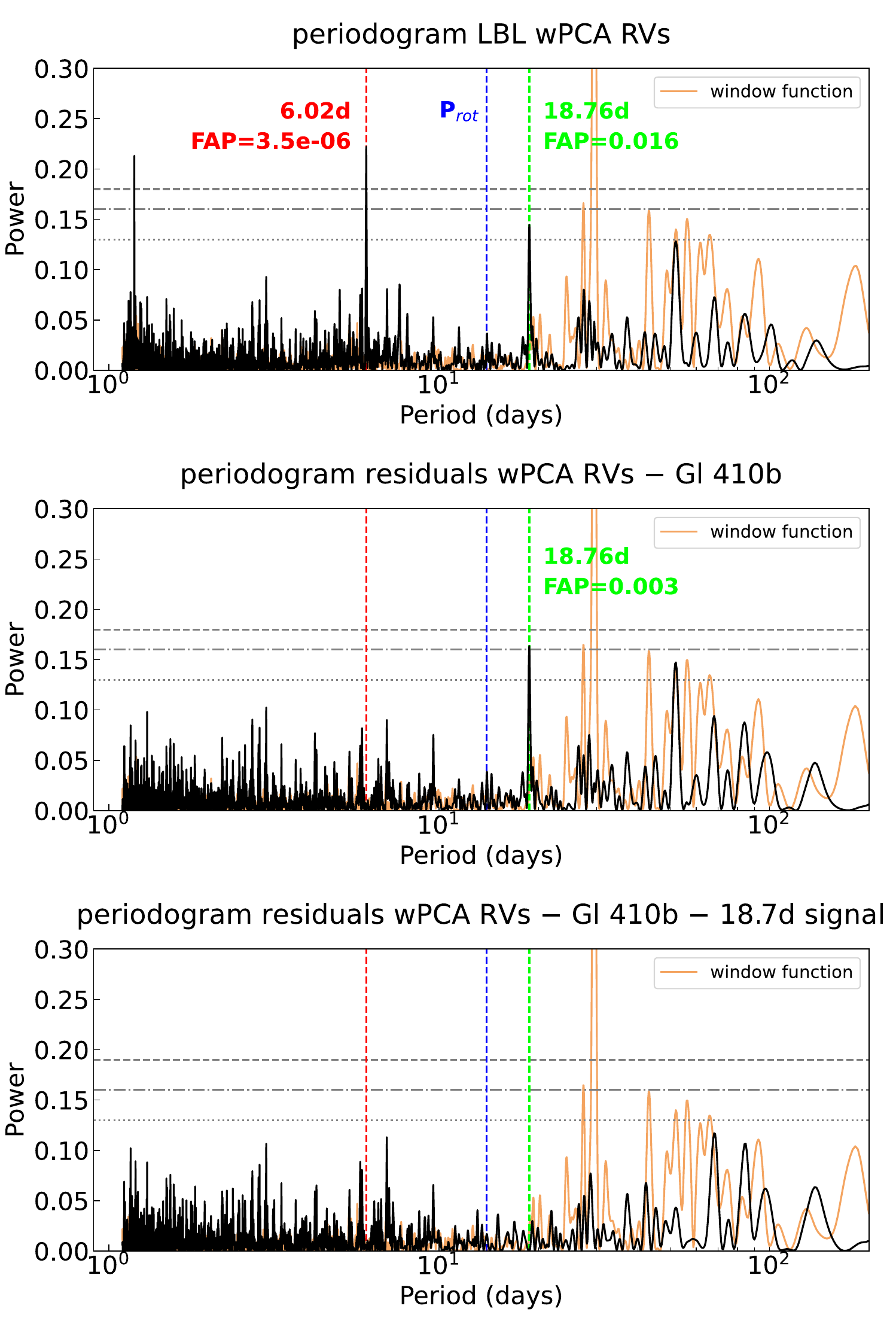}
    \caption{Periodograms of the SPIRou wPCA data
    after subtraction of the circular orbit models of Gl~410b and the candidate planetary signal at 18.7 days.
    }
      \label{GL410_wPCA_2planet}
\end{figure}

\section{Search for additional signals within the SPIRou wPCA data}

As presented in Sect.~\ref{Sec:SPIRou_radial_velocities} the GLS periodogram of the wPCA SPIRou RVs 
shows in addition to the 6.02 day peak a second peak at 18.76 days with FAP$_{\rm boostrap}$=1.6\%.
The presence of additional planetary signals in the data is encouraging as it is known from transit searches
that M-dwarfs often exhibit multi-planetary systems.
In this section we investigate the possibility of having additional planetary signals within the wPCA SPIRou RVs.

\subsection{Possibility of a planet at P=18.7 days}
If the 6.02-day Keplerian signal of Gl~410b (Table~\ref{planet-properties})
is subtracted from the  wPCA SPIRou time series and a new GLS periodogram is calculated on the residuals,
we find that the signal at $P=18.76$ days is still present.
Furthermore, its FAP$_{\rm boostrap}$ level improves to 0.3\%
(Fig.~\ref{GL410_wPCA_2planet}, middle panel).
This provides tentative evidence in favor for the presence of a second planet.

To further test the plausibility of the existence
of the signal at 18.7 days, we calculated a \texttt{RadVel}
MCMC run using a two-planet model.
For this test, 
we set broad uniform priors for
$P_b$ (0 to 20 days),
$P_c$ (0 to 30 days),
and   $K_b$ and $K_c$ (0 and 20 m s$^{-1}$).
Circular and eccentric orbit $(e<1)$ models were calculated.
The model priors and initial guesses are summarized in Table~\ref{Radvel_2planet_model_priors} in the Appendix.
Our goal was to determine which planet solution is favored by \texttt{RadVel}
and to compare the periods blindly retrieved by \texttt{RadVel} with those of the periodogram 
analysis.

The \texttt{RadVel} AICc model comparison  is presented in 
Table~\ref{tab_comp_2planet_models} in the Appendix.
\texttt{RadVel} finds the two-planet solution more likely compared to one-planet solution.
In fact, the models with one planet were ``ruled out'' by \texttt{RadVel}.
\texttt{RadVel} favors a model in which planet b has an eccentric orbit and planet c has a circular orbit
and a model in which both b and c have circular orbits. 
However, the two models are ``nearly indistinguishable'' for \texttt{RadVel}.
We favor the model in which the two planets are in circular orbits.

In the Appendix,
we provide the corner plots of the model with two planets in circular orbit (Fig~\ref{SPIRou_corner_plot_radvel_wPCA_2planet}),
a plot showing the phase-folded RVs, and the best fitting two-planet circular solution (Fig~\ref{2planet_radvel_wPCA}).
We also show a table summarising the retrieved orbital parameters and physical properties (Table~\ref{2planet}).

\texttt{RadVel} indicates for the candidate signal at 18.7 days a $M_c\sin (i)= 9.9\pm1.8$ M$_{\oplus}$ 
($K_c=3.6\pm0.7$ m s$^{-1}$)
and a period $P_c=18.76\pm0.03$ days ($a_c=0.113\pm0.001$ au).
This period, retrieved blindly by \texttt{RadVel}, corresponds 
to the peak in the periodogram once the RV signal of Gl~410b is removed from the data (middle panel of Fig.~\ref{GL410_wPCA_2planet}).
The properties of Gl~410b are similar to those retrieved in the one-planet model.
We note that $M_b\sin (i)$ slightly decreases to 7.9$\pm$1.2 M$_{\oplus}$,
which is within the 1$\sigma$ error of the one-planet model value.
The bottom panel of Fig.~\ref{GL410_wPCA_2planet}
displays the periodogram of the residuals of SPIRou wPCA RVs 
after subtraction of the signal of Gl~410b and the candidate signal at 18.7 days.
No peaks with a significance higher than FAP=10\% are left in the periodogram.
We also tested a \texttt{RadVel}  model on the combined SPIRou wPCA and SOPHIE$-$GP data set (not shown).
In this data set,
the model with two planets in circular orbit was favored by the AICc tests as well.  

Thus, Gl~410 has tantalising  evidence for hosting two sub-Neptune planets in circular 
orbits close to a 3:1 resonance. 
For curiosity, 
using the \texttt{RadVel} posterior distributions of the eccentric orbit models, 
we explored the stability of such a system (Sect.~\ref{dynamical_stability_analysis} in the Appendix).
We found that the system is safely stable for circular orbits and the nominal eccentricity values.

\subsection{$\ell_1$ and apodized sine periodogram analysis}

The results of the GLS periodogram and the two-planet \texttt{RadVel} modeling are promising.
To perform an independent assessment on the existence of periodicities within the SPIRou wPCA 
data set, we performed a $\ell_1$ and apodized sine periodogram (ASP) analysis. 
The $\ell_1$ periodogram searches for several signals simultaneously, 
and the ASP tests the consistency in frequency, phase, and amplitude of the candidate signals.   
We used the $\ell_1$ periodogram technique as defined in~\cite{hara2017}. 
This tool is based on a sparse recovery technique called the basis pursuit algorithm~\citep{chen1998}. 
It aims to find a representation of the RV time series as a sum of a small number of sinusoids whose frequencies are in the input grid. 
It outputs a figure that has a similar aspect as a regular periodogram but with fewer peaks due to aliasing. 
The peaks can be assigned an FAP, whose interpretation is close to the FAP of a regular periodogram peak.

The $\ell_1$ periodogram takes several parameters as input, 
in particular, a list of vectors, or predictor, 
that are fitted linearly along with the search for periodic signals. 
For this linear base model, 
we chose several stellar activity indicators available within the SPIRou data:
the longitudinal magnetic field ($B_\ell$) measurement,
the  LBL projection onto the second derivative of the spectrum (\texttt{d2v}),
the LBL projection onto the third derivative of the spectrum (\texttt{d3v}), 
the LBL differential line width (\texttt{dLW}),
and LBL the differential temperature measurement (\texttt{dTEMP}) with respect to a $T=4000$ K template.
The LBL activity indicators are calculated at the same time as the RVs \citep[][]{Artigau2022,Artigau2024}. 
We also added a periodic signal at the sidereal day
to account for apparent systematics at this period. 
The $\ell_1$ periodogram also necessitates to assume a certain form of the noise model. 
Following the method of~\cite{hara2020}, 
we tested a grid of noise models, 
and ranked them with cross-validation and Bayesian evidence, 
which we computed with the Laplace approximation. 
The detailed analysis is presented in Appendix~\ref{app:l1}.

\begin{figure}
    \centering
    \includegraphics[width=\linewidth]{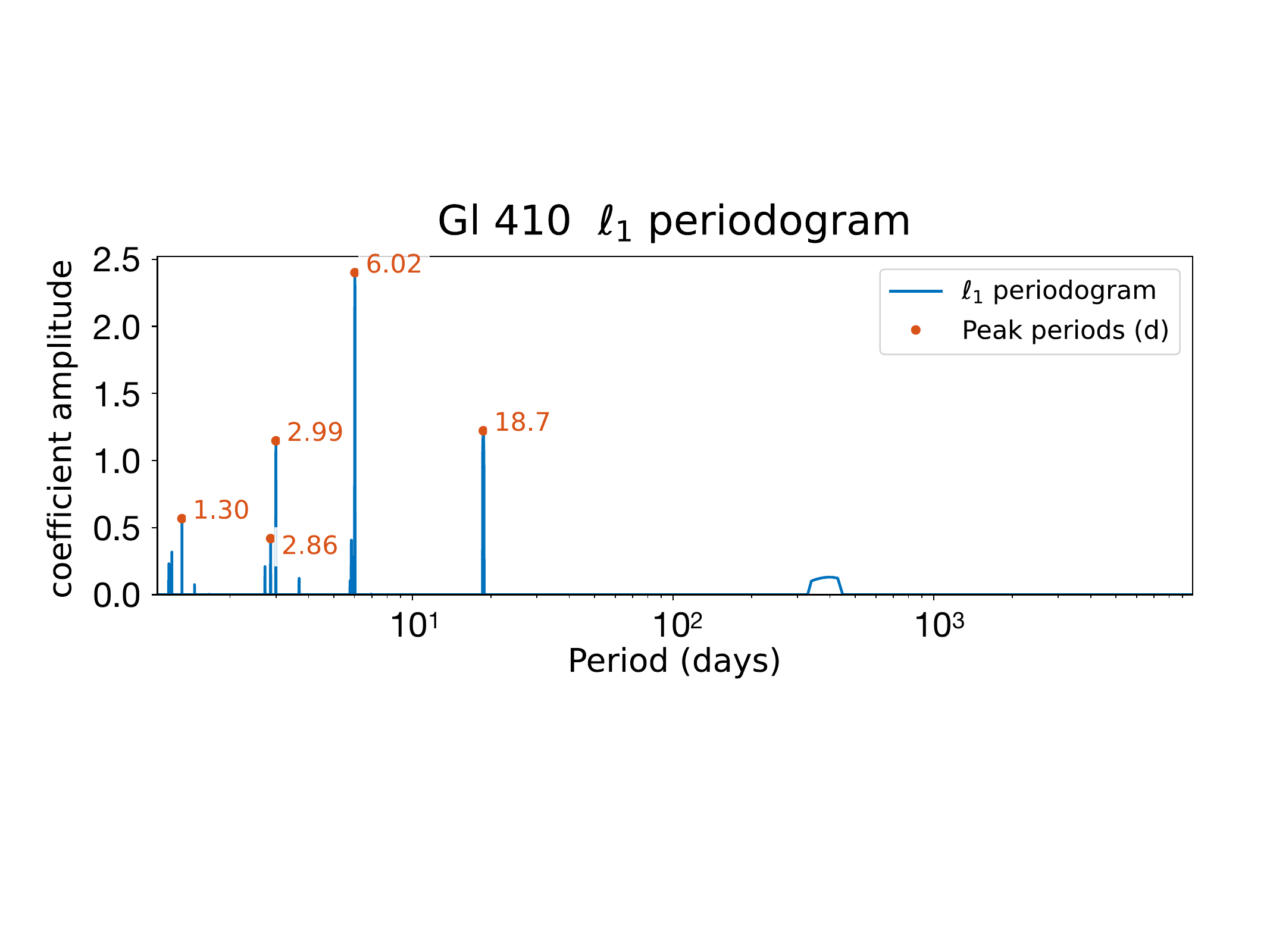}
    \caption{ $\ell_1$ periodogram of the model corresponding to the highest value of Bayesian evidence, calculated with the Laplace approximation. }
    \label{fig:l1_highestCV}
\end{figure}

 \begin{figure*}
   \centering  
    \includegraphics[width=0.65\textwidth]{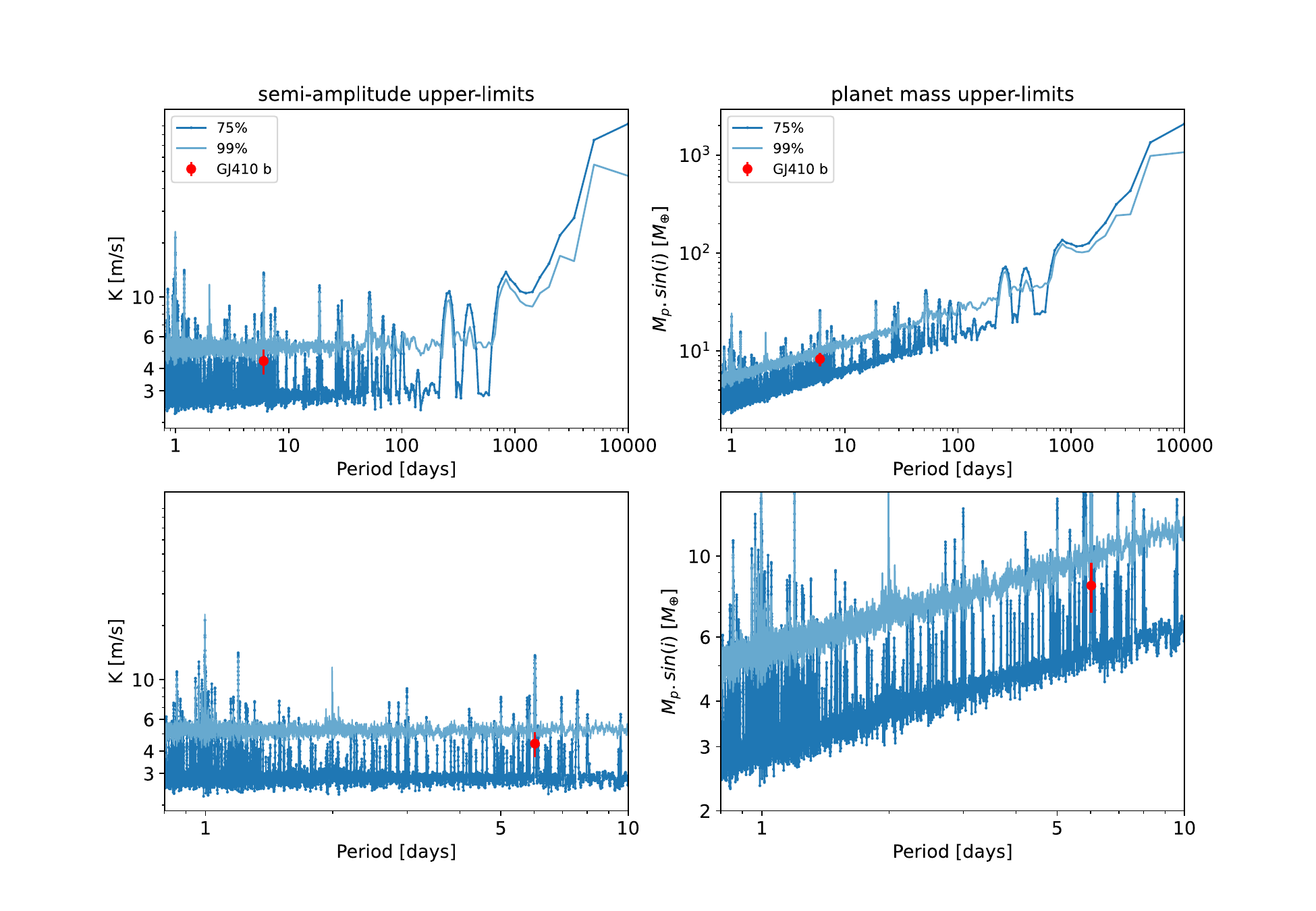}
    \caption{ Detection limits in the SPIRou wPCA RV time series.  
    The left panels show the upper limits on the semi-amplitude $K$ in meters per second.
    The right panels display the limits in terms of the projected mass m~{\it sin(i)}.
    A zoom-in of the period range  from 1 to 10 days is presented in the lower panels.
    Objects with a projected mass above the light blue and dark blue lines
    are ruled out with a confidence level of 99\% and 75\%, respectively. 
    The red dot indicates the location of Gl~410b.}
    \label{detection_limit}
\end{figure*}

In Figure~\ref{fig:l1_highestCV}, 
we represent the $\ell_1$ periodogram of the noise model with the highest cross-validation score. 
We found clear evidence for the 6.02-day signal with an FAP of $7.4\times10^{-9}$. 
The 18.7-day signal is also present with a $7.3\times10^{-3}$ FAP. 
We note the presence of  an additional 2.99-day signal with an FAP $3.8\times10^{-4}$. 
Finally, we found a marginally significant signal at 1.3 days. 
The signals at 6, 18, and 3 days are consistently found by the computed $\ell_1$ periodograms,
they present peaks in 100\% of the top 20\% of the models with highest cross-validation scores.     

To further refine the analysis, 
we tested whether the candidate signals are strictly periodic
or if their frequency, phase, and amplitudes show signs of variation. 
For this,
we employed the method of~\cite{hara2022b}, 
which consists in fitting apodized sinusoidal signals to the data set. 
These signals are wavelet-like, 
composed of a sine multiplied by a Gaussian function localized in time. 
In Appendix~\ref{app:l1}, 
we show that there is good evidence that the three signals found by the $\ell_1$ periodogram  are strictly periodic.

In summary,  
apart from the clear 6.02-day planet, 
we find in the $\ell_1$ and ASP analysis statistically significant candidates at 18.7 and 2.99 d, 
which are compatible with purely periodic signals. 
We note that the proximity of the 2.99 d planet to a 2:1 mean motion 
resonance with the 6-day planet further supports that it is indeed is a planet.
This third planet would have a mass of a few Earth masses.
Nonetheless, we do not label the two signals as confirmed planets, 
as their statistical significance is not completely sufficient. 

To conclude this section, 
we insist that although the detections of the signals at 18.7 and 2.99 days
pass several statistical tests,
at the present they are candidate signals because they are only retrieved in the wPCA SPIRou data.  
Further measurements and improvements on the methods to recover the data in the |V$_{\rm tot}$|<10 km s$^{-1}$ 
region are required to fully exploit all the measurements taken and to ultimately confirm 
or refute the planetary nature of these signals.

\subsection{Planet detection limits at long periods from the SPIRou wPCA radial velocities}
\label{upperlimits}
We computed the planet detection limits from the SPIRou RVs by performing an injection recovery test.
The method determines the upper limits by injecting signals
in the time series at given periods, calculating the GLS periodograms, 
and determining the maximum amplitude that can still be missed with a given probability.
For each period, 
the upper limit is given by the highest amplitude (at the worst phase) that 
could be present in the time series without creating a peak above the 1\% FAP level in the periodogram. 
The method consists of injecting for each period a sinusoidal Keplerian signal  (i.e., circular orbit) at 12 equi-spaced phases.
The amplitude (i.e., the planet signal strength) is then increased until the corresponding peak in the 
GLS periodogram reaches the 1\% FAP level.
The 1\% FAP confidence level was computed from the levels of the highest peaks of 1000 permuted data sets, 
and the 1\% upper limits was obtained by sorting the 
power of the peaks of all periodograms at each period.
We tested periods from 0.8 to 10000~days.

Figure~\ref{detection_limit} displays the projected mass as a function of the period 
at which 99\%  and 75\% of the planets injected would be detected in the GLS periodogram of the SPIRou data.
The degraded upper limits for the periods around 1, 2, 180, and 360 days are 
due to the data sampling and are an unavoidable consequence of observing at night from a single location 
and from the fact that SPIRou
is only mounted in bright time.
Our SPIRou RV data can detect with a 99\% confidence planets with masses larger than 
0.1, 1.0 and 4.7 M$_J$  at a periods of 100, 2560 and 5000  days (0.35, 3 and 5 au), respectively.  

\citet[][]{Kervella2022} 
performed a study aimed at detecting companions of Hipparcos catalog stars based on the proper motion anomaly  derived from Hipparcos and Gaia measurements.
Gl 410 is part of that study, and it displays a proper motion anomaly S/N of 0.13.
\citet[][]{Kervella2022} derived a companion mass of 0.02 M$_J$ (6.3 M$_\oplus$) in a 3 to 5 au orbit.
The mass of this companion candidate is significantly below the sensitivity capability of our current SPIRou RV measurements.

\section{TESS photometry analysis}
\begin{figure}
\centering
\includegraphics[width=0.49\textwidth]{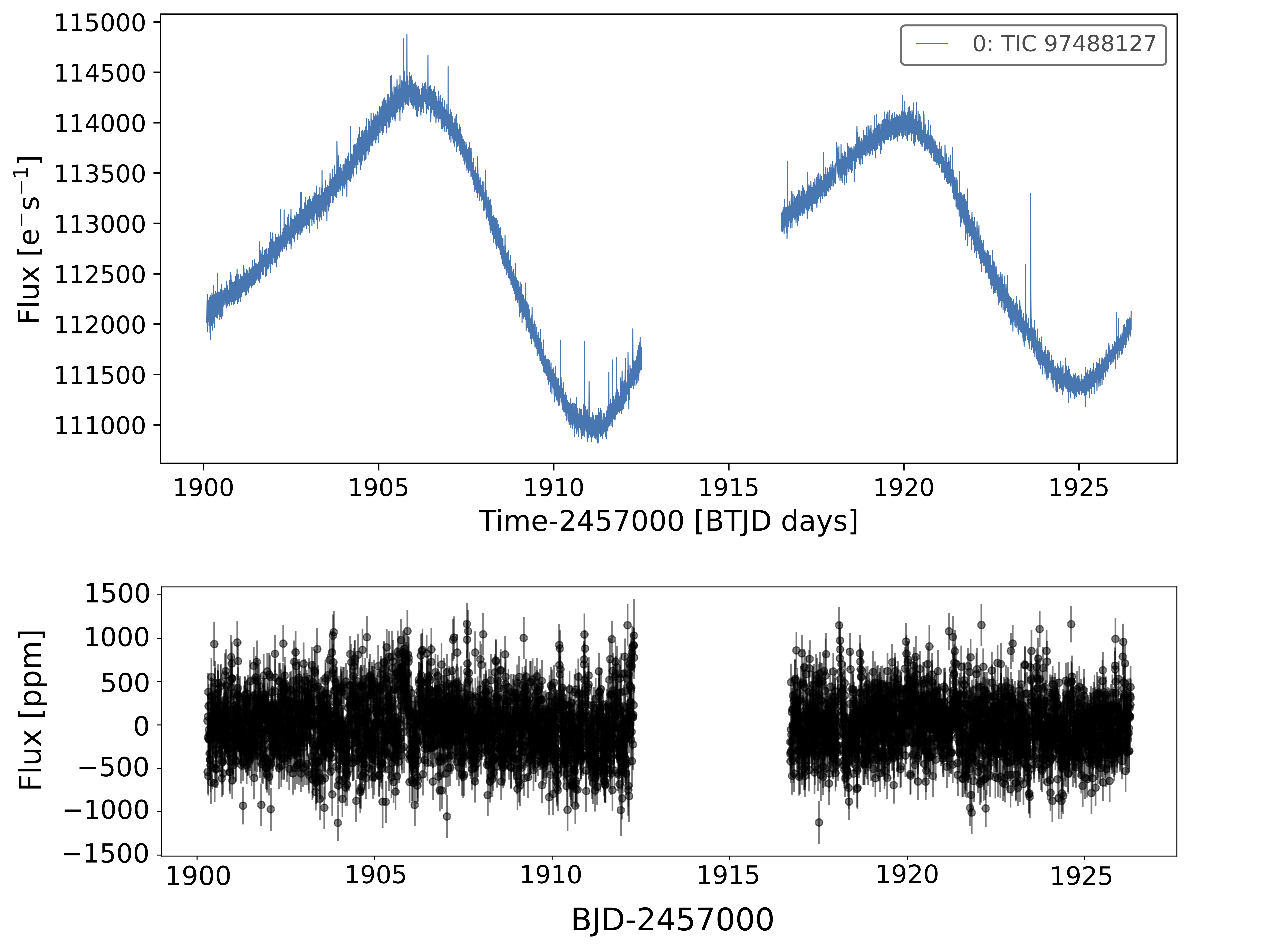}
\caption{
TESS photometry data for Gl~410. 
{\it Upper panel:} Presearch Data Conditioning flux time series processed by the TESS Science Processing Operations Center.
{\it Second panel:} \texttt{Wotan} \citep[][]{Hippke2019} detrended light curve.
}
\label{TESS_time_series}
\end{figure}

\begin{figure}
\includegraphics[width=0.48\textwidth]{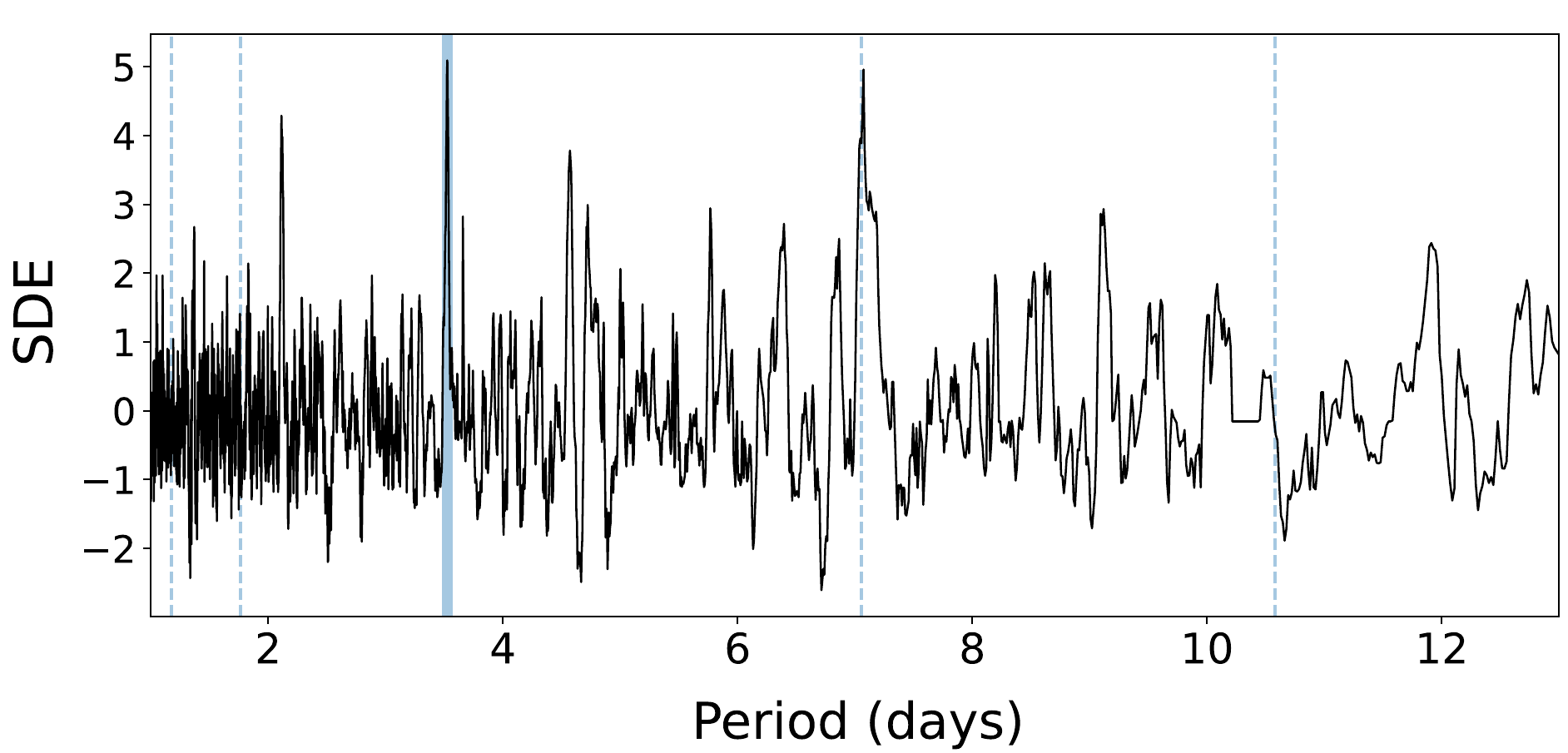}
\caption{\texttt{Wotan} signal detection efficiency as a function of the period.
The vertical light blue line at 3.5 days is the period with the highest SDE, 
and the dashed lines correspond to its harmonics. 
We note that the signal is not significant. 
}
\label{Wotan_SDE}
\end{figure}

The TESS photometry time series is shown in Fig.~\ref{TESS_time_series}.
One can clearly observe the modulation of the TESS light-curve by the stellar rotation.
However,  it is not possible to measure the rotation period with high accuracy
because the time span of the observations ($\sim$27 days) 
does not fully cover two complete rotation periods.
No transits were identified by the TESS automatic pipeline.
To further search for transit events, 
we applied a procedure customized for Gl~410. 
After  removing the outliers  using a 3$\sigma$ clipping procedure, 
we used the package \texttt{Wotan} \citep[][]{Hippke2019} to flatten the light 
curve and to remove the effects of variability due to stellar rotation.
The second panel of Fig.~\ref{TESS_time_series} 
displays the resultant detrended light curve,
which has a scatter of 403 ppm. 
Using the transit least squares algorithm of \citet[][]{HippkeHeller2019},
we searched for periodic transit events in the detrended light-curve, 
in particular  at the orbital period of planet b.
The algorithm retrieved periodic signals at 3.5 and 7 days with a signal detection efficiency (SDE) on the order of 5, 
which is not high enough to claim a detection (Fig.~\ref{Wotan_SDE}).  
Further inspection of those signals revealed thay they are artifacts.
The signals are close to the harmonics of the stellar rotation period and 
are likely related to residuals of the detrending procedure. 
No signal was detected at period of 6.02 days, suggesting 
that Gl~410b is not transiting.

\subsection{Transit detection limits}
\begin{figure}
\centering
\includegraphics[width=0.5\textwidth]{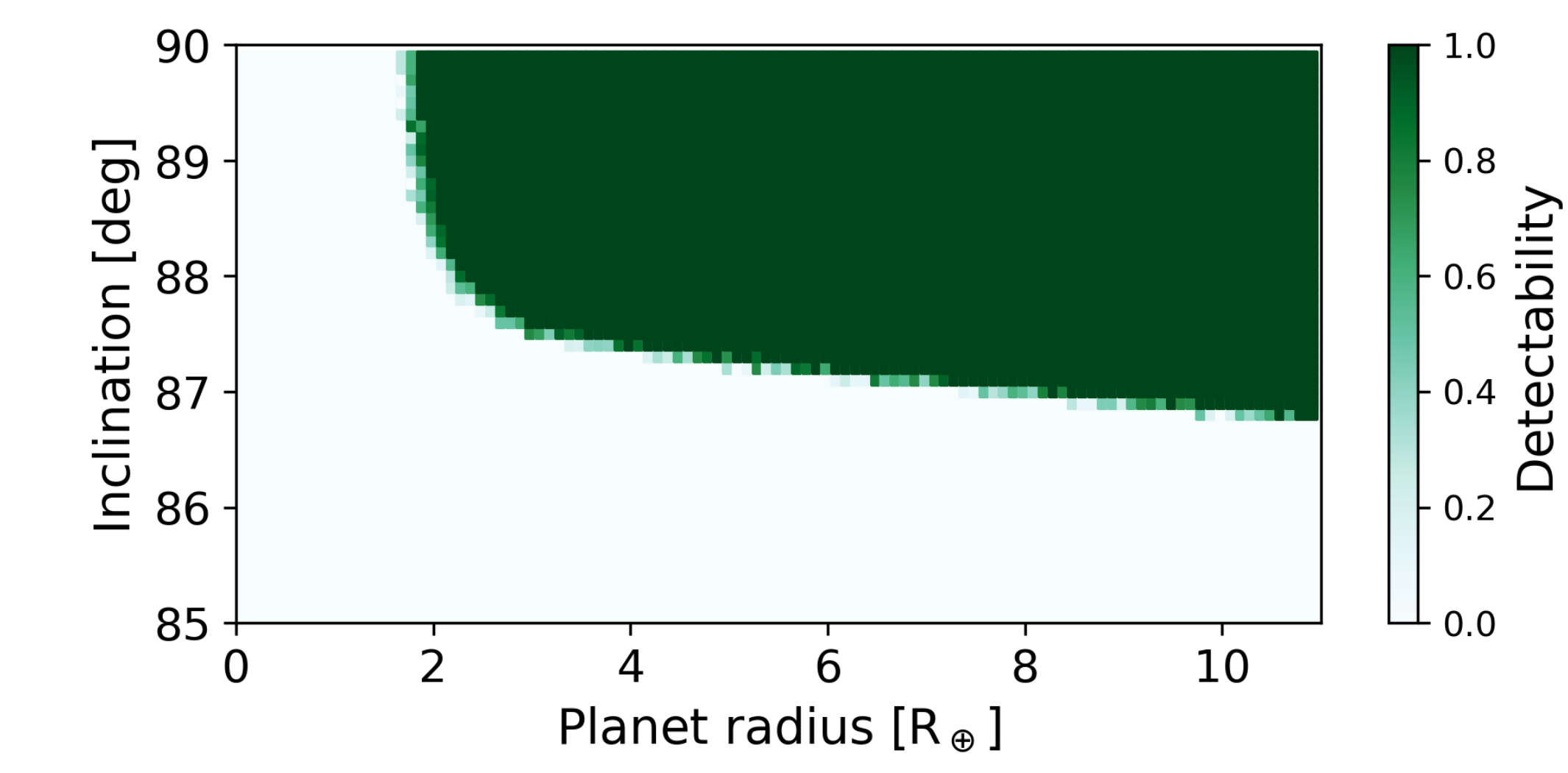}
\caption{Results of the injection recovery test in the TESS data. The figure shows the probability of detection of a transit for a given planet radius and orbit inclination. 
}
\label{injection-recovery_Gl410_TESS}
\end{figure}

To determine the detection limits in the TESS data,
we conducted an injection recovery test of transit events.
Following \citet[][]{Cortes-Zuleta2025},
using the stellar parameters of Gl~410 and the semimajor axis of 0.053 au,
we injected 50\,000 transits generated with the software \texttt{batman} \citep[][]{Kreidberg2015}, 
for which the orbital inclination and planet radius are drawn randomly from uniform distributions:
between 85$^{\circ}$ and 90$^{\circ}$ for the inclination 
and from 0.1 to 10\,$R_\oplus$ for the radius. 
For each synthetic transit,
we computed the box least squares periodogram\footnote{\url{https://docs.astropy.org/en/stable/timeseries/bls.html}} 
implemented in \texttt{astropy} and measured the power of the signal at a period of 6.02 days. 
Then, we measured the fraction of transit events that are detectable with the box least squares periodogram (i.e., power $>$ 1000). 
The results of the injection test are shown in Fig.~\ref{injection-recovery_Gl410_TESS}.
We fond that planets with a radius smaller than 1.8 R$_\oplus$ 
are not detectable in the current TESS data as well as 
planets in orbits with inclinations lower than 87 deg.

\section{Discussion}

\subsection{Gl~410b  in context}
Gl~410b sums up to the growing list of sub-Neptune mass planets detected around low-mass stars ($M<0.6$ M$_{\odot}$).
In order to place the Gl~410 planetary system in context,
in this section we concentrate on discussing the planet detection statistics of the stars in the 0.5$-$0.6 M$_\odot$ stellar mass bin,
that is,
the stars with a mass equal to that of Gl~410.
At the time of writing (1 July 2024), 
this stellar mass range,
according to the NASA exoplanets catalog\footnote{https://exoplanetarchive.ipac.caltech.edu},
has 212 planet candidates detected (of 5678 exoplanets).
Several of those planets are awaiting further characterization,
but 140 already have a reported radius measurement,
83 have a planetary mass measurement
and, 
30 have a mass constraint (i.e., $M_b\sin (i)$).

The period is always determined in transit and in RV planet detections.
However, in both methods, 
detection statistics are strongly positively biased to short periods,
as planets in short periods are the easiest to detect.
At the time of writing,
in the $0.5-0.6$ M$_\odot$ stellar mass bin,
of the  212 planet candidates, 
there are 161 detections at  $P_{\rm orbit}<$ 50 days,
with 80 \% of those detections at  $P_{\rm orbit}<$ 15 days.
These numbers illustrate the observational biases mentioned.
Perfoming a detailed study correcting for biases is out of the scope of this discovery paper;
however,
we can point out that planets in a 6-day period, such as Gl~410b, 
are not unusual in the $0.5-0.6$ M$_\odot$ stellar mass bin.

\begin{figure}
\centering
\includegraphics[width=0.45\textwidth]{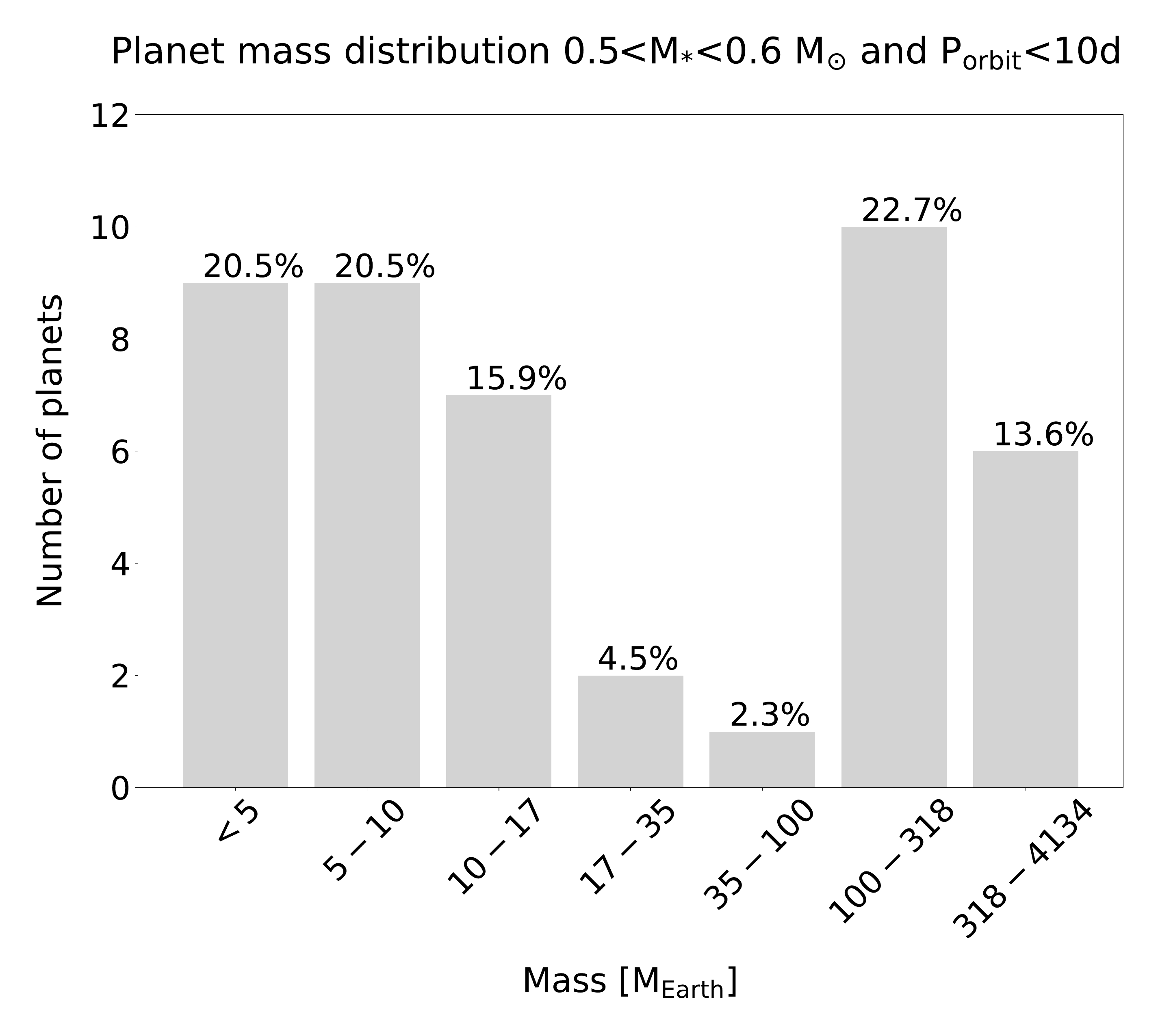}
\includegraphics[width=0.45\textwidth]{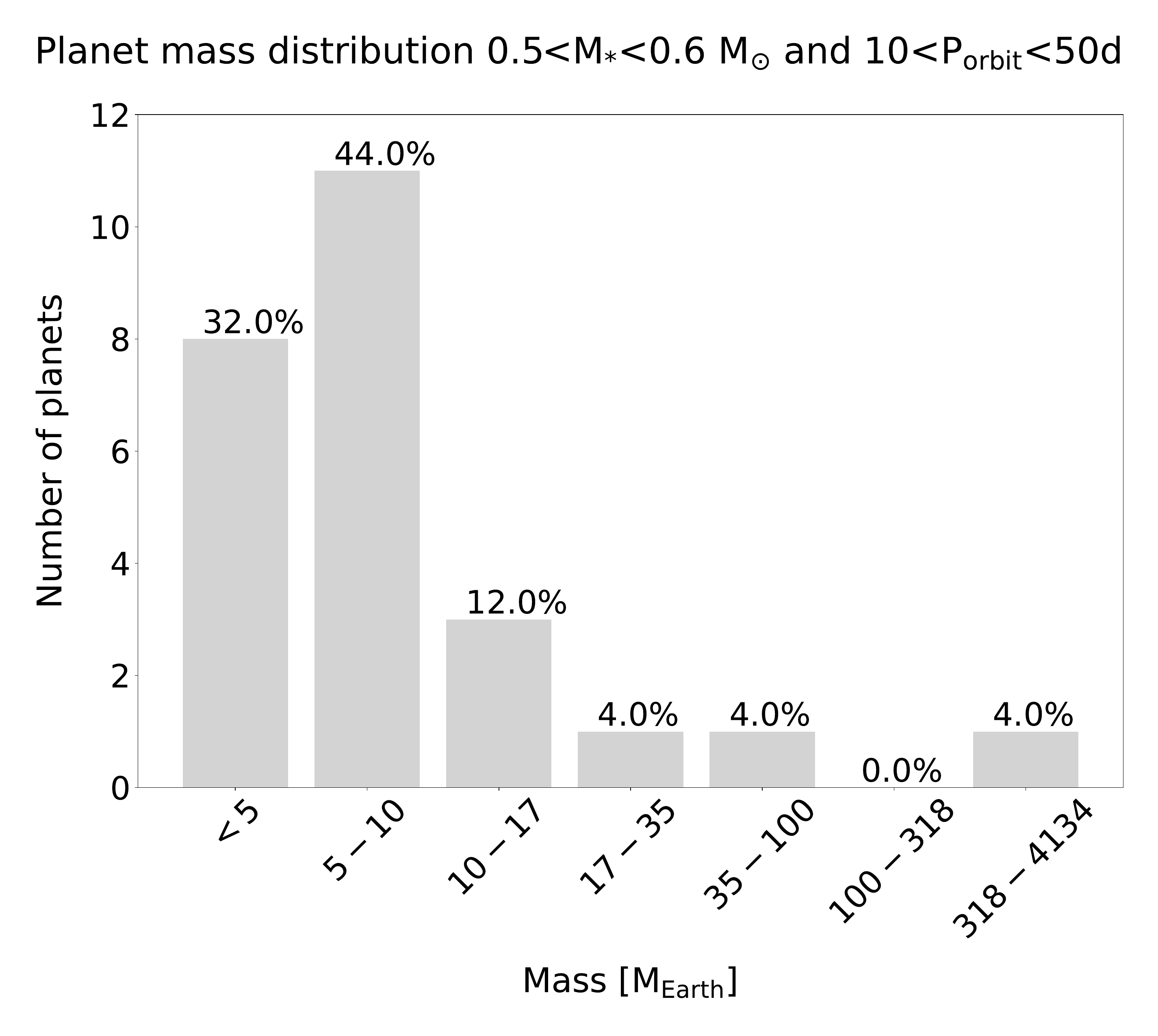}
\caption{Histograms of planet detections in the $0.5-0.6$ M$_\odot$ stellar mass bin.
Data from the NASA exoplanets catalog as of 1 July 2024.}
\label{histograms_planet_detections}
\end{figure}

Concerning the  planet mass, 
we are interested in discussing 
how abundant or scarce planets are in the mass range of Gl~410b (8.4$\pm$1.3 M$_\oplus$)
in orbits with periods  $P_{\rm orbit}<$ 50 days around stars in the $0.5 - 0.6$ M$_\odot$ stellar mass bin. 
We show in Fig.~\ref{histograms_planet_detections} 
the histograms of the planet mass distribution for a $P_{\rm orbit}<10$ days (top panel)
and for $P_{\rm orbit}$ between 10 and 50 days (bottom panel).
The histogram includes both mass measurements and $M_b\sin (i)$ lower limits.
In the subsample with $P_{\rm orbit}<10$ days,
the most abundant planets are those with masses less than 10 M$_\oplus$, 
with $\sim$40\% of the detections.
These are followed by planets with mass between 1/3 and 1 M$_{J}$, 
with 23\% of the detections.
Detection biases make planets with masses less than 10 M$_\oplus$ the hardest to detect.
The large fraction of detections of planets in the low mass range
means that they should  intrinsically be the most numerous.
The histograms show a significant decrease on the frequency of planets for the mass bin between 
17 M$_\oplus$ (Neptune mass)  and  100 M$_\oplus$ ($\sim$ 1/3 M$_{J}$).
As planets in this mass range are more easily  detected by RVs than planets with $M<$10 M$_\oplus$,
the decrease is real, and it is not a detection bias.
We find that Gl~410b is located in the bin with the highest probability of detection.

In the context of the subsample of planets with periods between 10 and 50 days,
we find that
{\it i)} the mass bin between 5 and 10 M$_\oplus$ has the highest fraction of detections,
{\it ii)} there is  a decrease in the number of detections of planets with $M<$ 5 M$_\oplus$,
and
{\it iii)} there are very few detections of giant planets.
The smaller fraction of low-mass planets could be an observational bias given
the fact that detecting planets with $M<$ 5 M$_\oplus$ around M dwarfs at longer periods is likely 
harder because the amplitude becomes smaller,
and it could be comparable to the activity jitter.
The very small fraction of giant planets, in contrast, is not an observational bias,
and it must be real, as giant planets are easier to detect through RVs. 
The tentative planet candidate signal at 18.7 days would thus belong to 
the category of the most frequently detected planets.

The very fact that Gl~410b is located in the bins with the highest number of detections
reinforces the current view that sub-Neptune mass planets are abundant around M dwarfs.
The decrease of the occurrence rate with increasing planetary mass
is a property that has already been highlighted in studies covering the entire M dwarf mass domain
(e.g., \citet[][]{Bonfils2013}, \citet[][]{Sabotta2021}, \citet[][]{Mignon2025}).
Current statistics raise the question of why 
planets with masses between Neptune and 1/3 M$_{J}$ are scarce
at periods shorter than 50 days around stars whose mass is in the $0.5-0.6$ M$_\odot$ mass bin.
Whether this could be a consequence of the planet formation process, 
or a product of the interaction of planetary atmospheres with energetic radiation and 
stellar wind particles due to the high activity levels of M dwarfs is an open question.

Since Gl~410b is not transiting, we do not have a stringent  constraint on its radius.
Based on the sample of planets in  the $0.5 - 0.6$ M$_\odot$ stellar mass bin
with $P_{\rm orbit}<50$ days and masses similar to Gl~410b  (7 to 10 M$_\odot$),
at present there are only a few planets for whom measurements of the radius are available, 
for example,
TOI-1725b \citep[$R_{\rm p}=1.8\pm0.1$\,R$_\oplus$,~$M_{\rm p}$=9.9$\pm1.3$\,M$_\oplus$,][]{Essack2023}
and 
TOI-2018b \citep[$R_{\rm p}=2.27\pm0.07$\,R$_\oplus$,~$M_{\rm p}$=9.2$\pm2.1$\,M$_\oplus$,][]{Dai2023}.

\subsection{Stellar activity and the evolution of Gl~410b}

Let us conclude this section by briefly discussing the implications of stellar 
activity in the atmospheric evolution of Gl~410b\footnote{Due to the pending the confirmation of the planet at 2.99 days with new measurements,
we not mention it in this discussion. However, the arguments exposed here for Gl~410b would be applicable to this planet given its shorter period.}.
As presented in Section \ref{gl410_properties}, Gl~410 is an active star.
Activity is the manifestation of the presence of stellar magnetic fields.
Magnetic fields heat the chromosphere and coronae and 
drive stellar winds \citep[e.g.,][]{CranmerWinebarger2019}.
High-energy radiation from stellar flares and X-ray/EUV coronal emissions
and particle fluxes from winds and coronal mass ejections 
interact with the magnetospheres and the atmospheres of planets in orbits close to the star,
thus driving thermal and non-thermal atmospheric escape processes \citep[for recent reviews see][]{Vidotto2022,Kubyshkina2024}.

As Gl~410b is  located very close to its host star at 0.05 au,
it is therefore expected to have an active interaction with the high-energy
radiation and the stellar wind particle fluxes from Gl~410.
Using the radius and $T_{\rm eff}$ given in Table~\ref{stellar-properties}, 
we obtained that Gl~410b receives 20.4$\times$ Earth's insulation.
It is out of the scope of this discovery paper to perform a detailed 
study on the conditions on which a planet with 9 Earth masses could retain its
atmospheres in the case that it is in the proximity of an active M dwarf such as Gl~410,
but it is an interesting topic that deserves further study.
Furthermore, 
it would be of extreme interest to investigate
the influence that stellar irradiation and activity could have in the chemical composition of the planet's atmosphere 
and the potential development of abundance anomalies due to the escape of volatile species \citep[e.g.,][]{Louca2023}
in single- and multi-planetary systems \citep[e.g.,][]{Acuna2022}.

\section{Conclusion} 
In this paper,
we have presented an RV study of Gl~410, a nearby ($d$=12 pc) early M dwarf 
with a mass of 0.55 M$_\odot$ and an age of 480$\pm$150 Myr\,\footnote{This work, see Appendix \ref{age_determination}.}. 
We monitored the star in the near-IR with the spectropolarimeter and velocimeter SPIRou
and in the optical with the velocimeter SOPHIE.
The SPIRou RVs show a robust periodic signal at $P$=6.020$\pm$0.004 days.
The signal is recovered in the raw LBL data and in the two PCA methods used to correct the LBL RVs
for systematics and telluric correction residuals.
In our baseline solution, 
the circular orbit model of the SPIRou wPCA data set (Table~\ref{planet-properties}),
the signal has a semi-amplitude of 4.4$\pm$0.7 m s$^{-1}$,
which indicates a sub-Neptune planet with $M$\,sin$(i)$ of  8.4$\pm$1.3 Earth masses 
at 0.0531$\pm$0.0006 au.
The same Keplerian signal is recovered in quasi-simultaneous SOPHIE RV measurements 
after correction for stellar activity using a GP that was trained
on the measurements of the longitudinal magnetic field ($B_\ell$) obtained with SPIRou.
The independent detection of Gl~410b in the SOPHIE RVs provides 
further evidence for the Keplerian nature of the periodic signal.

We searched the TESS archive and investigated whether Gl~410b could be a transiting planet. 
We found no evidence for the transit of Gl~410b.
A recovery analysis on the TESS photometry indicates that the transit would 
have been detected if Gl~410b's inclination were higher than 87 deg and its radius were larger than 1.8 R$_\oplus$.
Gl~410b is located very close to its host star,
and it receives 20.4 times Earth's insulation.
It would be of great interest to investigate
the influence that stellar irradiation and activity could have on the 
chemical composition of the planet's atmosphere
and in the potential atmospheric escape processes.

We find within the SPIRou wPCA RVs,
there is tentative evidence for two additional planetary signals at 2.99 and 18.7 days.
The 18.7-day signal (which is consistent with a sub-Neptune planet) is directly seen in the GLS periodograms 
and it is recovered in a \texttt{RadVel} two-planet model with broad priors for the period. 
The 2.99-day signal emerges together with the 18.7-day signal in an $\ell_1$ and apodize sine periodogram analysis,
which takes into account stellar activity indicators measured with SPIRou and a model for the correlated noise.
The planet at 2.99 days could be in a 2:1 mean motion resonance with the 6.02-day planet and could have a mass of a few Earth masses.

The detection of Gl~410b
shows that infrared RVs are a powerful tool for detecting 
planets around active stars,
objects which are often excluded from searches in optical RVs.
Care should be taken, however, to correct and filter 
systematics generated by residuals of the telluric correction or small structures in the detector plane. 
The LBL technique combined with PCA offers a promising way to reach this objective. 
This approach enabled us to use the RV measurements acquired when the star is at $\pm10$ km s$^{-1}$ 
of the center of the telluric absorption lines, 
which is important to detect additional planetary signals.
Further monitoring of Gl~410 is necessary.

\begin{acknowledgements}
Based on observations obtained at the Canada-France-Hawaii Telescope (CFHT) which is 
operated from the summit of Maunakea by the National Research Council of Canada, the Institut National des Sciences de l'Univers of the Centre 
National de la Recherche Scientifique of France, and the University of Hawaii. 
Based on observations obtained with SPIRou, an international project led by Institut de Recherche en Astrophysique et Plan\'etologie, Toulouse, France.
Based on observations obtained with the spectrograph SOPHIE 
at the Observatoire de Haute-Provence (OHP) in France,
operated by Institut National des Sciences de l'Univers of the Centre 
National de la Recherche Scientifique of France.
We acknowledge funding from the French ANR under contract number ANR\-18\-CE31\-0019 (SPlaSH).
This work is supported by the French National Research Agency in the framework of the Investissements d'Avenir program (ANR-15-IDEX-02), 
through the funding of the ``Origin of Life" project of the Grenoble-Alpes University.
JFD acknowledges funding from the European Research Council (ERC) under the H2020 research \& innovation program (grant agreement \#740651 NewWorlds).
This publication makes use of the Data \& Analysis Center for Exo-planets (DACE), which is a facility based at the University of Geneva (CH) dedicated to extrasolar planets data visualization, 
exchange and analysis. DACE is a platform of the Swiss National Centre of Competence in Research (NCCR) PlanetS, 
federating the Swiss expertise in Exoplanet research. The DACE platform is available at https://dace.unige.ch.
The project leading to this publication has received funding from the french government under the "France 2030" investment plan 
managed by the French National Research Agency (reference : ANR-16-CONV-000X / ANR-17-EURE-00XX) and 
from Excellence Initiative of Aix-Marseille University - A*MIDEX (reference AMX-21-IET-018). This work was supported by the "Programme National de Plan\'etologie" (PNP) of CNRS/INSU.
E.M. acknowledges funding from FAPEMIG under project number APQ-02493-22 and research productivity grant number 309829/2022-4 awarded by the CNPq, Brazil.
This work has made use of data from the European Space Agency (ESA) mission
{\it Gaia} (\url{https://www.cosmos.esa.int/gaia}), processed by the {\it Gaia}
Data Processing and Analysis Consortium (DPAC,
\url{https://www.cosmos.esa.int/web/gaia/dpac/consortium}). Funding for the DPAC
has been provided by national institutions, in particular the institutions
participating in the {\it Gaia} Multilateral Agreement.
The observations at the Canada-France-Hawaii Telescope were 
performed with care and respect from the summit of Maunakea which is a significant cultural and historic site.
 \end{acknowledgements}

\bibliographystyle{aa} % style aa.bst
\bibliography{Gl410_references.bib} % your references Yourfile.bib

\begin{appendix}
\onecolumn
\section{On the age of Gl~410}
\label{age_determination} 
\begin{figure}[h]
\begin{center}
\begin{tabular}{cc}
\includegraphics[width=0.52\textwidth]{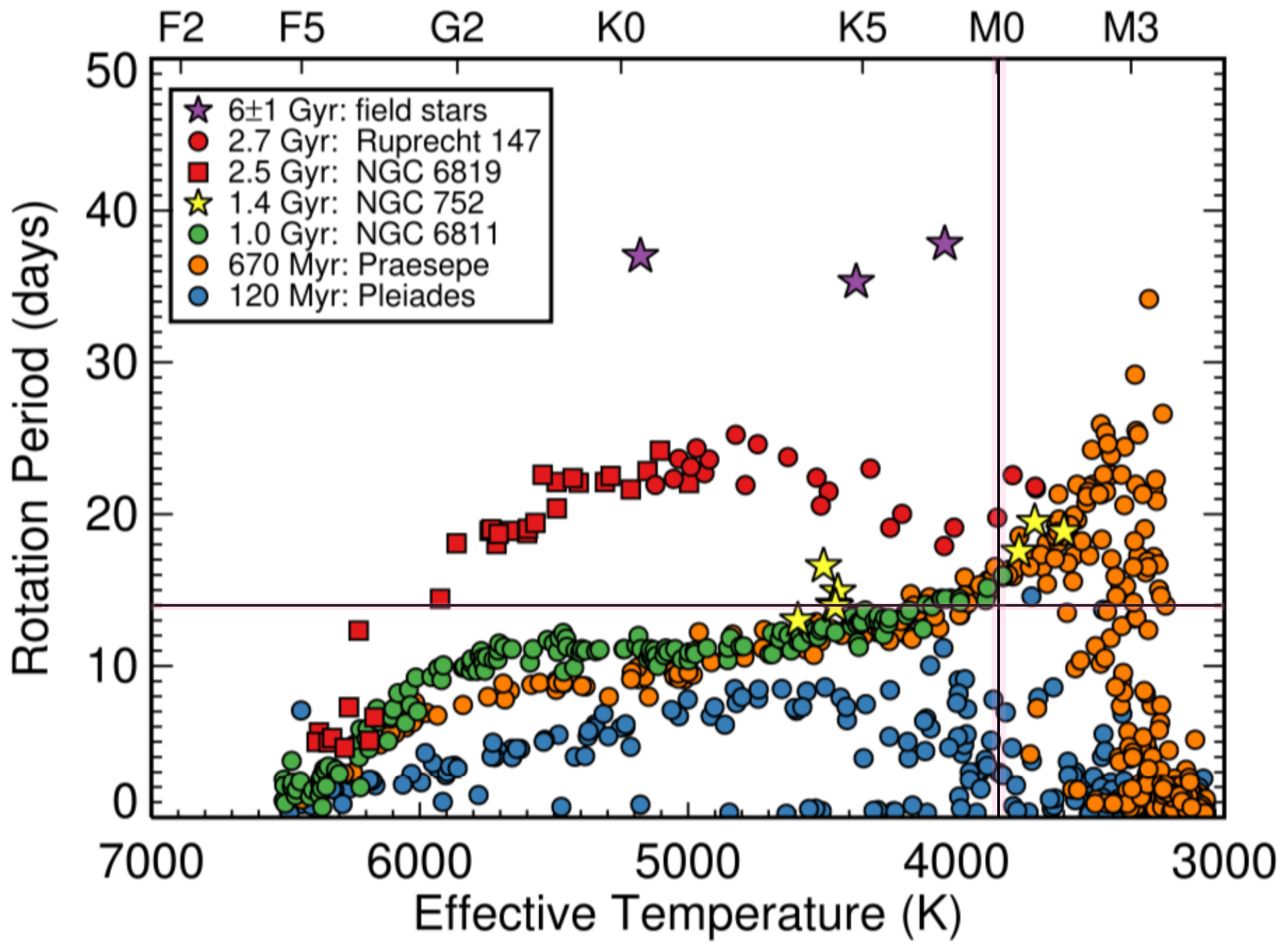} &
\includegraphics[width=0.46\textwidth]{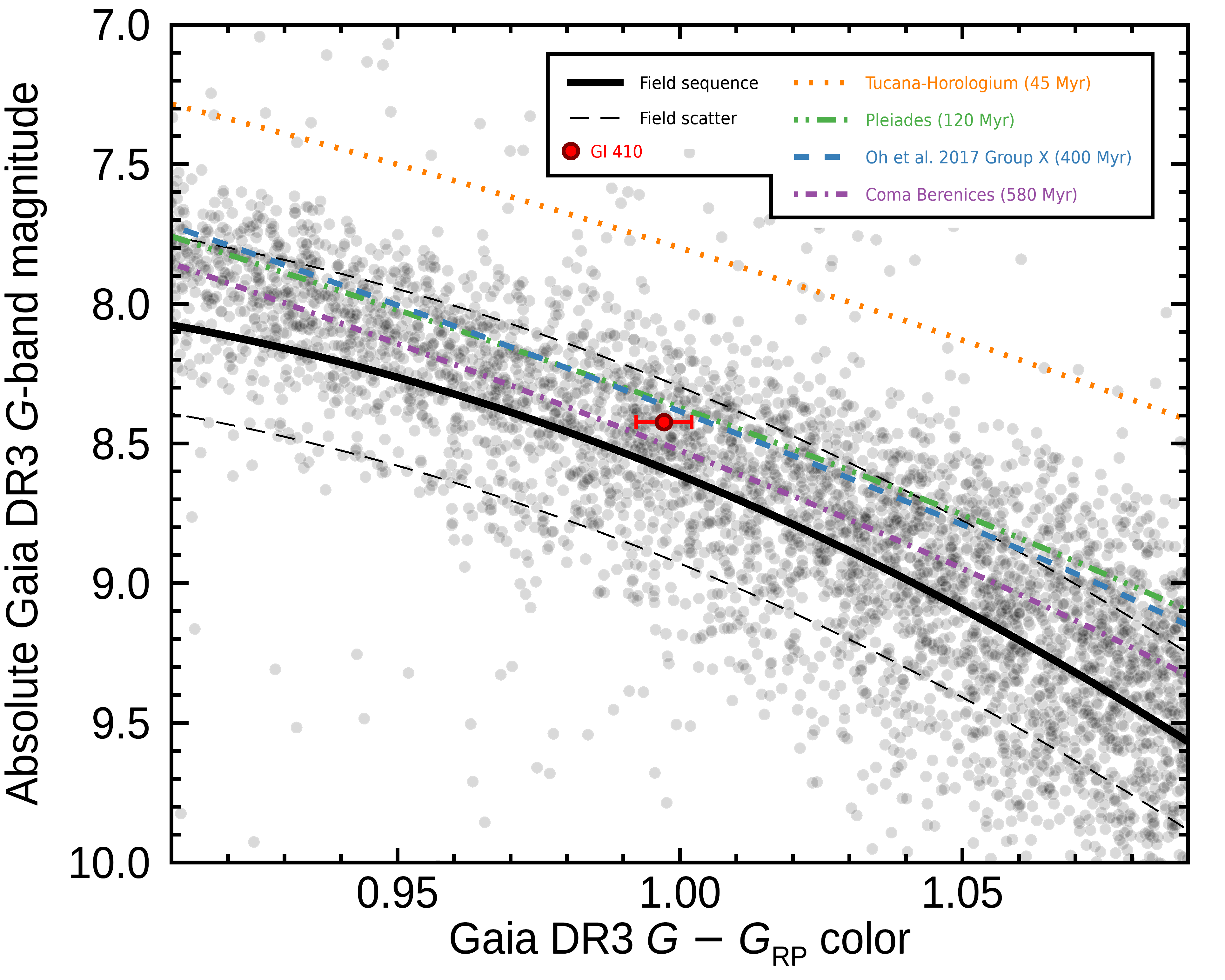}
\end{tabular}
\caption{{\it Left panel}: Rotational period vs effective temperature for stars in clusters of different ages from \citet[][]{Curtis2020}.
The location of Gl~410 is indicated in solid lines (error on $T_{\rm eff}$ is given in light red).
{\it Right panel}: Gaia DR3 color-magnitude diagram of young stars with calibrated ages (dots) and empirical age sequences from young associations \citep[][]{Gagne2021}. 
The position of Gl~410 is indicated with a red dot. Gl~410 is located between the 400 and 580 Myr sequences.
}
\label{fig:age_Gl410}
\end{center}
\end{figure}

Age determination using gyrochronology rely on empirical relations between the rotation period and the effective temperature or photometry of stars in clusters of known ages.
The age of 0.89$\pm$0.1 Gyr  for Gl~410 derived in \citet[][]{Fouque2023} 
was determined using the methods described in \citet[][]{Gaidos2023} 
which are calibrated on the empirical relations of \citet[][]{Curtis2020}.
In Fig.~\ref{fig:age_Gl410},
we display the $P_{\rm rot}$ vs $T_{\rm eff}$ diagram from  \citet[][]{Curtis2020},
indicating the location of Gl~410 in solid lines.
At the $T_{\rm eff}$  Gl~410,
the age-rotation relations of  Praesepe (670 Myr) and NGC 6811 (1 Gyr) sequences overlap.
The rotational period of Gl~410 locates the star below the  Praesepe sequence, 
which suggests that  Gl~410 could be younger than 670 Myr.
The \citet[][]{Gaidos2023} work targeted the vast majority of planet hosts which are middle-aged
and did not use calibrations below the Praesepe age so by definition no claim for a younger age 
could be done.

To investigate the possibility of a younger age for Gl~410,
we estimated its age using two alternative publicly available gyrochronology codes: 
\texttt{stardate} \citep[][]{Angus2019} and \texttt{gyro-inter} \citep[][]{Bouma2023}.
Both codes converge in a younger age estimate for Gl~410:
\texttt{stardate} suggests an age of 493$\pm$2 Myr;
\texttt{gyro-inter} indicates an age of 326$^ {+105}_{-56}$ Myr.
Furthermore, we checked the position of Gl~410 in the color-magnitude diagram (CMD). 
The right panel of Fig.~\ref{fig:age_Gl410}
displays the Gaia DR3  \citep[][]{Vallenari2023} CMD for Gl~410, field stars,
together with empirical CMD sequences from various young associations \citep[][]{Gagne2021}. 
Gl~410 is located between the 400 and 580 Myr sequences.
Interpolating in log space, an age of about 480 Myr is obtained for Gl~410.
We note, however, that the scatter around these empirical sequences is non negligible,
that at younger ages the age distribution is complex and likely non Gaussian,
and that the age is anchored on the ages assumed for Group X \citep{Oh2017} and Coma Ber.
From the empirical sequences,
we find that Gl~410 is slightly younger than Praesepe 
that is somewhat unlikely to be as young as the Pleiades ($\sim$120 Myr, although the probability is not 0\%), 
and very unlikely to be as young as the Tucana-Horologium association ($\sim$45 Myr).
Based on these considerations,
we propose an age of 480$\pm$150 Myr for Gl~410. 
Further dedicated work is necessary to determine the age of Gl~410 with a higher precision, 
but this study is beyond the scope of this paper.
A possible way to narrow-down the uncertainty, would be to map the distribution of reference age populations at this spectral type,
reverse the distribution using Bayes theorem and Gl~410's rotational period and  then obtain a non-gaussian age probability distribution for Gl~410.

A young age for Gl~410  is of interest because it suggest that Gl~410 might be related to the
Ursa Major (UMA) cluster which is approximately 400 Myr old.  
Gl~410 indeed has been previously suggested to be member of UMA by \citet[][]{Ammler-vonEiff2009},
member of a diffuse region associated with UMA by \citet[][]{King2003}
and member of the "UMA moving group" by \citet[][]{Montes2001}.
The Galactic $UVW$ velocities of Gl~410 appear consistent with the proposed Ursa Major corona \citep[][]{Gagne2020}.

\onecolumn

\section{Activity analysis}
\label{Sect_activity_analysis}
\begin{figure}[h]
\begin{center}
\includegraphics[width=0.96\linewidth]{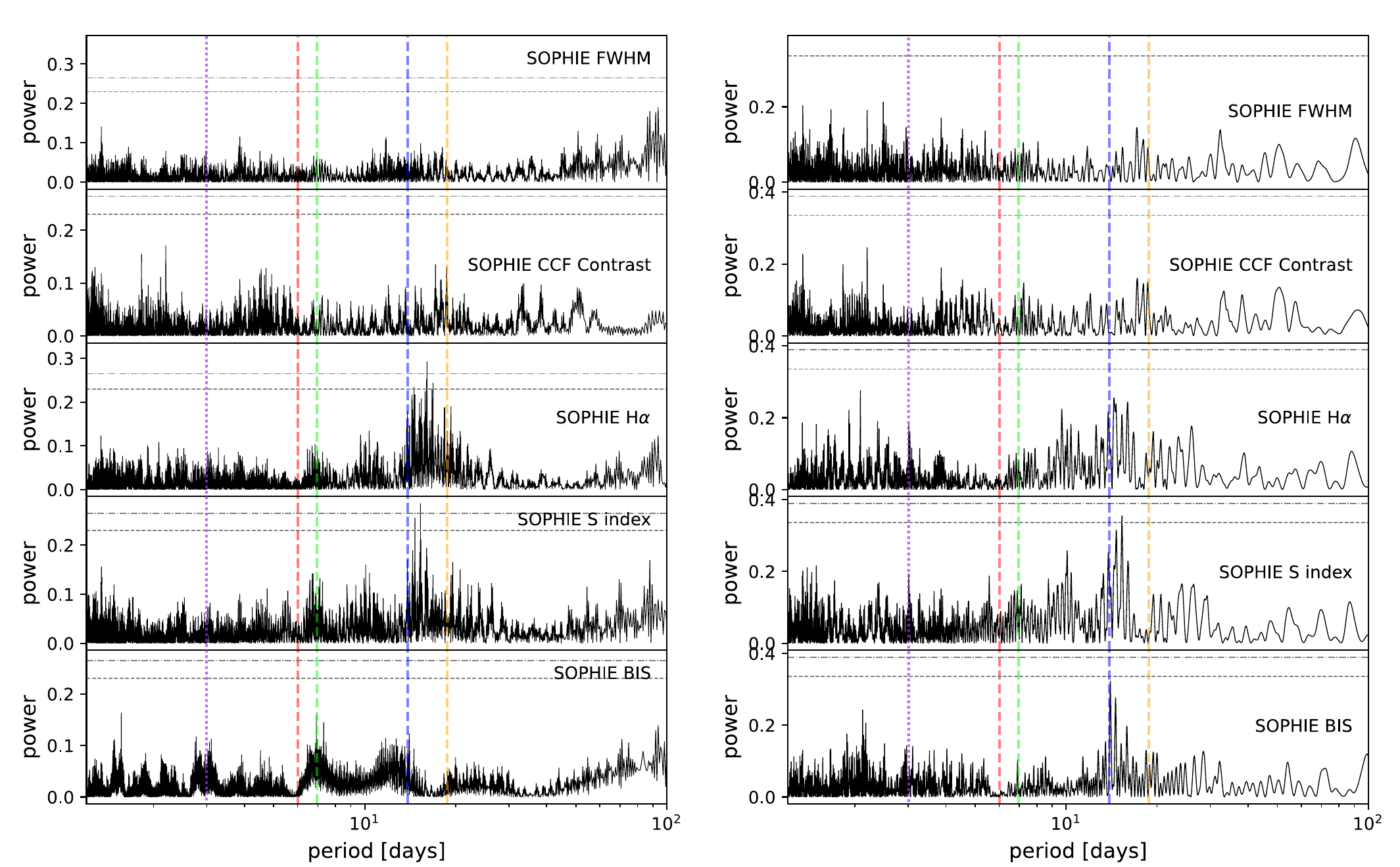}
\caption{Periodograms of the SOPHIE activity indicators: FWHM, CCF contrast, H$\alpha$, S index, and BIS. 
The left periodograms correspond to the whole data set, the right periodograms correspond  to the $2010-2012$ period. 
The vertical dashed lines depict: in red, 6.02 days; in blue,  the rotation period with its harmonic in green; 
in orange, 18.7 days, and, in purple dots, 2.99 days.
The horizontal dashed lines indicate the 1$\%$ and 0.1$\%$ FAP levels.}
\label{fig:SOPHIE_periodograms1}
\end{center}

\end{figure}

\begin{figure}
\begin{center}
\includegraphics[width=0.5\linewidth]{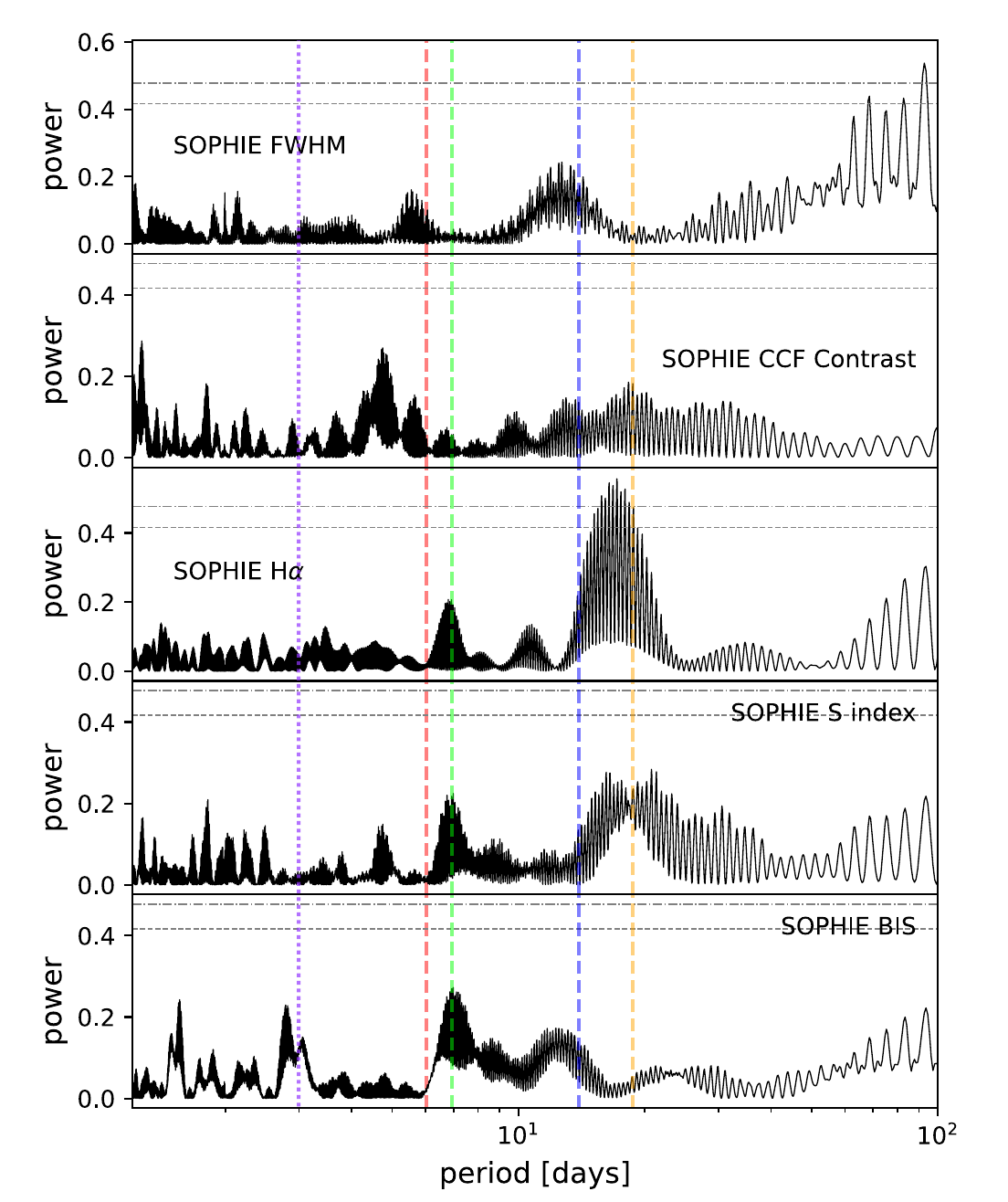} \\
\caption{Periodograms of the SOPHIE activity indicators: FWHM, CCF contrast, H$\alpha$, S index, and BIS for the $2021-2023$ period.
The vertical dashed lines depict: in red, 6.02 days; in blue,  the rotation period with its harmonic in green; 
in orange, 18.7 days, and, in purple dots, 2.99 days. The horizontal dashed lines indicate the 1$\%$ and 0.1$\%$ FAP levels.}
\label{fig:SOPHIE_periodograms2}
\end{center}

\end{figure}

We measured activity indicators in the SOPHIE and SPIRou spectra.
For the SOPHIE data set, 
the activity indicators were derived based on distortions of the cross-correlation function (CCF), 
the bisector (BIS), the full width half maximum (FWHM), and the CCF contrast, 
and on the spectral line indexes H$\alpha$ and S index \citep[based on the Ca II H\&K lines, calculated as described by][]{Boisse2009}. 
We investigated the temporal evolution of the activity indicators 
calculating GLS periodograms and searching for modulations related to the stellar rotation.
The periodograms of the SOPHIE activity indicators are shown in Fig.~\ref{fig:SOPHIE_periodograms1} and \ref{fig:SOPHIE_periodograms2}.
Considering the whole ensemble of SOPHIE activity indicators, 
none of their periodograms display a 
peak at the stellar rotation period (13.91 days),
with exception of H$_\alpha$ and the S index that show excess power below 0.1\% FAP around 15 and 16 days, 
periods which are relatively close to the stellar rotation period. 
To check for consistency,
we also computed GLS periodograms for the SOPHIE data set split in two periods: 2010$-$2012 and 2021$-$2023.
The periodograms of the 2010$-$2012 period do not display significant periodicities except for H$\alpha$ 
which exhibits several peaks at periods between 14 and 19 days, however, they are most likely be due to poor sampling.
In the case of the 2021$-$2023 period,  
the periodogram of the S index exhibits a significant periodicity (<1$\%$ FAP) at 15.3 
days and more importantly, the BIS shows excess power almost at 1$\%$ FAP at 14 days, which is consistent with the rotation period.
  
\begin{figure}[h]
\begin{center}
\includegraphics[width=0.65\linewidth]{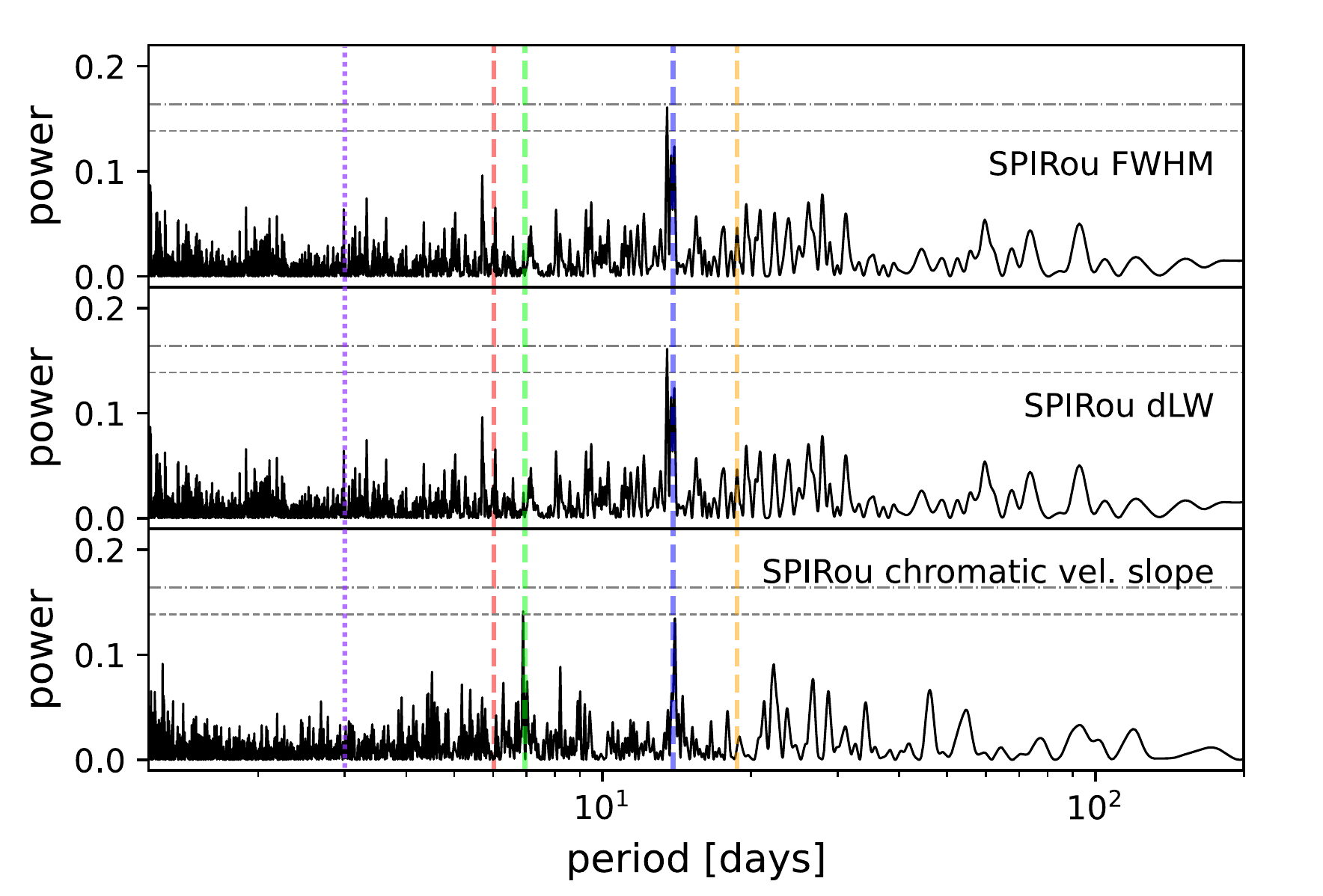}
\caption{Periodograms of the SPIRou activity indicators: FWHM, \texttt{dLW}, and chromatic velocity slope. 
The vertical dashed lines depict: in red, 6.02 days; in blue,  the rotation period with its harmonic in green; 
in orange, the 18.7 days, and, in purple dots, 2.99 days. The horizontal dashed lines indicate the 1$\%$ and 0.1$\%$ FAP levels.}
 \label{fig:SPIRou_periodograms}
\end{center}
\end{figure}

For the SPIRou data set, we analyzed the GLS periodograms of the differential line width (\texttt{dLW}), the FWHM, 
and the chromatic velocity slope calculated by the LBL algorithm.
The periodograms are shown in Fig.~\ref{fig:SPIRou_periodograms}.
We found excess power at $\sim$13 days in the \texttt{dLW} and chromatic velocity slope periodograms.
This periodic variability could be related to the stellar rotation. 
No periodicities are detected at 6.02, 18.7 nor 2.99 days neither in the SOPHIE nor SPIRou data sets.

\begin{figure}[h]
\begin{center}
\begin{tabular}{cc}
\includegraphics[width=0.44\textwidth]{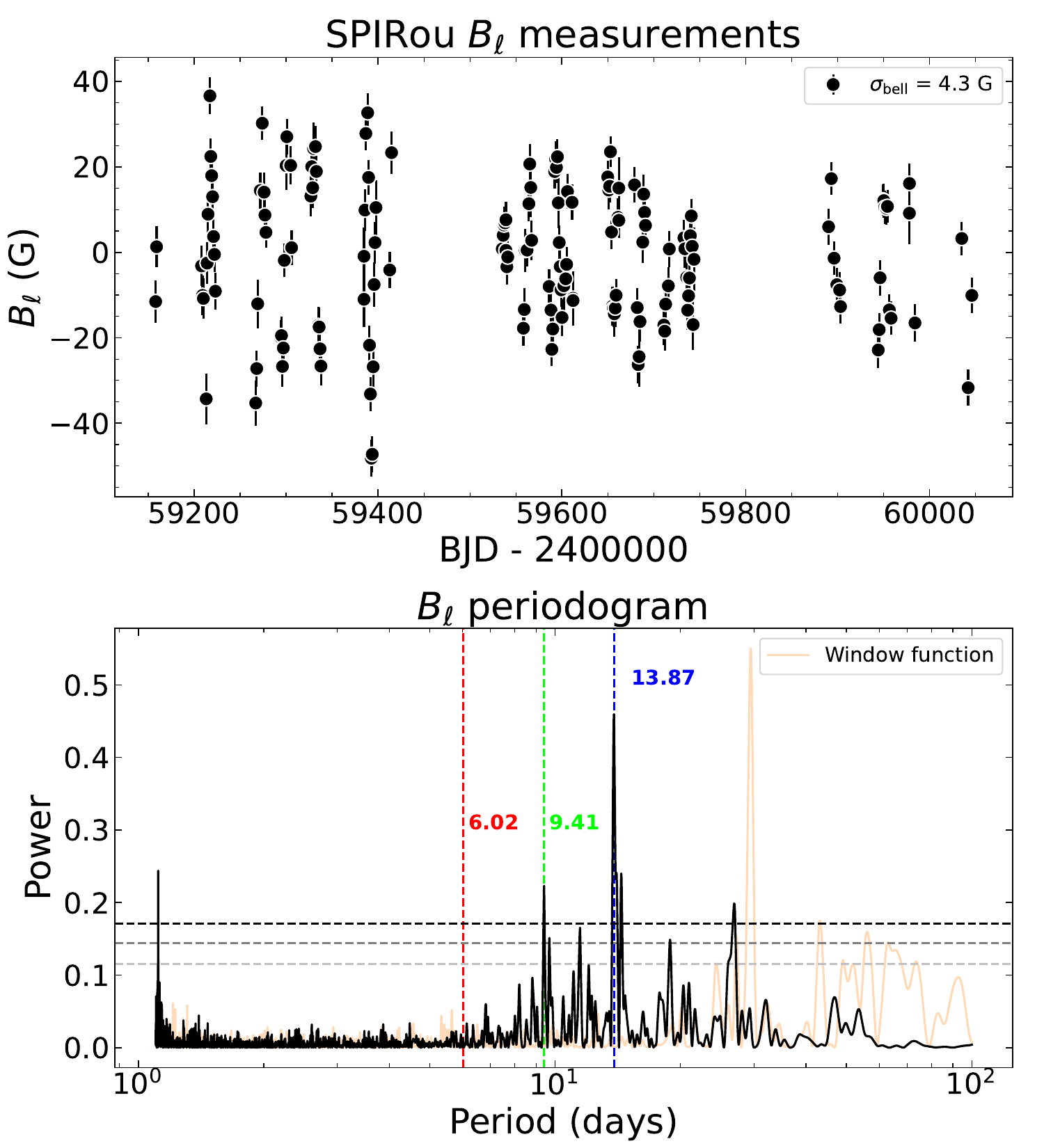} &
\includegraphics[width=0.52\textwidth]{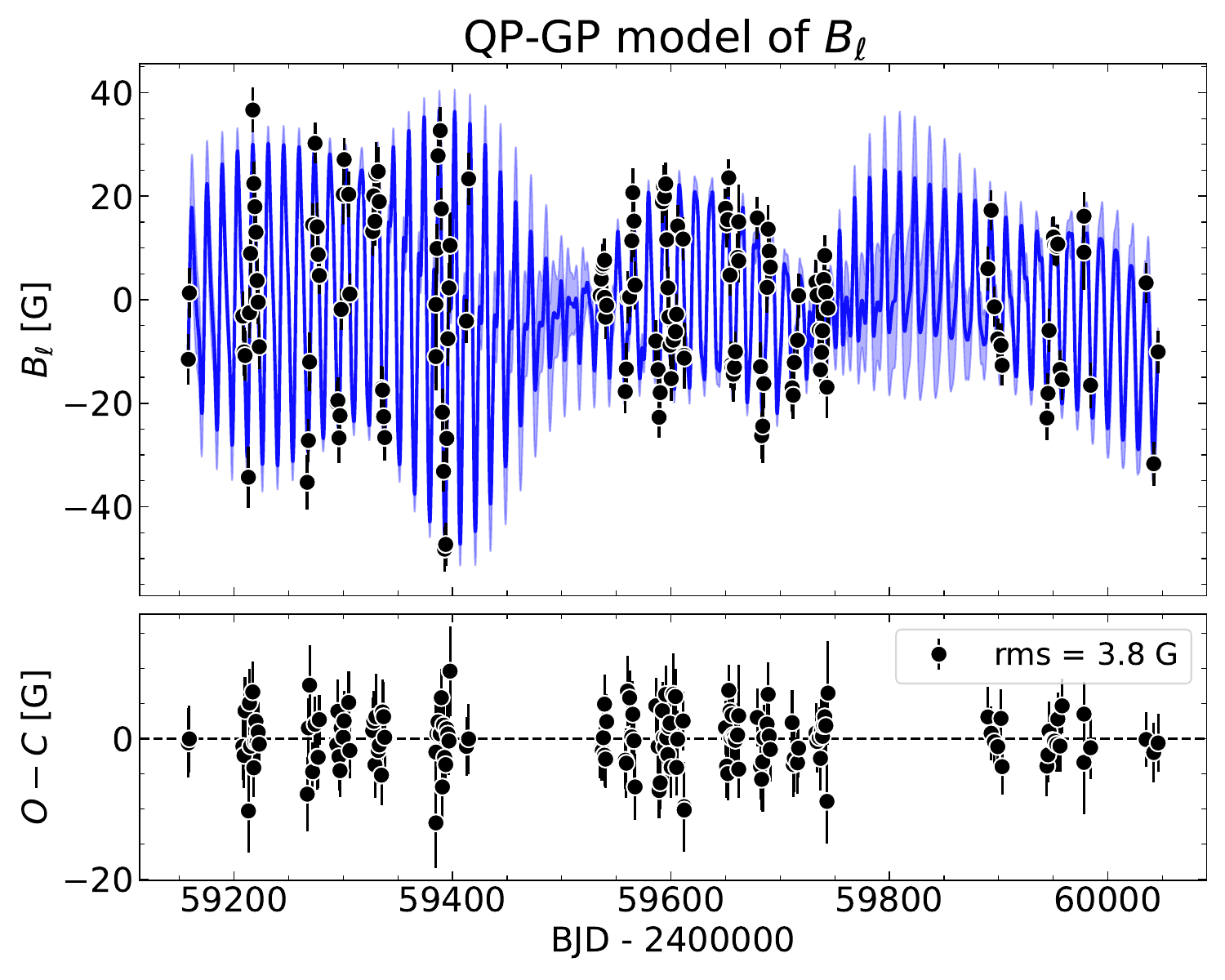}
\end{tabular}
\caption{ {\it Left panel:} SPIRou $B_\ell$ measurements from \citet[][]{Donati2023} and associated LS periodogram.
The horizontal dashed lines represent the FAP levels of 0.1\%, 1\%, and 10\% respectively.
A significant peak at $P=13.87$ days (and 1yr alias at 14.98 days) is clearly observed, 
which is associated with to 13.91$\pm$0.09 day rotational period of the star \citet[][]{Donati2023}.
An additional significant peak is seen at 9.41 days. No signals are seen at $P=6.02$ days. 
{\it Right panel:} quasi-periodic GP model fit (blue-line) of the SPIRou $B_\ell$ time series (black dots). 
Details of the GP are given in Table~\ref{GP_table}. 
The corner plot of the fit is given in Fig.~\ref{fig:corner_bell}.
}
\label{fig:Bl_periodogram_GP}
\end{center}
\end{figure}

\begin{figure}
    \centering
    \includegraphics[width=0.45\textwidth]{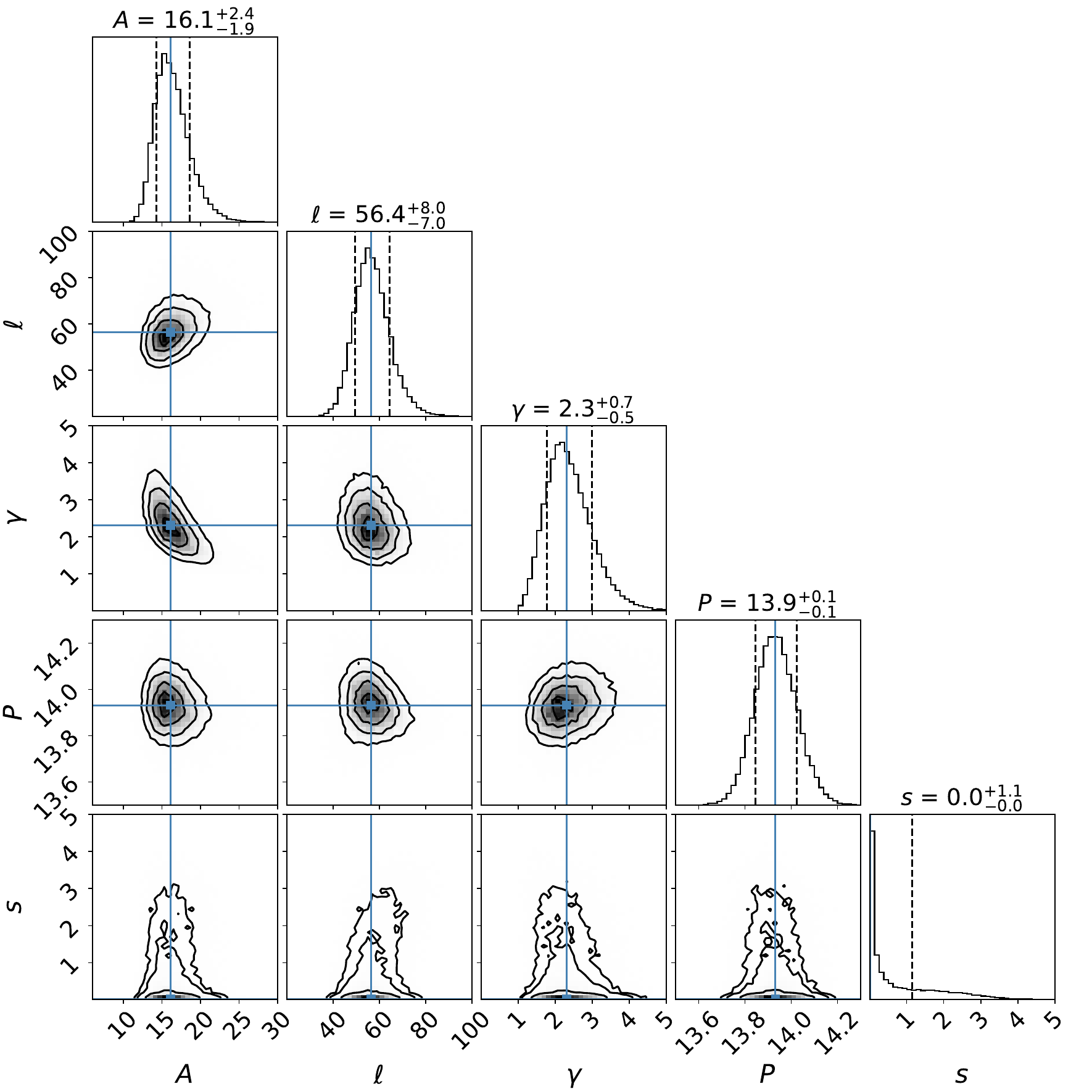}    
    \caption{Corner plot of the quasi-periodic GP fit of the SPIRou longitudinal magnetic field (B$_\ell$) time series of  Gl~410. 
    We provide the description of  the meaning of the variables in Table~\ref{GP_table}.} 
    \label{fig:corner_bell} 
   
\end{figure}

\clearpage
\begin{minipage}[t][0cm][t]{0.96\textwidth}
\section{Gaussian process modeling of SPIRou's  longitudinal magnetic field}
\label{Bl_GP_modeling}

Measurements of the longitudinal magnetic field ($B_\ell$) of Gl~410  taken with SPIRou 
have been previously presented and analyzed in \citet[][]{Fouque2023} and \citet[][]{Donati2023}.
Fig.~\ref{fig:Bl_periodogram_GP} displays the  LS periodogram of the $B_\ell$  data from \citet[][]{Donati2023}.
We retrieve the peak at $P=13.87$ days which is associated with to 13.91$\pm$0.09 day rotational period of the star
and its 1yr alias at 14.98 days. One additional peak with FAP level below 0.1\%, and not related to the window function, 
is detected at 9.41 days with its 1yr alias at 9.69 days. No peak is observed at 6 days.
Using the python module  \texttt{george}~\citep[][]{2015ascl.soft11015F}, 
we fitted a GP on the SPIRou $B_\ell$ time series
using a quasi-periodic kernel of the form:
\begin{equation}
    k\left(\Delta t\right) = A\exp{\left(-\frac{\Delta t^2}{2 l^2}\right)}
    \exp{\left(-\Gamma \sin^2{\left(\frac{\pi}{P}\Delta t\right)}\right)} 
    + \sigma^2 \delta\left(\Delta t\right),
\end{equation}
where $A$ is the amplitude, 
$\Gamma$ is the smoothing parameter, 
$l$ is the decay time in days, 
$P$ is the period (i.e., the cycle length) in days,
and $\sigma$ is the uncorrelated noise (s in the corner plots).
We used uniform distributions for the priors of the amplitude, decay time, smoothing, cycle length,
and uncorrelated noise as described in Table~\ref{GP_table}. 
The distribution of the hyperparameters was sampled using the Markov chain Monte Carlo (MCMC) 
implemented in the \texttt{EMCEE} tool \citep[][]{2013ascl.soft03002F}.
We used 100000 samples and 10000 burn-in samples. 
In Table~\ref{GP_table}  and Fig.~\ref{fig:Bl_periodogram_GP},
we summarize the results of the GP modeling of the $B_\ell$ time series.
Fig.~\ref{fig:corner_bell} present the corner plots of the fit.
As expected, we retrieve in our $B_\ell$ GP  the stellar rotational period of $13.93\pm0.09$ days
which is consistent with the previous determinations of the rotational period by 
\citet[][]{Donati2023} and \citet[][]{Fouque2023}. 
The values obtained for the amplitude, decay time and smoothing parameter are 
consistent to those of \citet[][]{Donati2023}.
\end{minipage}
 
\pagebreak
\onecolumn

\begin{table*}[]
\small
\centering
  \caption[]{Prior and posterior distributions of the quasi-periodic GP models fit to the SOPHIE RV time series and the SPIRou $B_\ell$ measurements.}
  \label{GP_table}
 \begin{tabular}{llllll}
    \hline\\[-10pt]
  \hline
  \multicolumn{2}{c}{\rm Parameter} & \multicolumn{1}{l}{\rm Prior distribution} & \multicolumn{2}{c}{\rm Posterior distribution} \\
         \hline
       \\[-8pt]
    \noalign{\smallskip}
       \multicolumn{5}{c} {\bf GP + Keplerian \texttt{RadVel} model of the SOPHIE RVs (circular orbit)}  \\
    \hline
    \\[-5pt]
    {\bf GP} \\
    Offset [m s$^{-1}$]          			& $V_0$            	&								&							& $4^{+16}_{-11}$ 	     	\\[1mm] 
    Amplitude  [m s$^{-1}$]   			& ${A}$         	 	& $\mathcal{U}\left(0,75\right)$ 			&							& $21^{+19}_{-8}$	     	\\[1mm]  
    Decay time [days]            			& $l$         		& $\mathcal{U}\left(0,2000\right)$		&							& $602^{+370}_{-230}$	\\[1mm] 
    Cycle length (period) [days]   		& $P$         	 	& Gaussian $(13.9\pm0.1)$  			&							& $13.84\pm0.02$    	 	\\[1mm] 
    Smoothing parameter            		& $\Gamma$	   	& $\mathcal{U}\left(0,2.0\right)$  		&							& $0.5^{+0.3}_{-0.2}$	\\[1mm] 
    Uncorr. Noise [m s$^{-1}$]   		    	& $\sigma$  	 	& $\mathcal{U}\left(0.0,15\right)$ 		&							& $6\pm1$ 	 	    	\\[2mm]
    {\bf Keplerian} \\
    Period [days]						& $P_b$		 	& $\mathcal{U}\left(0.0,20\right)$	        &							& $6.020^{+0.0051}_{-0.0056}$	  \\[1mm]
    Semi-amplitude [m s$^{-1}$ ] 		& $K_b$			& $\mathcal{U}\left(0.0,20\right)$  		&							& $7.6^{+1.3}_{-1.4}$		  \\[1mm] 
    Time of inferior conjunction [days]		& $T\rm{conj}_{b}$   & $\mathcal{U}\left(2459648.0\pm10\right)$    &							& 						      \\[1mm]   
    Time of periastron [days]			& $T\rm{peri}_{b}$    & 						           	&							& $2459646.53\pm0.19$	      \\[1mm] 
    Semimajor axis [au]				& $a_b$ 			& 								&							& $0.0531\pm0.0007$ \\[1mm]
    $M_b\sin (i)$ 	[M$_{\oplus}$] 	         & 				&								&							& $14.4\pm2.6$		  \\[1mm]
     \hline
      \hline\\[-8pt]
   \noalign{\smallskip}
    \multicolumn{5}{c}{\bf GP of SPIRou's $B_\ell$ measurements}\\
    \hline
        & & & \multicolumn{1}{l}{ln units} &  \multicolumn{1}{l}{linear units} \\
     \hline   
    \\[-5pt]
    Amplitude  [G]                         & $\ln{A}$               & $\mathcal{U}\left(2,6\right)$                                                             & $2.78^{+0.14}_{-0.12}$              & $16.1^{+2.4}_{-1.9}$              \\[1mm]
    Decay time [days]           	 & $\ln{l}$                 & $\mathcal{U}\left(3,5\right)$                                                             & $4.03^{+0.13}_{-0.13}$              & $56^{+8}_{-7}$            \\[1mm]
    Smoothing parameter             & $\ln{\Gamma}$    & $\mathcal{U}\left(0,2.0\right)$                                                          & $0.84^{+0.26}_{-0.26}$              & $2.31^{+0.68}_{-0.54}$          \\[1mm]
    Cycle length (period) ~[days]  & $\ln{P}$               & $\mathcal{U}\left(\ln{\left(12.91\right)},\ln{\left(14.91\right)}\right)$   & $2.6342^{+0.0067}_{-0.0063}$ & $13.93^{+0.09}_{-0.09}$ \\[1mm]
    Uncorr. Noise [G]                     & $\ln{\sigma}$      & $\mathcal{U}\left(-10,10\right)$                                                         & $-3.8^{+4.0}_{-4.2}$                  & $0.02^{+1.16}_{-0.02}$        \\[1mm] 
   \hline
  \\[-8pt]
    \noalign{\smallskip}
     \multicolumn{5}{c}{\bf GP of the SOPHIE optical RVs  using SPIRou's $B_\ell$ as an activity proxy}\\
    \hline
        \\[-5pt]
    Offset [m s$^{-1}$]          	& $V_0$            	& median(RV) + $\mathcal{U}\left(-10,10\right)$            &                                                            	& $-13833.6^{+3.8}_{-3.5}$     \\[1mm] 
    Amplitude  [m s$^{-1}$]   	& $\ln{A}$         		& $\mathcal{U}\left(-0.7,2.8\right)$                                 & $2.50^{+0.24}_{-0.32}$                      	& $12.2^{+3.2}_{-3.4}$    \\[1mm] 
    Decay time [days]            	& $\ln{l}$         		& $B_\ell$ Posterior                                                        & $4.08^{+0.13}_{-0.13}$                     	& $59.4^{+8.5}_{-7.1}$            \\[1mm]
    Smoothing parameter            & $\ln{\Gamma}$	& $B_\ell$ Posterior                                                        & $1.03^{+0.23}_{-0.25}$          	       	&  $2.8^{+0.7}_{-0.6}$         \\[1mm]
    Cycle length (period) [days]   & $\ln{P}$         	& $B_\ell$ Posterior                                                        & $2.6333^{+0.0055}_{-0.0051}$          	& $13.92^{+0.08}_{-0.07}$  \\[1mm]
    Uncorr. Noise [m s$^{-1}$]   & $\ln{\sigma}$  	& $\mathcal{U}\left(0,5\right)$                                                 & $2.03^{+0.24}_{-0.23}$                   	& $7.6^{+2.1}_{-1.6}$   \\[1mm] 
    \hline     
    \hline
   
 \end{tabular}
\label{tab:comp}
\end{table*}

\twocolumn
\begin{strip}%
 \centering
 \vspace*{-1.5cm}
\section{Corner plots and additional figures and tables of the one-planet model}
 \normalsize
\end{strip}

\begin{figure}
    \centering
    \includegraphics[width=0.47\textwidth]{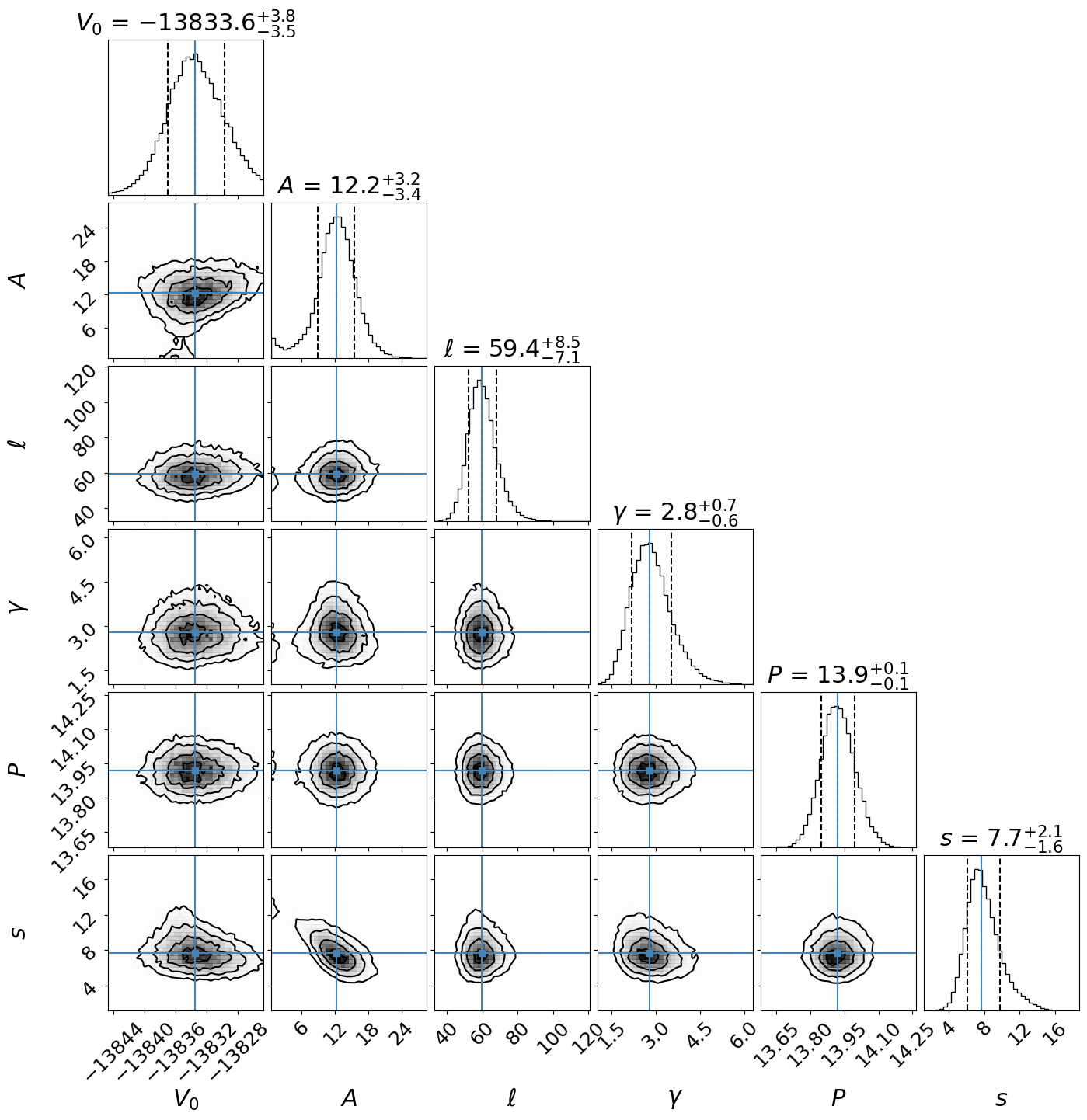} \\
    \caption{Corner plot of the quasi-periodic GP fit of the SOPHIE RVs using the posterior distributions of the decay time, smoothing length and $\gamma$ of the GP on the 
    SPIRou $B_\ell$ time series as priors. We provide the description of  the meaning of the variables in Table~\ref{GP_table}.}
   \label{fig:corner_sophie_rv} 
\end{figure}

 \begin{figure}
    \centering
    \includegraphics[width=0.48\textwidth]{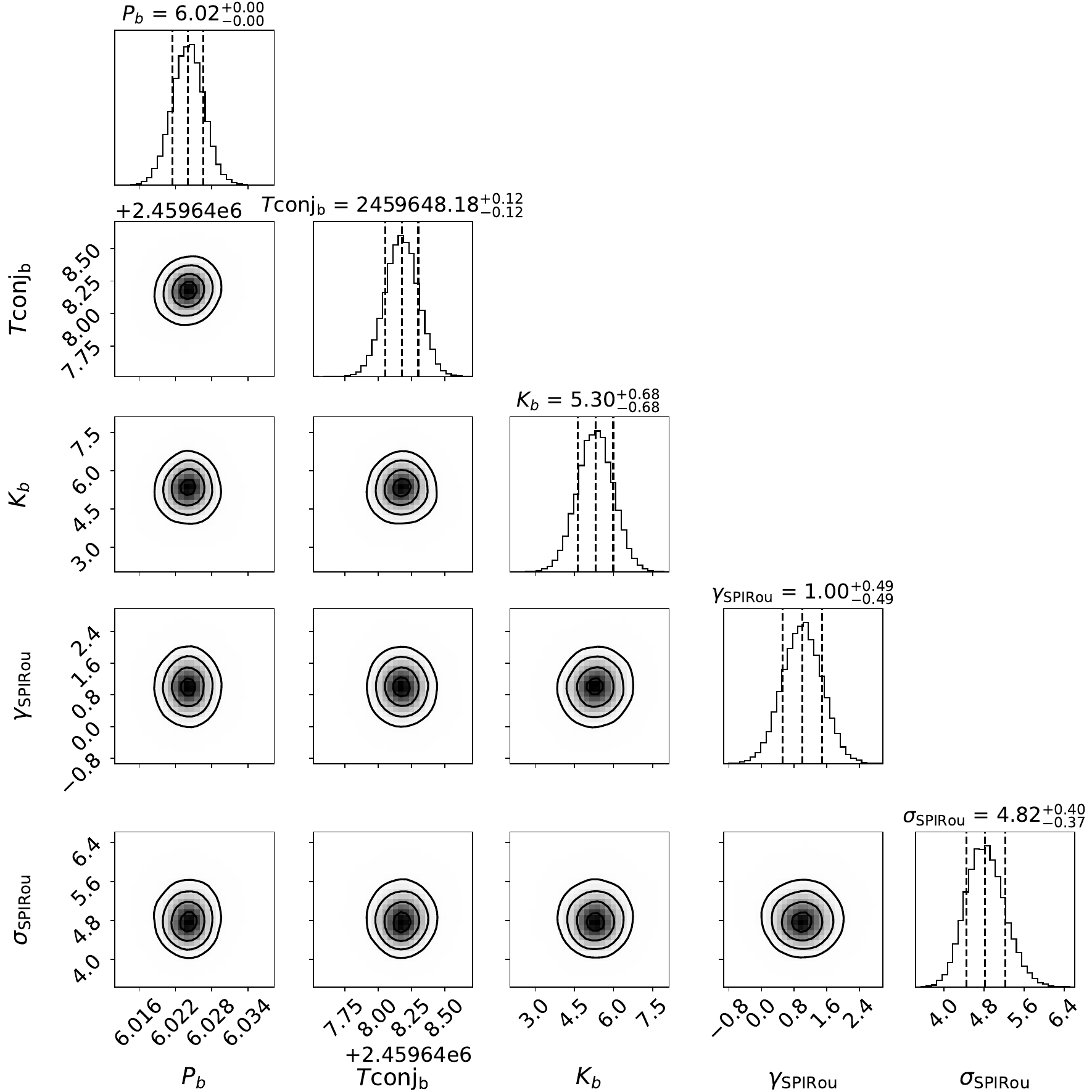}
    \caption{Corner plots of the posterior distributions of the  \texttt{RadVel} MCMC one-planet circular orbital model of the SPIRou RV time series 
    selecting the measurements at $|V_{\rm tot}|>10$  km s$^{-1}$.}
      \label{SPIRou_corner_plot_radvel}
\end{figure}

   \begin{figure}
    \centering
    \includegraphics[width=0.48\textwidth]{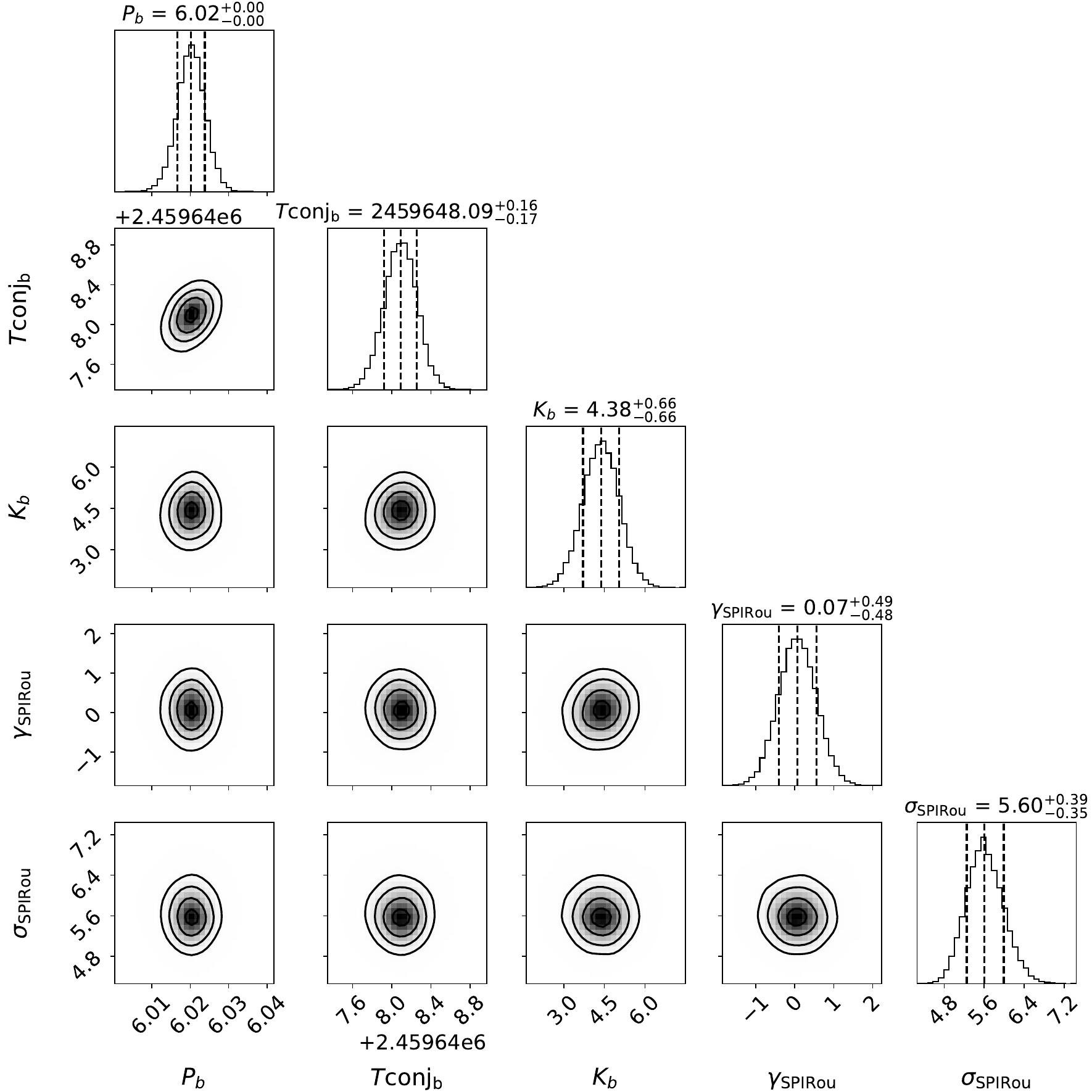}
    \caption{Corner plots of the posterior distributions of the  \texttt{RadVel} MCMC one-planet circular orbital model of the SPIRou wPCA RV time series}
      \label{SPIRou_corner_plot_radvel_wPCA}
\end{figure}

\clearpage
\onecolumn
   \begin{figure}
    \centering
    \includegraphics[width=0.5\textwidth]{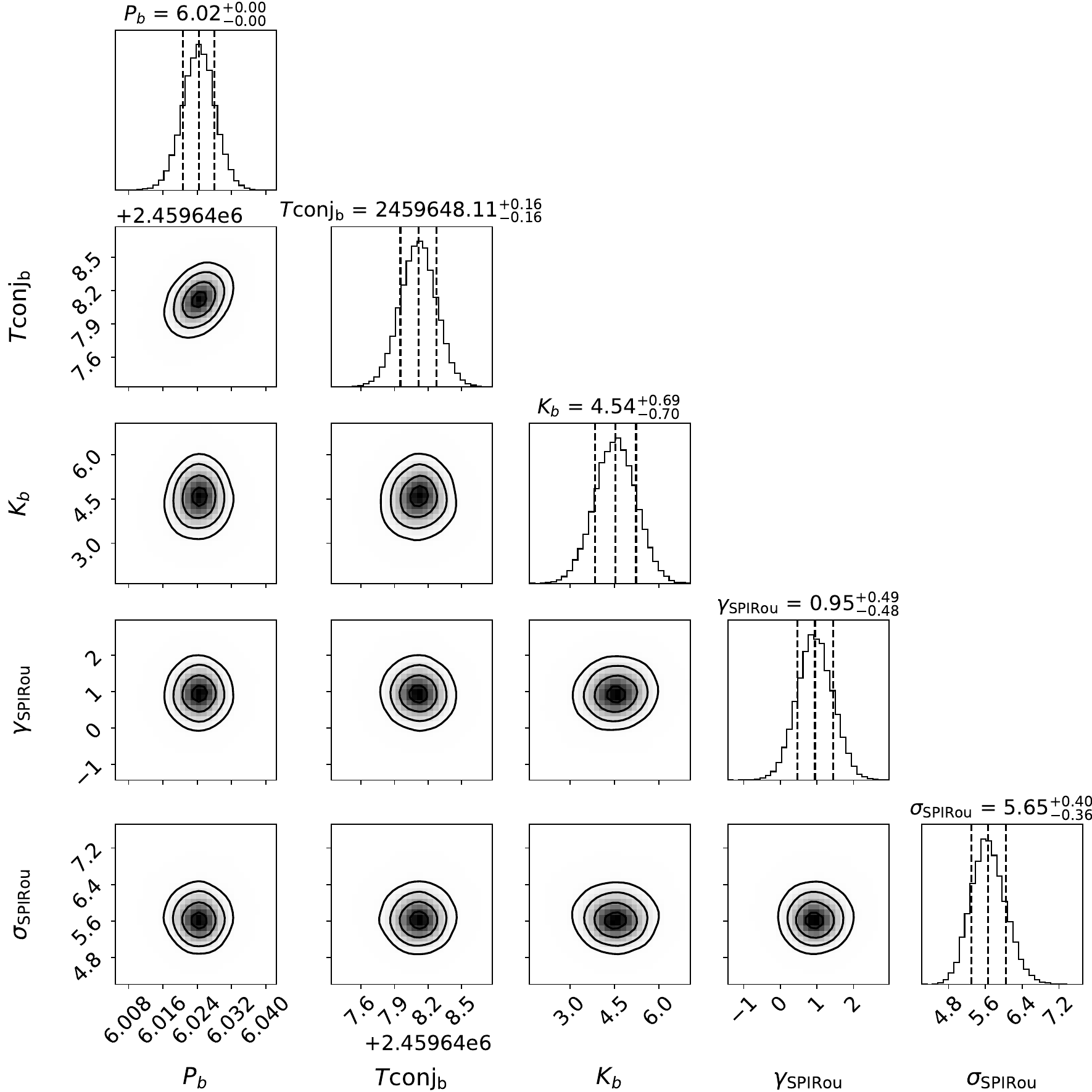}
    \caption{Corner plots of the posterior distributions of the  \texttt{RadVel} MCMC one-planet circular orbital model of the SPIRou Wapiti RV time series.}
      \label{SPIRou_corner_plot_radvel_wapiti}
\end{figure}

\begin{figure}
\centering
\includegraphics[width=0.55\textwidth]{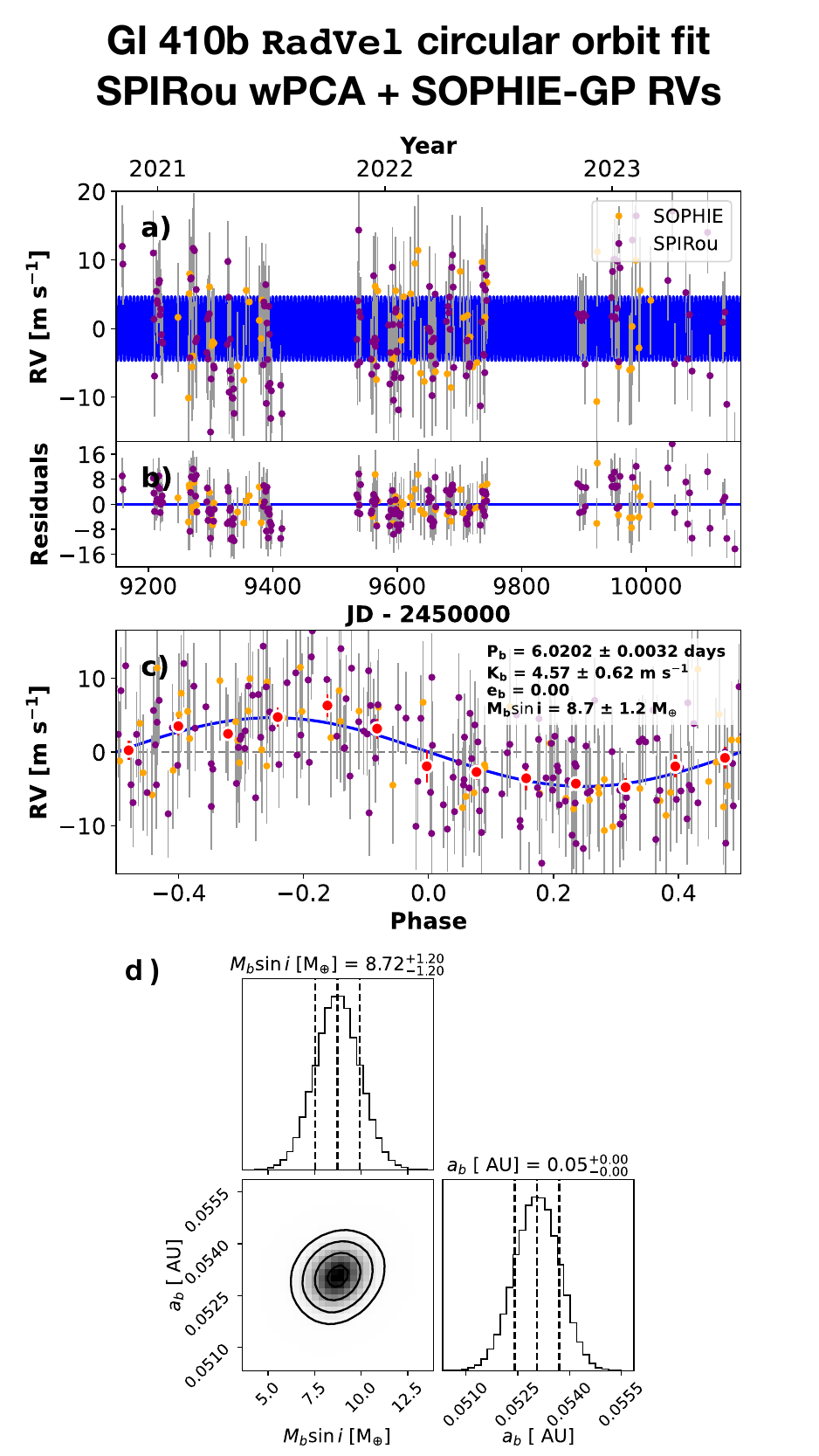}
\caption{Best-fitting one-planet \texttt{RadVel} circular orbit model for the data set combining 
the SPIRou wPCA RVs and the SOPHIE RVs corrected for activity using $B_\ell$ as an
activity proxy. The $T{\rm peri}_{b}$  retrieved is 59646.6$\pm$0.15. The uncorrelated noise
of the fit are $\sigma_{\rm SPIRou}=5.6\pm0.4$ m s$^{-1}$ 
and $\sigma_{\rm SOPHIE-GP}=0.09^{+1.9}_{-0.9}$ m s$^{-1}$.
}
\label{SOPHIE_wPCA_1planet}
\end{figure} 

\clearpage
\twocolumn
\begin{strip}%
 \vspace*{-1.5cm}
\section{Corner plots and additional figures and tables of the \texttt{RadVel} two-planet model}
 \normalsize
\end{strip}

\begin{table}[b]
\begin{center}
\caption{Priors and initial guess values of the \texttt{RadVel} two-planet model run on the SPIRou wPCA RV time series.}
\small
\begin{tabular}{lllllll}
    \hline
    \hline
    \\[-5pt]							
   Parameter				& \multicolumn{1}{c}{Prior}				& \multicolumn{1}{c}{Initial guess} 	 	\\[1mm]
    \hline
     \\[-3pt]	
     {\bf Gl~410b}\\[3pt]	
     $P_b$	[days]	 		& $\mathcal{U}\left(0.0,20\right)$	        		& 6.02							\\[1mm]
     $K_b$	[m s$^{-1}$ ] 		& $\mathcal{U}\left(0.0,20\right)$  			& 4.4								\\[1mm] 
     $T\rm{conj}_{b}$ [JD]  		& $\mathcal{U}\left(59648.0\pm10\right)$ 		& 59648.0							\\[1mm]   
     $\omega_b$   			& $\mathcal{U}\left(0.9\pm \pi \right)$ 		& 0.9								\\[1mm]
     $e_b$   				& $\mathcal{U}\left(0,1\right)$				& 0.0								\\[3mm]
     {\bf candidate signal at 18 days}\\[3pt]
     $P_c$	[days]	 		& $\mathcal{U}\left(0.0,30\right)$	        		& 18.8$^\dagger$					\\[1mm]
     $K_c$	[m s$^{-1}$ ] 		& $\mathcal{U}\left(0.0,20\right)$  			& 3.8								\\[1mm] 
     $T\rm{conj}_{c}$ [JD]  		& $\mathcal{U}\left(59652.0\pm10\right)$ 		& 59652.0						\\[1mm]   
     $\omega_c$   			& $\mathcal{U}\left(0.1\pm \pi \right)$ 		& 0.1								\\[1mm]
     $e_c$   				& $\mathcal{U}\left(0,1\right)$ 				& 0.0								\\[1mm]
     \\[3pt]
     $\sigma_{\rm SPIRou}$	& $\mathcal{U}\left(0.0,10\right)$			& 0.0								\\[1mm]
\\[3pt]
\hline
\end{tabular}
\tablefoot{ 
Epochs given in JD - 2400000.0. 
$^\dagger$We tested several initial guesses for $P_c$, 10, 18.8, 20, 30 (increasing the $P_c$ interval to 50) days.
The final $P_c$ converged in all the cases to the 18.75 to 18.78 days range.
}
\label{Radvel_2planet_model_priors} 
\end{center}
\end{table}

\begin{table}
\begin{center}
\caption{\texttt{RadVel} model comparison table of the one and two-planet models for the SPIRou wPCA data set.}
\small
\begin{tabular}{llrrrrrrrr}
\hline
\hline
\\[-5pt]
& \bf{$N_{\rm free}$} & {rms} & {BIC} & {AICc}  & {$\Delta$AICc}\\
\\[-5pt]
\hline
\\
\multicolumn{5}{l}{\bf SPIRou wPCA RVs }  \\ 
\\[-8pt]
\hline
\\[-5pt]
  \multicolumn{2}{l}{AICc Favored Model} \\
 $e_{b}$, $K_{b}$, $K_{c}$, {$\sigma$}, {$\gamma$} 		& 10 	&  5.3   	& 1018 	& 988.5  	& 0.0 \\[5pt]
  \multicolumn{2}{l}{Nearly Indistinguishable} \\
 $K_{b}$, $K_{c}$, {$\sigma$}, {$\gamma$} 				& 8 	&  5.4   	& 1013	& 989.8  	& 1.3 \\[5pt]
  \multicolumn{2}{l}{Somewhat Disfavored}\\
  $e_{b}$, $K_{b}$, $e_{c}$, $K_{c}$, {$\sigma$}, {$\gamma$}  & 12  & 5.2    	& 1026 	& 991.6  	& 3.0 \\[1pt]
  $K_{b}$, $e_{c}$, $K_{c}$, {$\sigma$}, {$\gamma$} 		& 10 & 5.3    	& 1021 	& 992.4 	& 3.8 \\[5pt]
  \multicolumn{2}{l}{Ruled Out} \\ 
   {$\sigma$}, {$\gamma$} 							& 2  & 6.8 	    	& 2422 	& 2416 	& 1427 \\[1pt]
   $e_{b}$, $K_{b}$, {$\sigma$}, {$\gamma$} 				& 7  & 20   	& 2447 	& 2426 	& 1438 \\[1pt]
   $e_{c}$, $K_{c}$, {$\sigma$}, {$\gamma$} 				& 7  & 49   	& 2448 	& 2427 	& 1439 \\[1pt]
   $K_{b}$, {$\sigma$}, {$\gamma$} 						& 5  & 346   	& 2482 	& 2467 	& 1479 \\[1pt]
   $K_{c}$, {$\sigma$}, {$\gamma$} 						& 5  & 352    	& 2583 	& 2469 	& 1480 \\[1pt]
\hline
\end{tabular}
\label{tab_comp_2planet_models} 
\end{center}
\end{table}

\begin{figure}
\centering
\includegraphics[width=0.5\textwidth]{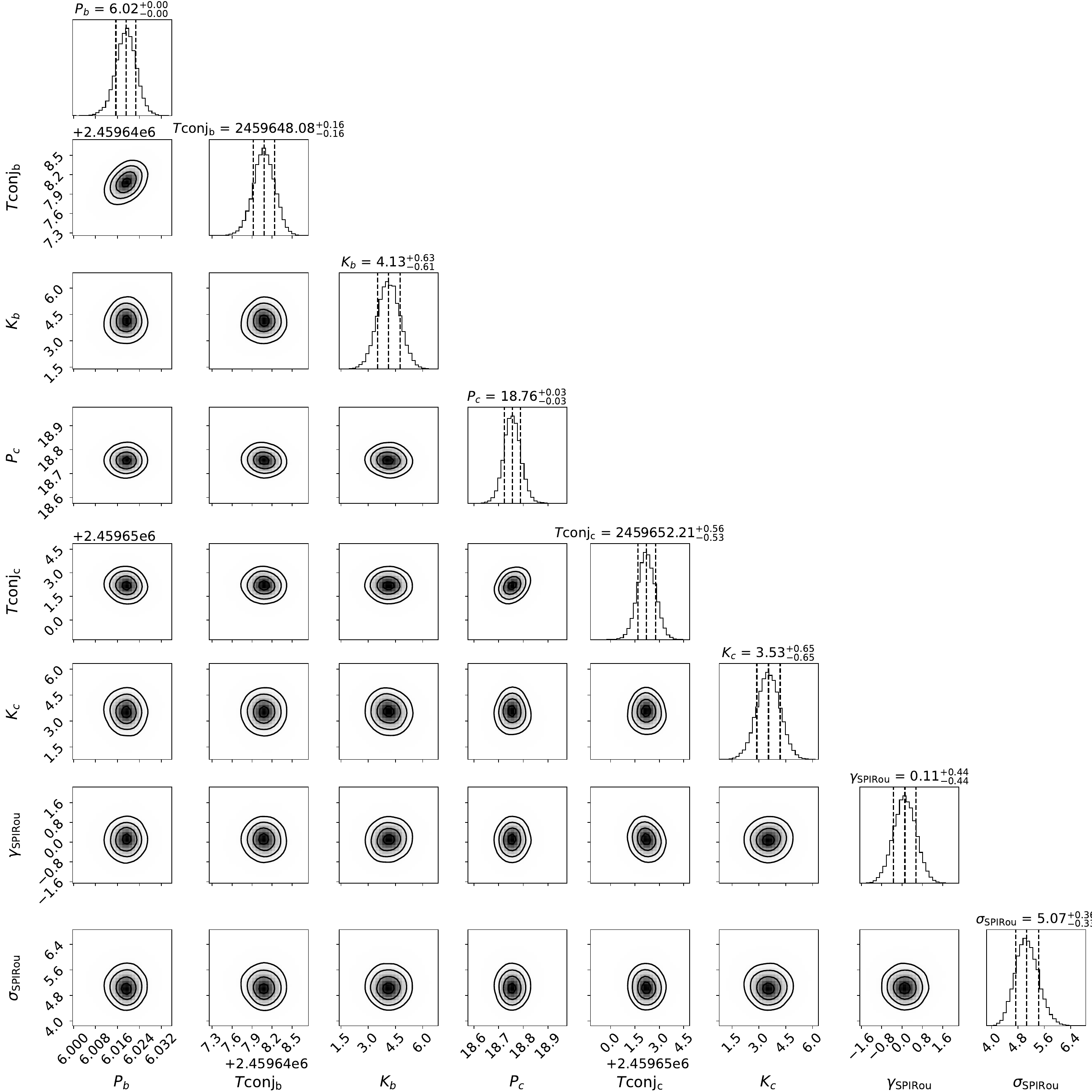}
\includegraphics[width=0.5\textwidth]{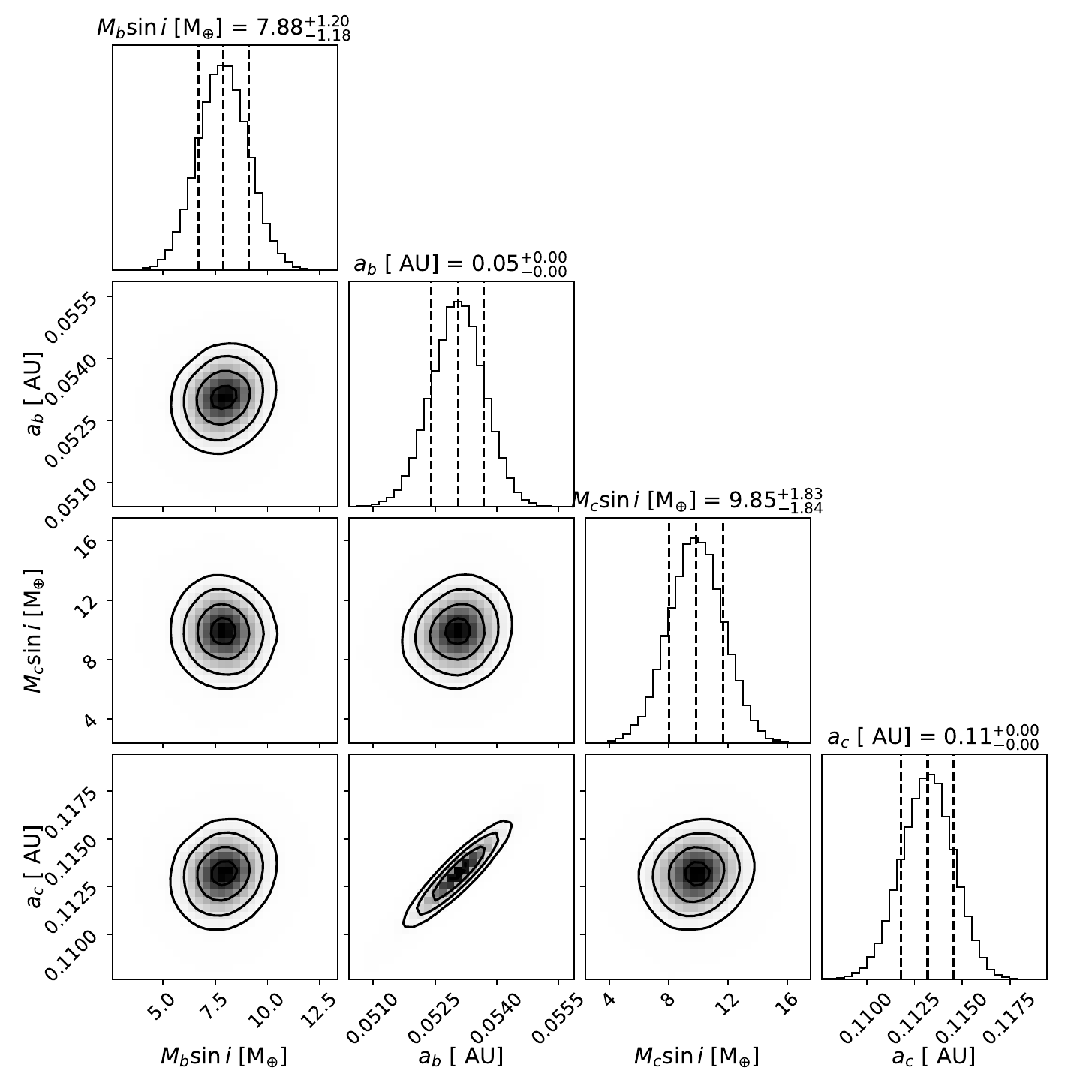}
\caption{Corner plots of the posterior distributions of the  \texttt{RadVel} MCMC two-planet circular orbital model of the SPIRou wPCA RV time series.}
\label{SPIRou_corner_plot_radvel_wPCA_2planet}
\end{figure} 

\clearpage

\begin{table}
\small
\begin{center}
\caption{Orbital parameters and planet physical properties derived 
from a \texttt{RadVel} MCMC two-planet in circular orbit model of the SPIRou wPCA data.}
\begin{tabular}{lcc}
\hline
\hline
\\[-5pt]
\multicolumn{1}{l}{SPIRou wPCA } 								& \multicolumn{1}{c}{Gl~410b}				& \multicolumn{1}{c}{candidate signal at 18.7 days}\\[5pt]
\hline
\\[-5pt]
\multicolumn{1}{l}{\bf Orbital parameters}\\[3pt]
$P$	[days] 						&	$6.019\pm0.004$		&	$18.76\pm0.03$ 	\\[3pt]
$K$	[m s$^{-1}$] 					& 	$4.1\pm0.6$			&	$3.5\pm0.7$	 	\\[3pt]
$T{\rm peri} $ [JD]  					&	$59646.57\pm0.16$		&	$59647.52\pm0.55$	\\[3pt]
\\[1pt]
\multicolumn{1}{l}{\bf Planet parameters}\\[1mm]
$M\sin (i)$  [M$_{\oplus}$]  			& 	$7.9\pm1.2$			& 	$9.8\pm1.8$		\\[3pt]
$a$ [au]				     			& 	$0.0531\pm0.0007$ 		& 	$0.113\pm0.001$ 	\\[3pt]
\hline
\end{tabular}
\tablefoot{ 
Epochs given in JD - 2400000.0. 
The value of the uncorrelated noise $\sigma_{\rm SPIRou}$ retrieved is 5.1$\pm$0.4 m s$^{-1}$.
}
\label{2planet}
\end{center}
\end{table}

\begin{figure}
\centering
\includegraphics[width=0.5\textwidth]{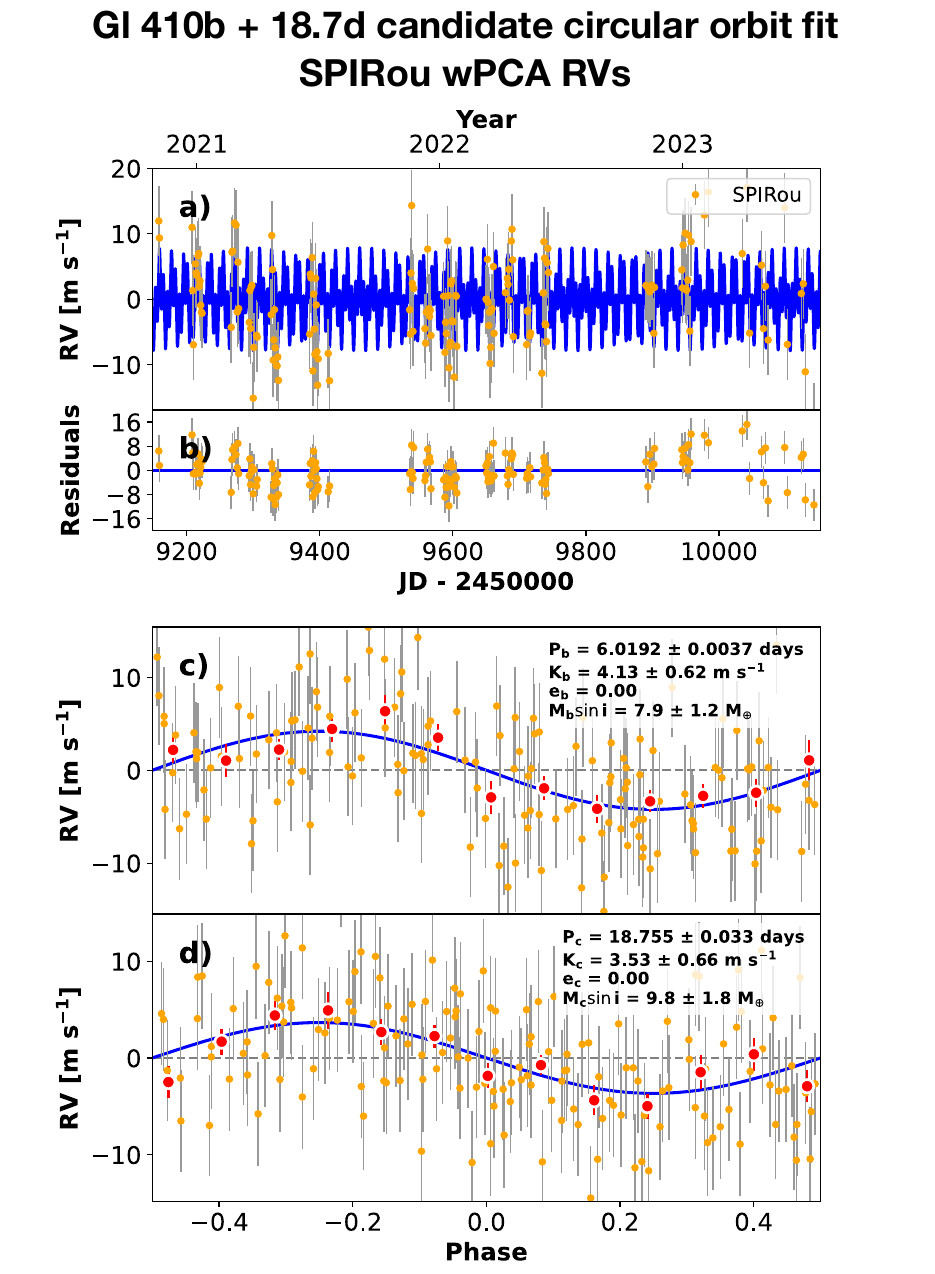}
\caption{\texttt{RadVel} best fitting two-planet in circular orbits model derived from the wPCA SPIRou RVs.
}
\label{2planet_radvel_wPCA}
\end{figure}

\section{Dynamical stability analysis}
\begin{figure}[h]
\centering
\includegraphics[width=0.4\textwidth]{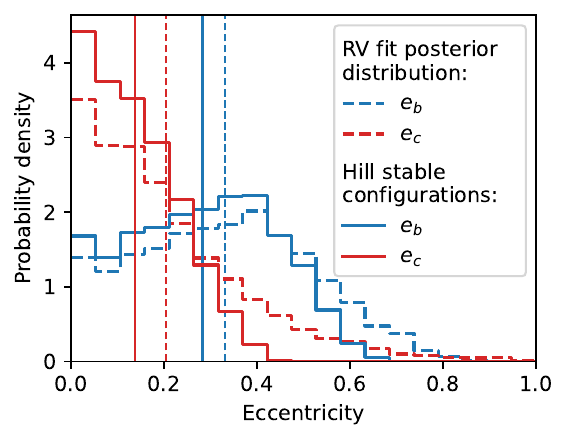}
\caption{Eccentricity distributions of Gl~410b and the 18.7-day candidate signal with the distributions for Hill stable configurations.
The vertical lines indicate the mean of the respective distributions.}
\label{dynamical_analysis}
\end{figure}

\label{dynamical_stability_analysis}
While somehow limited constraints on the orbital configuration can be obtained from the RV measurements, 
we performed a dynamical stability assessment for the two-planet system Gl~410b and 18.7-day candidate 
to consolidate the picture on the system architecture.
The period ratio (larger than 3) is wide and Gl~410 b and the 18.7-day candidate  are small enough 
that the system is safely stable for circular orbits and the nominal eccentricity values. 
Indeed the orbital separation in Hill radii is of $a_c-a_b = 23 \pm 1.5\ R_H$ which 
is far beyond the stability limit of $3.46~R_H$ \citep[][]{Gladman1993}.
Although we favor a model with circular orbits,
we can derive some constraints on the maximum stable eccentricities by computing the general Hill stability criterion as described in \citet[][]{Petit2018}.
We take the \texttt{RadVel} posterior and compute the Angular Momentum Deficit (AMD) of the system as well as the critical value ensuring Hill stability 
(in other words, the maximum AMD such that close encounters are always forbidden).
In Fig.~\ref{dynamical_analysis}, 
we compare the posterior eccentricity distribution resulting from the \texttt{RadVel} RV fit
and the posterior eccentricity distribution only keeping the stable configurations.
Rejecting Hill unstable configurations from the posterior, 
we obtain an upper limit for $e_c$ at 0.4 but no additional constraints on $e_b$.
Stable configurations are on average more circular.
A similar analysis on the masses show that no stringer constraints can be drawn from the stability on this system.

The Hill stability can also constrain the maximum mutual inclination. 
To obtain this constraint, we assume that the system invariant plan is aligned with the line of sight\footnote{As the main constraint comes from the mutual 
inclination and not the impact on the real planetary masses, this is a reasonable hypothesis.} and we 
assign a random mutual inclination to each configuration in the posterior. 
Computing the critical AMD in this case, we find that systems with a mutual inclination smaller than 40 degrees are stable.
As for the eccentricities, this is mainly an upper limit. 
The system is fully consistent with coplanar orbits.

\clearpage
\twocolumn

{ 
\section{Additional radial velocity analysis of the SPIRou wPCA data}
\subsection{$\ell_1$ periodogram analysis} 
\label{app:l1}

The $\ell_1$ periodogram has several inputs: the RV time series to be analyzed, 
its nominal uncertainties, a grid of frequencies $\boldsymbol{\omega}$, a linear model assumed 
to be in the data $\mathbf{M}_0$, an assumed covariance matrix $\mathbf{V}$ of the noise as input. 
Overall, the model of the RV time series $y(t)$ can be described by  the equation

\begin{equation}
   \boldsymbol{y}  = \mathbf{M}_0 \boldsymbol{x} +  \sum\limits_{k=1}^p A_k\cos(\omega_k \boldsymbol{t}) + B_k\sin(\omega_k \boldsymbol{t}) + \boldsymbol{\epsilon} \label{eq:l1model}
\end{equation}
where $\boldsymbol{\epsilon}$ is a Gaussian noise with a certain covariance. 
Overall, the free parameters of the models are the coefficients of the linear model, $\boldsymbol{x}$, 
the coefficients of periodic signals $A_k, B_k$, and the covariance parameters, which describe the properties of the noise. In practice, one chooses a 
parametrization of the covariance matrix of the noise $\mathbf{V}(\boldsymbol{\theta})$, where the element of $V$ at index $k,l$  depends on $|t_k-t_l|$ 
and a vector of parameters $\theta$. In the following analysis, the parametrization chosen for $\mathbf{V}$ is such that its element at index $k,l$ is
\begin{equation}
\begin{split}
V_{kl}(\theta )  =  &\delta_{k,l} (\sigma_{k}^2 + \sigma_{W}^2)  +   \sigma_{R}^2 \mathrm{e}^{-\frac{(t_k-t_l)^2}{\tau_R^2}}  \\
&+ \sigma_{QP}^2\mathrm{e}^{-\frac{(t_k-t_l)^2}{\tau_{QP}^2}} \frac{1}{2} \left(1 + \cos\left(\frac{2\pi(t_k-t_l)}{P_\mathrm{act}} \right)  \right)
\end{split}
\label{eq:kernel}
\end{equation}
where $\sigma_k$ is the nominal measurement uncertainty, $\sigma_W$ is an additional white noise, $\sigma_{C}$ is a calibration noise, and where $c$ 
equals one if measurements $k$ and $l$ are taken within the same night and zero otherwise. 
The quantities $\sigma_{R}$ and $\tau_R$ parametrize a correlated noise, which might originate from the star or the instrument. 
Additionally,  $\sigma_{QP}, \tau_{QP} $, and $P_\mathrm{act}$ parametrize a quasi-periodic covariance, potentially resulting from spots or faculae. 

The $\ell_1$ periodogram looks for a small number of frequencies such that the model in Eq.~\eqref{eq:l1model} approximates the data. 
Depending on the different assumptions on the noise model and base linear model, the $\ell_1$ periodogram might find signals at different frequencies. 
To test the robustness of the results to a change in the model, we adopt the methodology of~\citet{hara2020}. 
We consider a grid of potential noise models. 
For each of them, we computed the $\ell_1$ periodogram, and kept the signals that have a false alarm probability below 0.1. 
For each noise model in the grid, we then have candidate periodic signals. 
To each couple noise model/periodic signals, 
we attribute a cross-validation score. 
The model is fitted on 60\% of the data points (the training set), 
and the likelihood of its prediction on the remaining 40\% (the test set) is computed. 
This operation is repeated for 300 random draws of selection of training sets, 
and the final cross-validation score is computed as the average of the likelihood of the test set. 
We also rank the model according to their Bayesian evidence, evaluated with the Laplace approximation~\citep[see][]{hara2020, nelson2020}. 

\begin{figure}
    \centering
    \includegraphics[width=0.5\textwidth]{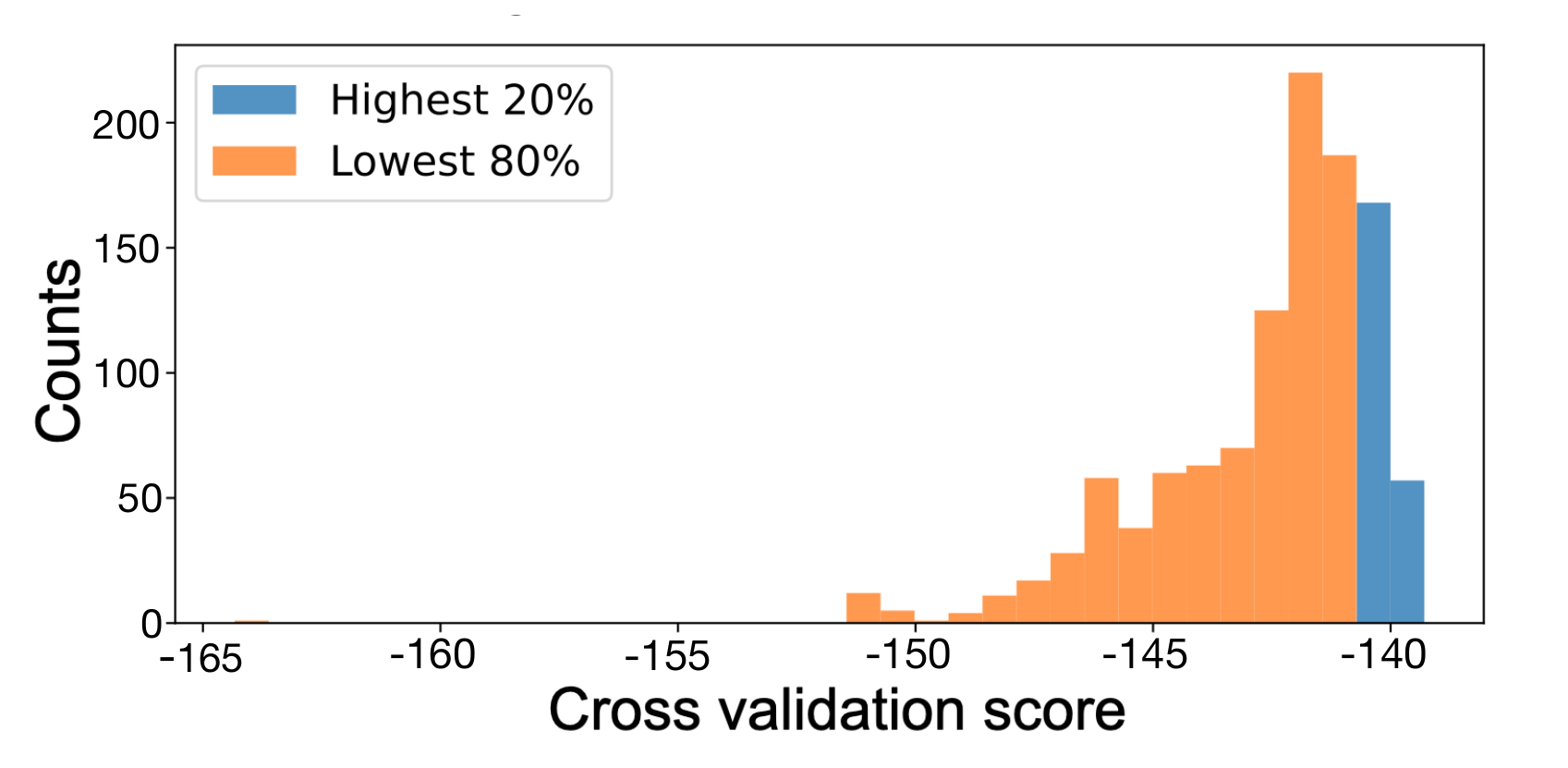}
    \caption{Histogram of the cross-validation score of the tested models. The blue histogram represents the counts of the 20\% highest models. }
    \label{fig:histogram_cv}
\end{figure}

\begin{figure}
    \centering
    \includegraphics[width=\linewidth]{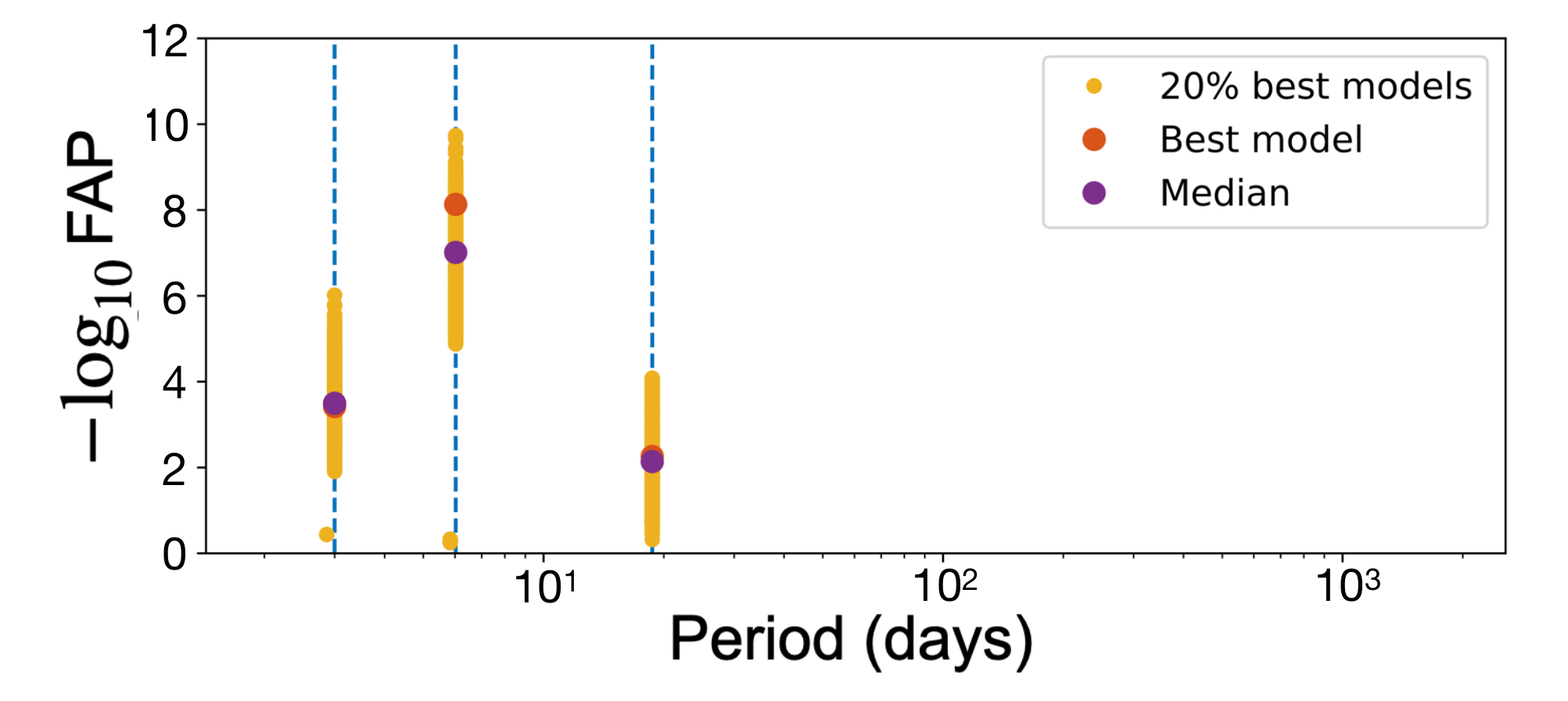}
    \caption{False alarm probability (FAP) of the signals appearing in the 20\% of models with the highest Bayesian evidence (Laplace approximation). 
    Each yellow point represents the FAP of a signal included in the model. Purple points represent the median values of the FAPS, and the red ones indicate the FAP of the highest ranked model.   }
    \label{fig:cvfaps}
\end{figure}

In our analysis, we adopt as $\mathbf{M}_0$ an eight-column matrix. 
Its columns are: 
an offset,
the spectral shape indicators (i.e., stellar activity proxies): 
the SPIRou longitudinal magnetic field ($B_\ell$),
and the LBL-derived~\citep{Artigau2022, Artigau2024}
projection onto the 2nd derivative of the spectrum (\texttt{d2v}), 
projection onto the 3rd derivative of the spectrum (\texttt{d3v}), 
differential line width (\texttt{dLW}), 
and differential temperature with respect to a  T=4000 K template (\texttt{dTEMP}$_{4000}$) 
and, 
finally, 
a cosine and sine term at the sidereal day (period equal to 0.997 days). 
The grid of values chosen for the noise model is given in Table~\ref{tab:cv}. 
All the possible combinations of the values of its components were generated, 
and the corresponding covariance matrices were created according to Eq.~\eqref{eq:kernel}. In~\ref{fig:histogram_cv}, 
we represent the histogram of cross-validation scores. 

The $\ell_1$ periodogram corresponding to the highest ranked model is represented in Fig.~\ref{fig:l1_highestCV}. It shows four clear peaks at 6, 3, 18 and 1.3 days. 
To evaluate the robustness of these detections to the noise model, for each the 20\% highest ranked models, 
we report the false alarm probabilities of the peaks found. 
The false alarms of the highest ranking models are reported in Fig.~\ref{fig:cvfaps}. 
The percentage of cases where the 6, 3, 18 and 1.3 days signals were found in the 20\% best models, 
the FAPs corresponding to the highest ranked model, and the median FAPs are reported in Table~\ref{tab:faps_cv}. 
We find that the 6, 3, 18 days signal are present in a 100\% of the highest ranked models, with median false alarm, 
probabilities of $7.5\times10^{-8}$, $1.5\times10^{-4}$ and $5.6\times10^{-2}$. 
The highest ranked models by Bayesian evidence give a similar picture, although they seem to provide more statistical significance for the signals 
(see Table~\ref{tab:faps_laplace}). They can be considered respectively as a confirmed planet, a very strong candidate, and a strong candidate.

\begin{table} 	
\caption{Inclusion of periodicities in the $\ell_1$ periodogram}
\centering
\begin{tabular}{p{1cm}|p{1.5cm}| p{1.5cm}|p{1.5cm}} 
Period (d) & FAP (best fit) & Inclusion in the model & $\mathrm{CV}_{20}$ median FAP \\ 
\hline
\hline
2.997& $1.54\times10^{-4}$ & 100.0\% & $1.19\times10^{-3}$\\ 
6.021& $7.47\times10^{-8}$ & 100.0\% & $1.60\times10^{-7}$\\ 
18.70& $5.64\times10^{-2}$ & 100.0\% & $3.98\times10^{-3}$\\ 
\label{tab:faps_cv}
\end{tabular}
\end{table}

\begin{table} 	
\caption{Inclusion of periodicities in the $\ell_1$ periodogram. The models are ranked by Bayesian evidence.} 
\centering
\begin{tabular}{p{1cm}|p{1.5cm}|p{1.5cm}|p{1.5cm}} 
Period (d) & FAP (best fit) & Inclusion in the model & $\mathrm{BE}_{20}$ median FAP \\ 
\hline \hline 
2.997& $3.81\times10^{-4}$ & 100.0\% & $3.18\times10^{-4}$\\ 
6.021& $7.42\times10^{-9}$ & 100.0\% & $9.78\times10^{-8}$\\ 
18.70& $5.53\times10^{-3}$ & 100.0\% & $7.31\times10^{-3}$\\ 
\end{tabular}\label{tab:faps_laplace} \end{table}

        \begin{table}
                \caption{Value of the parameters used to define the grid of models tested. }
                \label{tab:cv}
                \centering
                \begin{tabular}{p{1.5cm}|p{3.1cm}|p{1.cm}|p{1.cm}}
                        Parameter  & Values & Highest CV score & Highest evidence\\   \hline  \hline
                        $\sigma_W$ (m/s) &1, 1.5, 2, 2.5, 3, 3.5, 4  & 2.5  & 2  \\
                        $\sigma_{R}$ (m/s) & 0, 1 ,1.5, 2 ,2.5, 3, 3.5, 4  & 3 & 2.5 \\ 
                        $\tau_R$ (d)& 0, 3, 6  &  6  & 6 \\
                        $\sigma_{QP}$ (m/s)& 0, 1 ,1.5, 2 ,2.5, 3, 3.5, 4  &  3   & 2.5 \\ 
                        $\tau_{QP}$ (d) & 13.91, 27.82, 41.73, 55.64 &  13.91  &  27.82  \\
                        $P_\mathrm{act}$ 13.91 & 13.91 &  13.91 &  13.91 \\
                \end{tabular}
        \end{table}

\subsection{Apodized sine periodogram analysis}
\label{app:asp}

To further examine the properties of the signals, 
we apply the method of \cite{hara2022b}, 
which aim at testing whether the signals have a 
consistent amplitude, phase and frequency over time, 
which are the properties expected for planetary signals. 
For this we compute apodized sine periodograms (ASPs), defined as follows. 

We compared the $\chi^2$ of a linear base model $H$, and the $\chi^2$ of a model $K(\omega, t_0,\tau)$ 
defined as the linear model of $H$ plus an apodized sinusoid $e^{-\frac{(t-t_0)^2}{2\tau^2}}  (A \cos \omega t + B \sin \omega t )$.
The ASP is defined as the $\chi^2$ difference between the two models
\begin{align}
	z(\omega, t_0,\tau) &= \chi^2_H - \chi^2_{K(\omega, t_0,\tau)}.\label{eq:z}
\end{align}
 We used the same base model as in our analysis with the $\ell_1$ periodogram, 
 and used the covariance that corresponds to the noise model with highest Bayesian evidence. 
 
 To compute the ASP, 
 we consider a grid of timescales $\tau = 10\cdot T_{obs}, T_{obs}/3, T_{obs}/9$ and $T_{obs}/27$, where $T_{obs}$ is the total observation timespan of the data. 
 In Fig. \ref{fig:ASP}.a, left panel, we represent $z(\omega, t_0,\tau)$  as defined in Eq.~\eqref{eq:z}, maximized over $t_0$ for the values of $\tau$ in the grid. 
 In other words, we always select the center of the time window which best fits the data. 

We searched iteratively for signals. 
Once the maximum peak of the ASP was found, 
we then included the corresponding model in the base linear model $H$, 
and repeated the process. 
The four first iterations are shown in Fig.~\ref{fig:ASP}, a, b, c, and d respectively. 
The left panels represent the periodograms, the middle panels a zoom in on the highest peak, 
 and the right hand panel represents a statistical significance test, explained below. 
 
In Fig.~\ref{fig:ASP}.a, the dominating peak has a period of 6.02 days. 
We want to test whether the signal is statistically compatible with a constant one (i.e., with $\tau = 10\cdot T_{obs}$). 
Denoting by $t_{(\tau,\omega)}$ the value of $t_0$ maximizing the value of the periodogram~\eqref{eq:z} for a given $\omega$ and $\tau$,  
we compute the distribution of 
\begin{align}
   D_z = z(\omega, t_{(\tau,\omega)}, \tau) - z(\omega, t_{(\tau^\prime,\omega)}, \tau^\prime) 
   \label{eq:dz}
\end{align}
 with the hypothesis that the model $K(\omega, t_{(\tau,\omega)}, \tau, A^\star, B^\star )$ is correct, 
 where the fitted cosine and sine amplitudes $A^\star, B^\star$ are obtained by fitting model $K$ to the data. $D_z$ 
 can easily be expressed as a generalized $\chi^2$ distribution, 
 with mean and variance is given by an analytical expression \citep{hara2022b}.

 We find that even if the best fitting model for the 6.02 signal is not the purely periodic signal, 
 it is still compatible with a purely periodic one. 
 In Fig.~\ref{fig:model_as}.b and c we see that the candidate signals at 2.99 and 18 days are compatible with strictly sinusoidal signals. 
 In the case of the former, other signal shapes are excluded.

 In the next iteration, in Fig.~\ref{fig:model_as}.d, 
 we find that the best fitting model is a short term signal at 5.78 days. In Fig.~\ref{fig:model_as}, we show the corresponding model, which points to a transient signal. possibly caused by outliers or stellar variability.

Since the \texttt{d3v} vector, used in our linear base model $\mathbf{M}_0$, 
has shown in other SPIRou data set to be less reliable\footnote{https://lbl.exoplanets.ca/output-data} 
we re-did the whole analysis with the $\ell_1$ and ASP periodograms without it. 
The only notable difference is that it seems to slightly decrease the stability of the 18.7 day signal in frequency, amplitude and phase. 
We note that the 2.99 day planet could be in 2:1 mean motion resonance with the 6 d planet. 
In \cite{hara2019}, it has been shown that eccentric planets can mimic 2:1 mean motion resonances. 
However, it is very unlikely the case here, as the 2.99 d signal is significant and sharp. 
In conclusion, both 18.7 and 2.99 d planets are viable candidates, 
but the evidence for the latter is stronger. 

\begin{figure}
    \centering
    \includegraphics[width=0.5\textwidth]{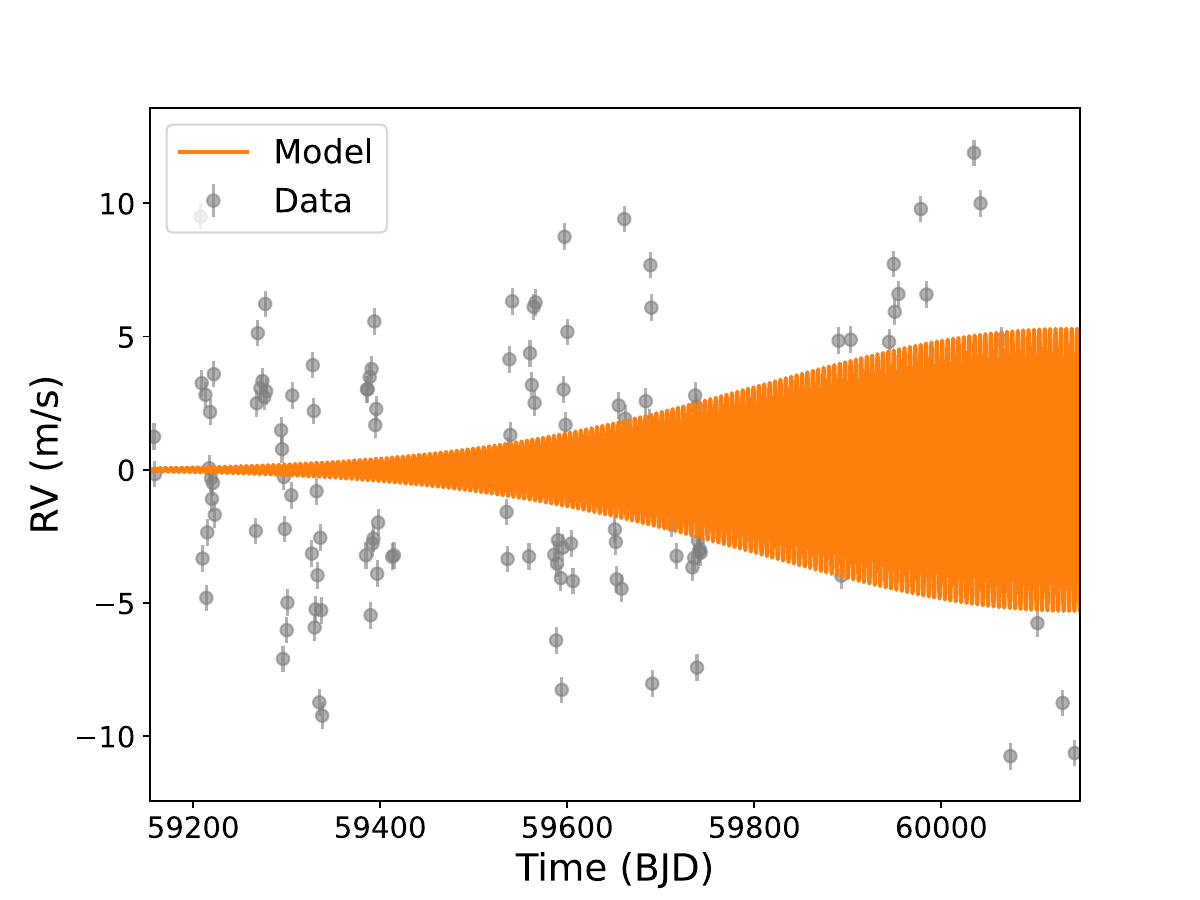}
    \caption{Best-fitting apodized sine model after three iterations.  }
    \label{fig:model_as}
\end{figure}
}

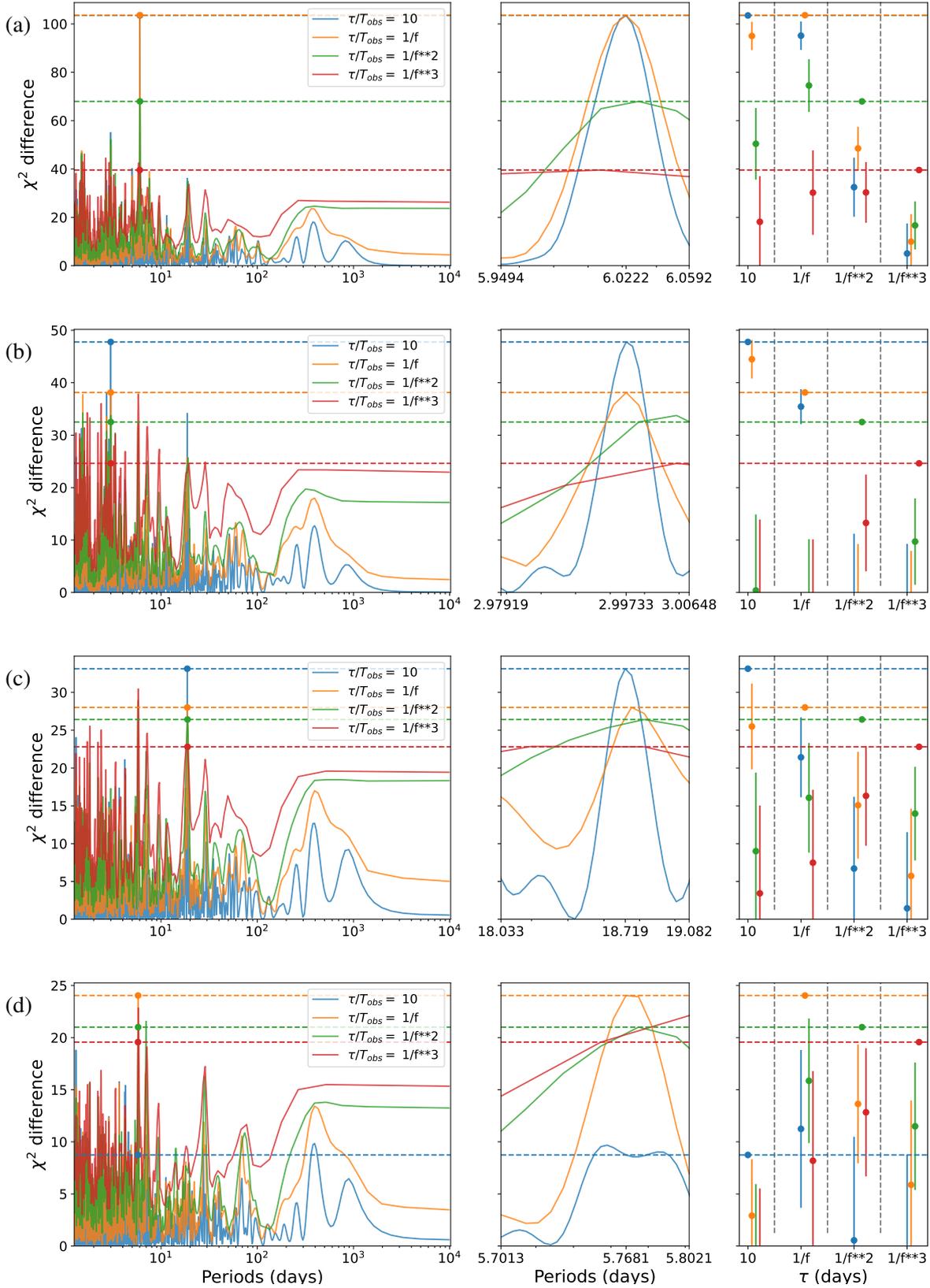
\begin{figure*}
	\noindent
	\centering
	\hspace{-1,5cm}
	\begin{tikzpicture}		
\path (0,0) node[above right]{\includegraphics[width=\linewidth]{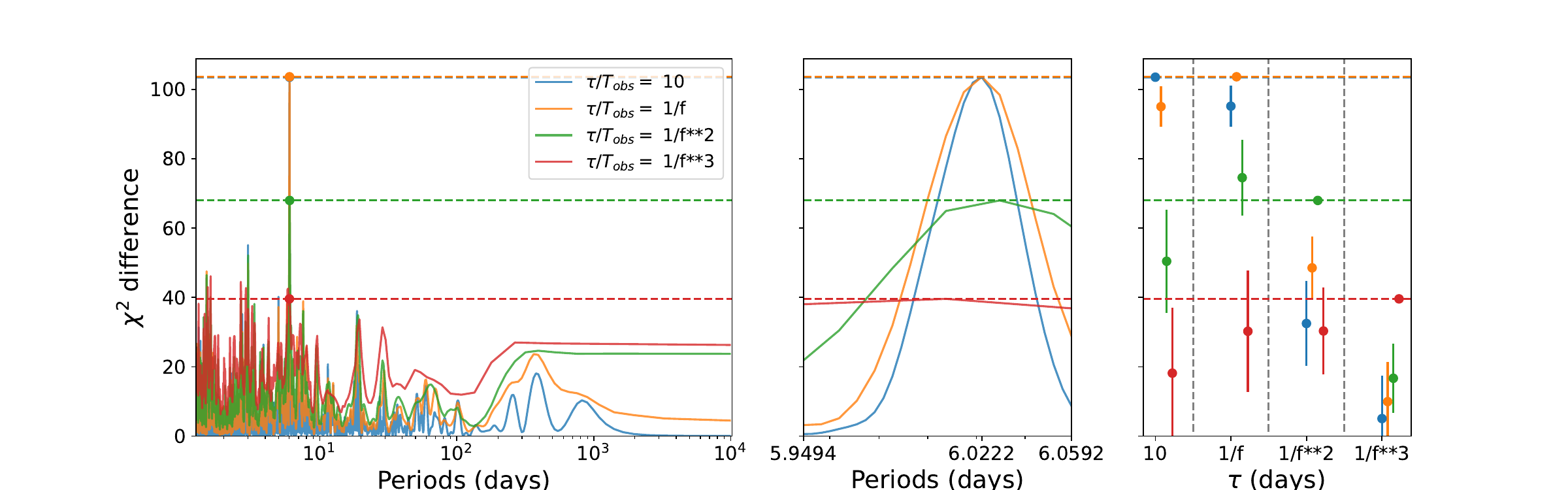}};
	\path (1.15,4.5) node[above right]{\large(a)};
	\begin{scope}[yshift=-5.5cm]
\path (0,0) node[above right]{\includegraphics[width=\linewidth]{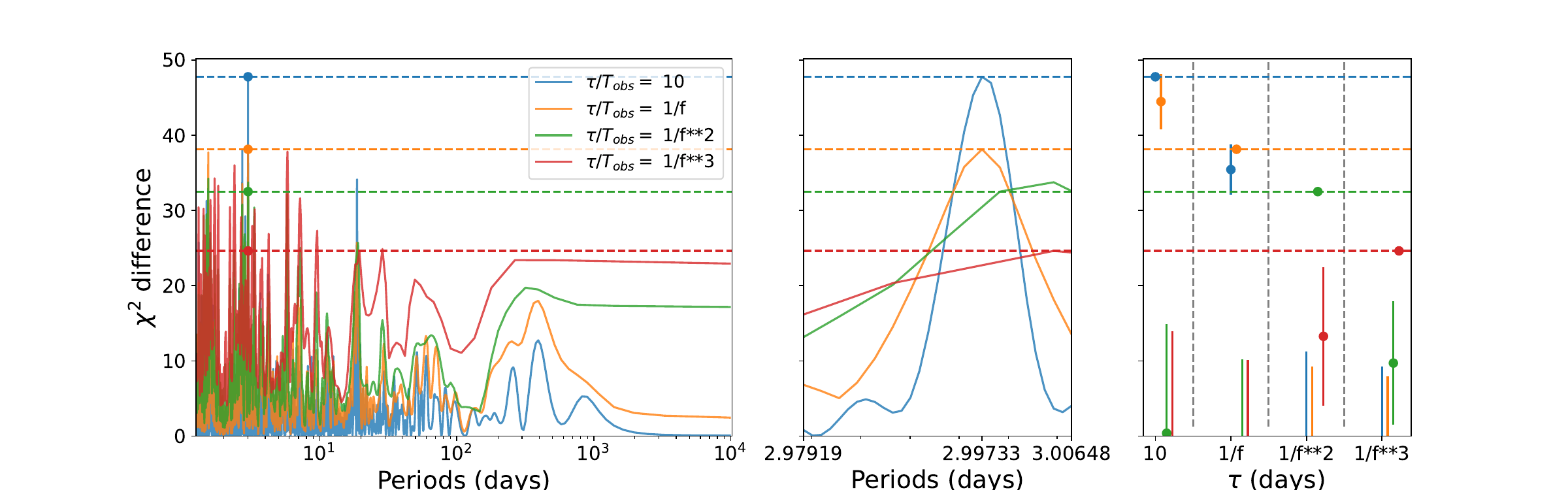}};
	\path (1.15,4.5) node[above right]{\large(b)};
	\end{scope}
	\begin{scope}[yshift=-11cm]
\path (0,0) node[above right]{\includegraphics[width=\linewidth]{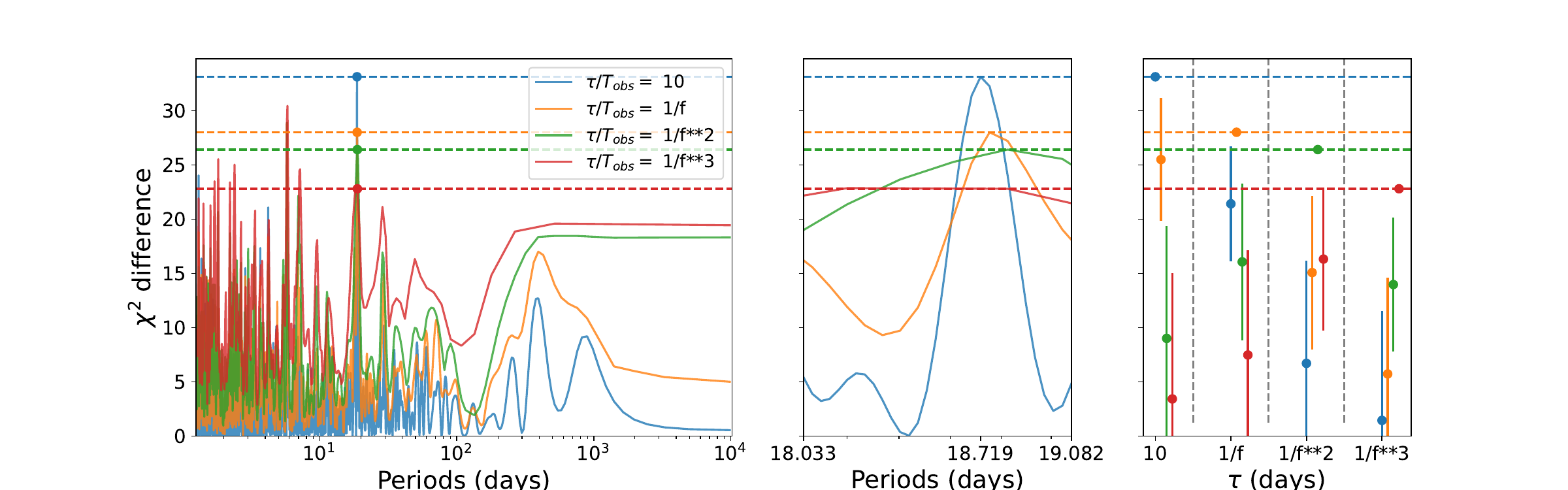}};
	\path (1.15,4.5) node[above right]{\large(c)};
	\end{scope}
	\begin{scope}[yshift=-16.5cm]
\path (0,0) node[above right]{\includegraphics[width=\linewidth]{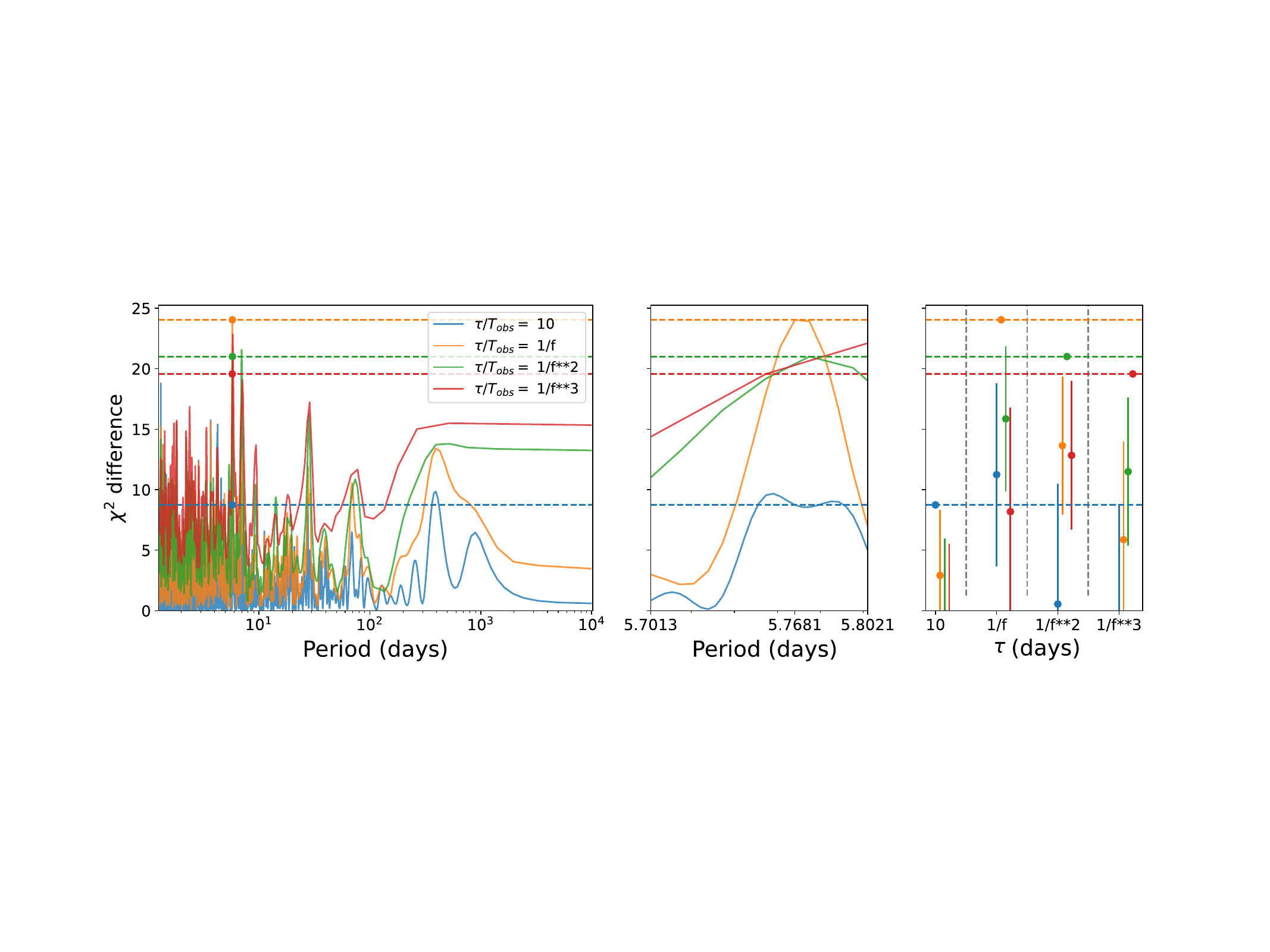}};
	\path (1.15,4.5) node[above right]{\large(d)};
	\end{scope}
	\end{tikzpicture}
	\caption{Four first iterations of the ASPs method. Models correspond to the maximum of the ASPs. 
	The left panels represent the periodograms, the middle panels a zoom in on the highest peak, 
 	and the right hand panel represents a statistical significance test (see text for details).}
	\label{fig:ASP}
\end{figure*}

\end{appendix}

\end{document}